\documentclass[aps,prb,reprint,groupedaddress,amsmath,amssymb,longbibliography]{revtex4-2}

\usepackage{graphicx} 
\usepackage{bm}
\usepackage[pdfstartview=FitH, CJKbookmarks=true, bookmarksnumbered=true, bookmarksopen=true, colorlinks=true, pdfborder=0 0 1, citecolor=blue, linkcolor=blue, linktocpage=true] {hyperref}
\usepackage{epstopdf} 

\begin{document}

\title{One-dimensional $\mathbb{Z}$-classified topological crystalline insulator under space-time inversion symmetry}

\author{Ling Lin$^{1,2}$}
\email{Email: linling@szu.edu.cn}

\author{Yongguan Ke$^{1,2}$}
\email{Email: keyg@szu.edu.cn}

\author{Chaohong Lee$^{1,2}$}
\email{Email: chleecn@szu.edu.cn, chleecn@gmail.com}

\affiliation{$^{1}$Institute of Quantum Precision Measurement, State Key Laboratory of Radio Frequency Heterogeneous Integration, College of Physics and Optoelectronic Engineering, Shenzhen University, Shenzhen 518060, China}
\affiliation{$^{2}$Quantum Science Center of Guangdong-Hong Kong-Macao Greater Bay Area (Guangdong), Shenzhen 518045, China}

\date{\today}

\begin{abstract}
We explore a large family of one-dimensional (1D) topological crystalline insulators (TCIs) classified by $\mathbb{Z}$ invariants protected by space-time inversion symmetry.
This finding stands in marked contrast to the conventional classification of 1D band topology protected by inversion symmetry and characterized by $\mathbb{Z}_2$-quantized polarization (Berry-Zak phase). 
Such kind of enriched topological phases relies on imposing restriction on tunneling forms.
By considering the nontrivial relative polarization among sublattices (orbitals), we introduce the inversion winding number as a topological invariant for characterizing and categorizing band topology.
The bulk-edge correspondence with regard to the inversion winding number is discussed.
%
Leveraging real-space analysis, we discover disorder-induced topological Anderson insulators and propose to experimentally distinguish band topology through relative polarization of edge states or bulk states. 
Our comprehensive findings present a paradigmatic illustration for the ongoing investigation and classification of band topology in TCIs.
\end{abstract}

\maketitle

\section{Introduction}
Crystalline symmetry plays an important role in classifying topological insulators~\citep{PhysRevLett.106.106802, slager2013space, PhysRevX.7.041069,  bradlyn2017topological, po2017symmetry,PhysRevB.76.045302, PhysRevB.99.075105,wu2019non}, complementary to the three intrinsic symmetries (i.e. chiral, time-reversal, and particle-hole symmetries).
Tremendous topological states protected by crystalline symmetries, known as topological crystalline insulators (TCIs), have been explored~\cite{slager2013space,SLAGER201924,PhysRevX.7.041069, PhysRevLett.106.106802, bradlyn2017topological,po2017symmetry, PhysRevLett.125.053601, PhysRevB.76.045302, PhysRevB.99.075105,wu2019non,ando2015topological, PhysRevLett.133.093404}.
In recent years, systematic search for TCIs has been achieved by analyzing high-symmetry points in momentum space and chemical orbitals in real space~\cite{slager2013space, PhysRevX.7.041069, PhysRevB.100.195135, bradlyn2017topological,po2017symmetry}.
TCIs exhibit nontrivial topological properties such as higher-order topology~\cite{Feng2017,benalcazar2017quantized,PhysRevB.96.245115}, non-Abelian topological charges~\citep{wu2019non, bouhon2020non,PhysRevLett.125.053601, jiang2021experimental, guo2021experimental,peng2022phonons, li2023floquet,PhysRevA.110.023321,slager2024non}, and period-multiplied Bloch oscillations~\cite{Holler2018,di2020non}.
These properties show that crystalline symmetries introduce richer topological structures, and may have promising applications in spintronics, quantum computing, and novel electronic devices.

Among crystalline symmetries, the inversion symmetry combined with time-reversal symmetry, dubbed the space-time inversion symmetry or parity-time symmetry, is of fundamental importance in studying TCIs.
It introduces constraints on the band topology and leads to quantization of topological invariants~\citep{PhysRevB.83.245132,PhysRevB.83.245132,  PhysRevB.89.155114, PhysRevB.90.165114, PhysRevX.4.021017, PhysRevB.100.205126, PhysRevB.108.085116}.
Specifically, in one dimension (1D), given a gapped band, the polarization $p$ related to the Berry-Zak phase is widely used as a $\mathbb{Z}_2$-quantized invariant to characterize different 1D TCIs~\citep{slager2013space, PhysRevX.7.041069,PhysRevLett.45.1025,PhysRevB.23.2824, PhysRevLett.48.359,PhysRevLett.62.2747,PhysRevB.100.205126, Song_sciadv_2018, PhysRevResearch.2.023277}, though it may fail to predict the bulk-edge correspondence \citep{PhysRevB.97.115143}.
For composite multiple bands, one can utilize the inversion eigenvalues at high-symmetry points or the Wilson loop to identify a non-negative integer invariant $\mathbb{Z}^{\geq}$~\citep{ PhysRevB.83.245132, PhysRevB.89.155114,PhysRevB.96.245115,PhysRevResearch.2.013300}.
In the presence of chiral symmetry, it is well known that the band topology in 1D transcends $\mathbb{Z}_2$ classification and falls into the $\mathbb{Z}$ classification, which can be characterized by the chiral winding number \citep{RevModPhys.88.035005, Ryu2010}.
Notably, it has been discovered that the chiral winding number can be equivalently expressed as the polarization difference between the two sublattices (or orbitals)~\citep{PhysRevLett.113.046802, Science362, PhysRevB.103.224208}, offering a more fundamentally physical perspective on comprehending 1D chiral-symmetric topological insulators.
The number of in-gap zero modes is proportional to the chiral winding number, meaning that nontrivial relative polarization between sublattices can give rise to nontrivial boundary states.
Therefore, in the absence of chiral symmetry, a natural question arises: \textit{Can nontrivial relative inter-sublattice polarization exist as a result of crystalline symmetry and further categorize the band topology in 1D}?

In this work, we present a theory demonstrating that, under space-time inversion symmetry, non-trivial topological structures can exist in a large family of 1D TCIs.
Upon adding restrictions on tunnelings, the band topology is turned from $\mathbb{Z}_2$ to $\mathbb{Z}$ classification.
The space-time inversion symmetry refers to the combination of inversion symmetry and time-reversal symmetry, which is expressed by
\begin{eqnarray}
    \label{eqn:inversion_relation}
{\cal I \cal T}H\left( k \right) {\left({{\cal I \cal T}}\right)^{ - 1}} = H\left( k \right),
\end{eqnarray}
where $k$ is the quasi momentum, $\mathcal{I}$ denotes the space-inversion operator and ${\cal T}$ represents the time-reversal operator.
The space-time inversion symmetry imposes constraints on the Bloch Hamiltonian and its eigenvectors locally within the momentum space.
This phenomenon brings internal topological structures to the Bloch states, enabling a substantial classification of band topology within 1D TCIs.
Our work gives insights into systematic exploration of TCIs with hidden topological structures in higher dimensions.
The article is organized in the following structure.
In Sec.~\ref{sec:three_band_model_and_inversion_windng_number}, we introduce a general form of three-band model and the Bloch state satisfying the space-time inversion symmetry.
Based on the symmetric structure and motivated by the winding number in chiral-symmetric systems, we define the inversion winding number to capture the relative polarization between sublattices.
We show that the inversion winding number is topologically stable when some specific tunnelings are ruled out.
The bulk-edge correspondence is also discussed.
In Sec.~\ref{sec:Robustness_winding_number}, we study the real-space formula of the inversion winding number and numerically show that the inversion winding number is robust and topologically stable stable against disorder given that the space-time inversion symmetry is not broken and specific tunnleing terms are not present.
We observe the disorder-induced topological transition and find that bulk-edge correspondence still persists.
Based on the real-space formula, we propose a scheme for experimental measurement of the inversion winding number.
In Sec.~\ref{sec:Summary}, we summarize our findings and discuss potential applications to other systems.
\section{Three-band model and inversion winding number}
\label{sec:three_band_model_and_inversion_windng_number}
We consider a simple but general three-band model with space-time inversion symmetry~\eqref{eqn:inversion_relation}. 
Its Hamiltonian matrix and inversion operator are given by
\begin{equation}
\label{eqn:Ham_k_3band}
H\left( k \right) = \left( {\begin{array}{*{20}{c}}
{{\Delta _{k}^{(1)}}}&{{g_k}}&{{h_k}}\\
{{g_k}^*}&{ {\Delta _{k}^{(2)}}}&{{g_k}}\\
{{h_k}^*}&{{g_k}^*}&{{\Delta _{k}^{(1)}}}
\end{array}} \right),  \;\;{\cal I} = \left( {\begin{array}{*{20}{c}}
{0}&{0}&1\\
{0}&1&{0}\\
1&{0}&{0}
\end{array}} \right),
\end{equation}
with $g_k, h_k \in \mathbb{C}$ being complex parameters and $\Delta_k^{(1,2)} \in\mathbb{R}$ being real parameter.
The time-reversal operation on the Bloch Hamiltonian here is the complex conjugation operation: ${\cal T}H(k){\cal T}^{-1} = {H(k)}^*$. 
Clearly, this model has no chiral symmetry.
Due to the space-time inversion symmetry, the eigenvector of $H(k)$ can be written as a general form
\begin{equation}
\label{eqn:eigenvector_3bands}
{\boldsymbol{u}_k} = {\left( {\begin{array}{*{20}{c}}
{{\alpha _k}{e^{ - i{\xi _k}}}}&{{\beta _k}{e^{ - i\frac{{{\xi _k}}}{2}}}}&{{\alpha _k}^*}
\end{array}} \right)^{\rm{T}}}
\end{equation}
with $\xi_k \in [0, 2\pi)$, $\alpha_k \in \mathbb{C}$ and $\beta_k \in \mathbb{R}$.
For simplicity, we have omitted the band index and assumed that the target band is gapped.
Under the inversion transformation, we have ${\left( {{\cal I}{\boldsymbol{u} _k}} \right)^*} = {e^{  i{\xi _k}}}{\boldsymbol{u} _k}$.
In particular, the first and third sublattices are related by a space-time inversion operation, and the phase factor associated with $\xi_k$ cannot be gauged out. 
This leads to the possible presence of nontrivial topological structures among the sublattices.

It is important to note that, in 1D chiral-symmetric topological insulators, the polarization difference between two sublattices can lead to nontrivial polarizations and charge accumulations at the edge \citep{PhysRevLett.113.046802, PhysRevB.103.224208, lin2024probing}. 
This phenomenon is associated with the chiral winding number belonging to the $\mathbb{Z}$ class.
In general, the single-gand topology in 1D TCIs without chiral symmetry fall into a $\mathbb{Z}_2$ classification characterized by the total polarization.
It is of interest to generalize the idea of polarization difference to the TCI with space-time inversion symmetry and study if there are more topological phases.
Thus, motivated by the chiral winding number, we define a gauge-invariant winding number via the sublattice polarization difference between the first and the third sublattice,
\begin{eqnarray}
\label{eqn:winding_number_3bands_momentum}
\nu  = \frac{1}{{\pi i}}\int_0^{2\pi } {dk\;\left[ {\frac{{\langle {u_k}|\frac{\partial }{{\partial k}}\left( {{{\hat P}_3} - {{\hat P}_1}} \right){u_k}\rangle }}{{\langle {u_k}|{{\hat P}_1} + {{\hat P}_3}|{u_k}\rangle }}} \right]} ,
\end{eqnarray}
where ${{\hat P}_\sigma } = \sum\nolimits_k {|k,\sigma \rangle \langle k,\sigma |} $ is the projector onto the $ \sigma = 1,2,3$ sublattice with $|k,\sigma\rangle$ being the basis of the momentum space, and ${\langle {u_k}|{{\hat P}_1}+{{\hat P}_3}|{u_k}\rangle }$ denotes the normalization factor.
Due to the inversion symmetry, one can find ${\langle {u_k}|{{\hat P}_1}|{u_k}\rangle } = {\langle {u_k}|{{\hat P}_3}|{u_k}\rangle }$.
This gauge-invariant quantity is quantized to integers $\nu \in \mathbb{Z}$ and it is directly related to the quantized polarization such that $p=\nu/2 \mod 1$ (see Appendix.~\ref{appendix:relation_between_inv_winding_and_polarization} for proof).
Alternatively, the inversion winding number can be equivalently expressed as a more familiar form
\begin{equation}
\label{eqn:winding_number_qk_form}
\nu  = \frac{1}{{2\pi i}}\int_0^{2\pi } {dk\;\left[ {q{{\left( k \right)}^{ - 1}}{\partial _k}q\left( k \right)} \right]} ,
\end{equation}
where $q\left( k \right) = \langle {u_k}|{{\cal P}_{1,3}}|{u_k}\rangle \in \mathbb{C}$, ${{\cal P}_{1,3}} = \sum\nolimits_k {|k,1\rangle \langle k,3|} $ reflects the phase difference between $\sigma=1,3$ sublattices.
It is worth noting that the inversion winding number shares some similarities with the mirror Chern number, where the difference of the Chern number between the two sectors having different mirror eigenvalues constitutes the mirror Chern number \citep{PhysRevB.78.045426, hsieh2012topological}.
In our study, the inversion winding number is obtained via projection onto the two inversion-symmetric sublattices.
We also note that Ref.\citep{PhysRevB.110.125424} proposed the sublattice winding number to capture topological nature in the absence of chiral symmetry recently.
A crucial difference is that the sublattice winding number is based on the point-gap topology, while we define the inversion winding number based on the space-time inversion symmetry and the sublattice polarization.

Below we discuss the topological stability of the defined inversion winding number~\eqref{eqn:winding_number_3bands_momentum}.
It can be found that the inversion winding number only becomes ill-defined when $\alpha_k=0$ in Eq.~\eqref{eqn:eigenvector_3bands}.
To change the value of the inversion winding number, the system must pass this singular point.
To find in what condition that $\alpha_k = 0$, let us focus on the eigenvalue equation $H(k) u_k  = E_k u_k$ using Eq.~\eqref{eqn:Ham_k_3band} and Eq.~\eqref{eqn:eigenvector_3bands}.
There is 
\begin{eqnarray}
    \Delta _k^{\left( 1 \right)}{\alpha _k}{e^{ - i{\xi _k}}} + {g_k}{\beta _k} + \Delta _k^{\left( 1 \right)}{\alpha _k}^* = {E_k}{\alpha _k}{e^{ - i{\xi _k}}} .
\end{eqnarray}
Let $\alpha_k = 0$, we immediately have $\beta_k = 1$ and $g_k = 0$.
Therefore, we find that $g_k = 0$ is necessary for $\alpha_k = 0$.
It can be concluded that, to change the inversion winding number, one must either break the space-time inversion symmetry, close the spectral gap, or introduce certain terms to realize $g_k = 0$.
Generally speaking, the topological protection in TCIs manifests that the target system cannot be adiabatically connected to other distinct TCI phases without closing the spectral gap and breaking the underlying symmetry \citep{PhysRevLett.106.106802, ando2015topological,PhysRevB.100.205126}.
In this sense, the inversion winding number here seems unstable, as it may possibly change through tuning $g_k = 0$.
Nevertheless, the inversion winding number is related to the polarization via $p=\nu/2 \mod 1$, see proof in Appendix.~\ref{appendix:relation_between_inv_winding_and_polarization}.
Strictly speaking, the parity of the inversion winding number is a $\mathbb{Z}_2$ topological invariant.
Although the inversion winding number becomes ill-defined when $\alpha_k = 0$, it is still possible to introduce perturbations to ensure that $g_k \neq 0$, thereby avoiding this singular point.

Intriguingly, the condition $g_k \neq 0$ exists in many realistic systems.
As the parameter ${g_k} \propto \sum\nolimits_{j \le j'} {t_{j,j'}^{\sigma ,\sigma '}{e^{ik\left( {j - j'} \right)}}}$ describes the coupling strength between $\sigma=\{1,3\}$ and $\sigma'=2$ sublattices in momentum space, one can find that $g_k = 0$ physically requires some very special tunneling terms in real space.
Here, $t_{j,j'}^{\sigma, \sigma'}$ represents the tunneling strength from the $\sigma$ sublattice in the $j$th cell to the $\sigma'$ sublattice in the $j'$th cell in real space.
There are two cases where $g_k = 0$ occurs under the space-time inversion symmetry. 
The first one is to simply set $g_k = 0$ for all $k \in [0,2\pi)$, making the sublattice $\sigma = 2$ totally isolated from other sublattices.
The second one is to introduces more than two different types of tunneling processes between $\sigma = {1,3}$ and $\sigma = 2$, which should include long-range tunneling across cells.
Otherwise, there must be $g_k \ne 0$ for all $k \in [0,2\pi)$.
Therefore, by imposing constraints that $g_k \neq 0$, the inversion winding number is topologically stable.
In this sense, the band topology falls into $\mathbb{Z}$ classification.
This argument remains valid for a large family of realistic systems, offering a more fruitful classification for space-time inversion-invariant systems.
Such kind of stability can resist the symmetry-preserving disorder.
Later, we will provide some evidences to support this point.

\begin{figure}
\includegraphics[width = 0.95\columnwidth ]{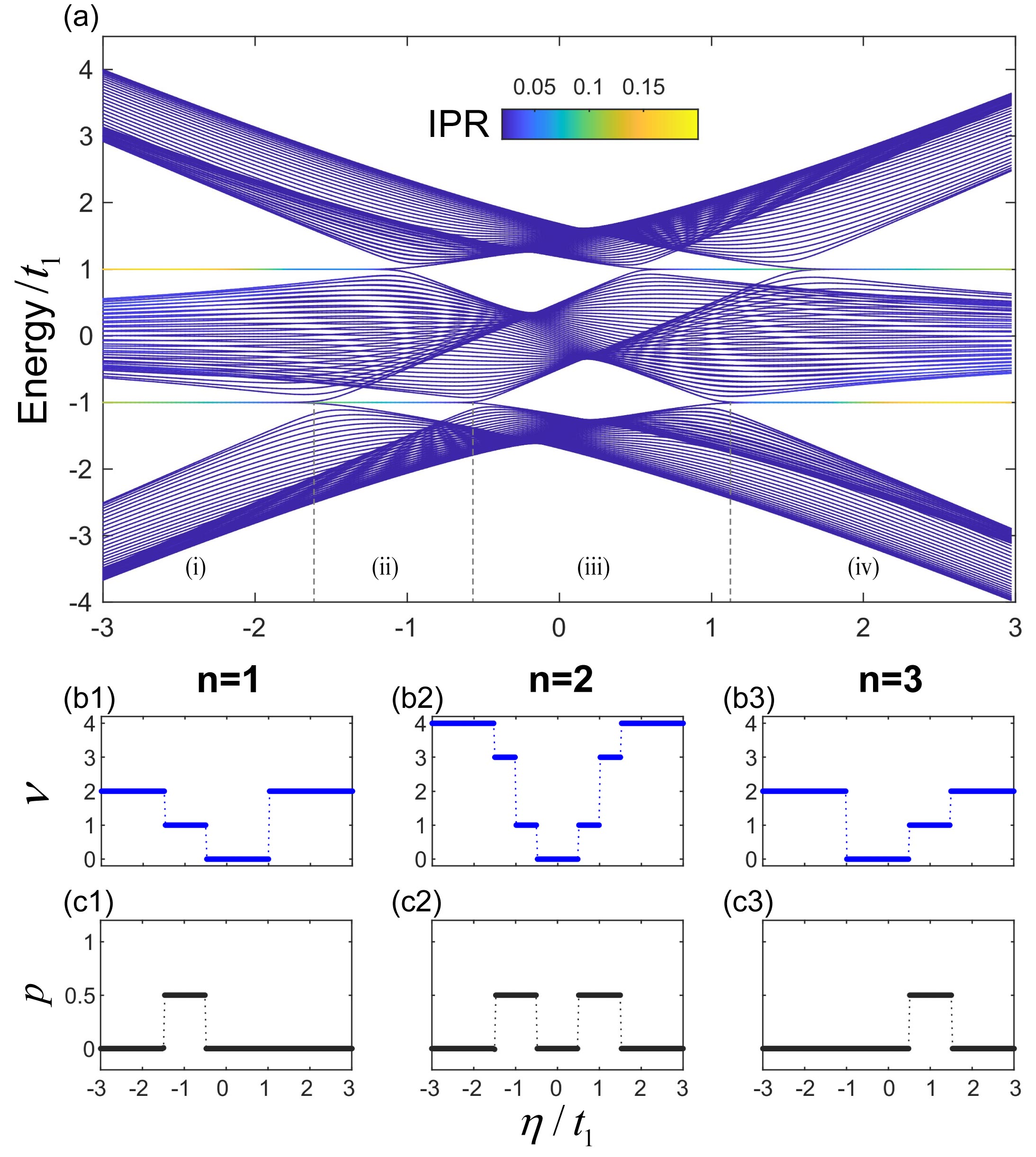}
\caption{\label{fig:FIG_three_band_spectrum}
(a) Energy spectrum under the OBC as a function of $\eta/t_1$.
The color of each line corresponds to the inverse participation ration of the eigenstate: $\mathrm{IPR} = \sum_{j,\sigma} |\langle \psi_n|j,\sigma\rangle|^4$, which reflects the localization property of an eigenstate.
Grey-dashed lines roughly indicate the gap-closing point between the first and second bands and divide into four regions (i-iv).
(b1-b3) Inversion winding number for the three bands ($n=1,2,3$) as functions of $\eta/t_1$ under PBC.
(c1-c3) Polarization of the three bands under PBC.
Other parameters are chosen as $t_2/t_1 = 1/2$.
}
\end{figure}

\subsection{Prototypical model}
To show the validity of the defined inversion winding number and demonstrate its $\mathbb{Z}$ classification, we choose $g_k = -t_1$, $h_k = -t_2e^{-ik} - \eta e^{-2ik}$ and $\Delta_k = 0$, which corresponds to a system with long-range tunneling in real space.
The real-space formulation of this model is demonstrated in Appendix.~\ref{appendix:model_and_experimental_realization}.
By varying $\eta/t_1$, the spectrum under open boundary condition (OBC) is shown in Fig.~\ref{fig:FIG_three_band_spectrum} (a), comprising three bands labeled as $n=1,2,3$ from the bottom to the top.
In both the first and second band gaps, several in-gap states appear.
These in-gap states correspond to edge states that are exponentially localized near the two boundaries.
To better elucidate the results, we concentrate on the lowest gap and designate four regions (i-iv) separated by gap-closing points and delve into its topological properties.
Intriguingly, there are three different phases.
We find that the in-gap edge states exhibit a four-fold degeneracy in regions (i) and (iv), and a two-fold degeneracy in region (ii).
In region (iii), there is no in-gap edge state.
This suggests that these three regions are three distinct phases.

Subsequently, we calculate the inversion winding number~\eqref{eqn:winding_number_3bands_momentum} under the periodic boundary condition (PBC), see Figs.~\ref{fig:FIG_three_band_spectrum} (b1-b3).
Remarkably, the inversion winding number identifies more distinct phases and its variation coincides with the locations of gap-closing points.
Under the OBC, the number of edge states closely relates to the inversion winding number of the adjacent bands.
{For instance, the numbers of in-gap edge states in the first and second gaps are respectively proportional to the inversion winding numbers of the first ($n=1$) and third ($n=3$) bands: $N_{{\rm{edge}}}^{\# 1} = 2|{\nu _1}|$ and $N_{{\rm{edge}}}^{\# 2} = 2|{\nu _3}|$, where $N_{{\rm{edge}}}^{\# m}$ and $\nu_n$ respectively denote the number of edge states in the $m$th gap and the inversion winding number of the $n$th band.}
Regarding the second band ($n=2$), its inversion winding number is the sum of the first and third bands: $\nu_{2} = \nu_{1} + \nu_{3}$, and the change of this inversion winding number requires the closing of either the first or the second band gap.
It can be further deduced that the nontrivial topology of the second band contributes to the emergence of in-gap edge states in both band gaps.

For comparison, we calculate the polarization for each band, as shown in Figs.~\ref{fig:FIG_three_band_spectrum} (c1-c3).
Generally, the emergence of in-gap edge states is expected as a result of non-zero band polarization, which can be calculated using the Berry-phase formalism under PBC \citep{PhysRevB.47.1651, RevModPhys.66.899}: $p_n =1/2\pi \int_{{\rm{B}}.{\rm{Z}}.} {dk\;\langle {u^n_k}|i\partial {u^n_k}\rangle } \mod 1$, with $n$ being the band index.
The polarization provides insights into the center of Wannier function within a unit cell \citep{RevModPhys.84.1419}.
Owing to the space-time inversion symmetry, the polarization for each gapped band is quantized to either $p_n=0$ or $p_n=1/2$ \citep{Ahn_2019}, which has been widely employed to distinguish the band topology.
However, as shown by the numerical results, the polarization is unable to explain the presence of the four-fold degenerate edge states observed above.
Therefore, unlike the defined inversion winding number, the polarization fails to identify the rich band topology.

\begin{figure}
\includegraphics[width = \columnwidth ]{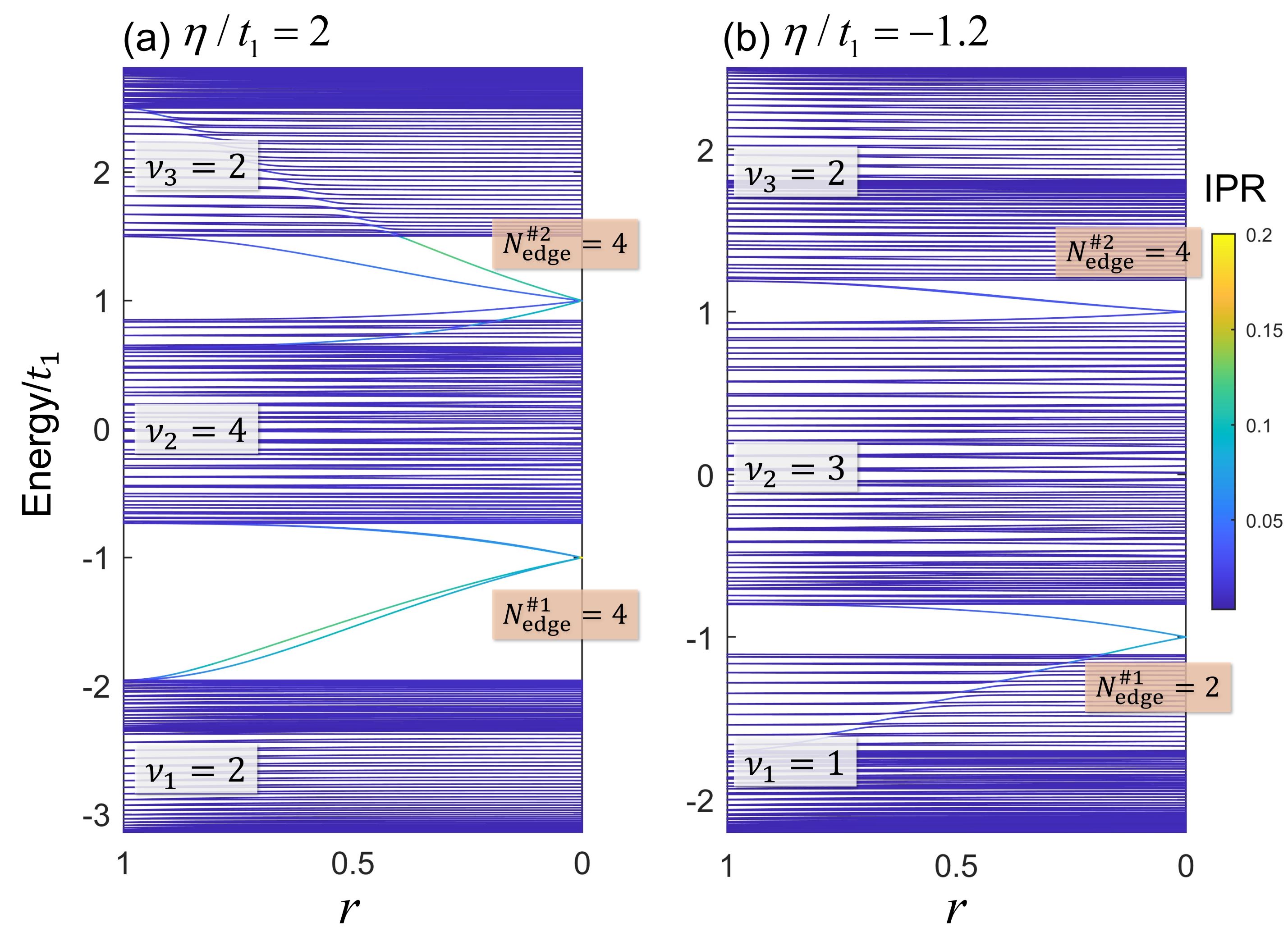}
\caption{\label{fig:FIG_Bulk_edge_Correspondence}
{Energy spectrum versus the real-valued parameter $r\in [0,1]$.
Here, $r=1$ corresponds to the PBC, while $r=0$ corresponds to the OBC.
The inversion winding number ($\nu_n$) of each band ($n=1,2,3$) under PBC and the number of  edge states ($N_{\rm{edge}}^{\# m}$) in each gap ($m=1,2$) under OBC are explicitly labelled accordingly, which clearly manifests the bulk-edge correspondence.}
}
\end{figure}

\subsection{Discussion on the bulk-edge correspondence}
\label{subsec:bulk_edge_correspondence}
To elucidate the bulk-edge correspondence more profoundly, we consider a continuous connection between the PBC and the OBC.
This transition can be effectively modeled by introducing a real-valued boundary parameter $r \in [0,1]$ to the tunneling: $t_{j,j'}^{\sigma ,\sigma '} \to rt_{j,j'}^{\sigma ,\sigma '}$ when $j,j'$ crosses the boundary under PBC.
Thereby, $r=1$ corresponds to the PBC, while $r=0$ corresponds to the OBC.
As depicted in Fig.~\ref{fig:FIG_Bulk_edge_Correspondence} (a-b), edge states arise from the two adjacent bands, with their number related to the inversion winding number.
It can also be observed that the number of edge states arising from the second band is equal to the number of edge states arising from the other two bands.
To relate the inversion winding number to the edge state, one of the prominent approaches is through the twisted boundary condition (TBC), which has been applied in many topological systems to study the bulk-edge correspondence \citep{PhysRevB.74.045125, PhysRevB.103.224208,lin2024probing}.
Under TBC, wavefunctions satisfy the relation: $\psi \left( {x + L} \right) = \exp \left( {i\theta } \right)\psi \left( x \right)$, in which $L$ is the number of cells.
It can be verified that the space-time inversion symmetry is preserved: $\hat{\mathcal{I}} \hat{H}(\theta) \hat{\mathcal{I}}^{-1} =  \hat{H}(\theta)^{*}$.
In analogy to the momentum-space method [Eq.~\eqref{eqn:winding_number_qk_form}], and motivated by the chiral-symmetric system, we define the $L\times L$ matrix:
\begin{equation}
\label{eqn:q_matrix_real_space}
{q}\left( \theta  \right) = {{\bf{\Psi }}_\theta }^\dag {\hat{\cal P}_{1,3 }}{{\bf{\Psi }}_\theta },
\end{equation}
where ${{\bf{\Psi }}_\theta } = \left( {|{\psi _{{m_1}}}\left( \theta  \right)\rangle , \cdots ,|{\psi _{{m_L}}}\left( \theta  \right)\rangle } \right)$ is a $3L\times L$ matrix forming by $L$ eigenvectors on the target band under TBC and ${\hat{\cal P}_{1, 3 }} = \sum\nolimits_x {|x,1\rangle \langle x,3|} $.
The inversion winding number can be expressed through the twisted boundary condition:
\begin{equation}
\label{eqn:SM_TBC_winding_number}
    \nu  = -\frac{1}{{2\pi i}}\int_0^{2\pi } {d\theta \;{\rm{Tr}}\left[ {q{{\left( \theta  \right)}^{ - 1}}{\partial _\theta }q\left( \theta  \right)} \right]} .
\end{equation}
%
%
The $q$ matrix in Eq.~\eqref{eqn:q_matrix_real_space} can be written as a product of two $L\times L$ matrices: $q(\theta) = u^\dagger(\theta) v(\theta)$.
Their matrix elements reads as
\begin{eqnarray}
{\left[ {{u }}(\theta) \right]_{x,m}} = \langle x,1|{\psi _m} (\theta)\rangle ,\;{\left[ {{v (\theta)}} \right]_{x,m}} = \langle x,3|{\psi _m}(\theta)\rangle,
\end{eqnarray}
in which $m$ is the index of eigenstate.
It is worth noting that matrices $u(\theta)$ and $v(\theta)$ respectively corresponds to $\sigma=1,3$ sector of ${{\bf{\Psi }}_\theta }$.
As we are focusing on a single gapped band, the number of eigenstates here is $L$ as well.
It can be understood that the $q$ matrix encodes the phase difference between two sublattices.
When the system is under PBC ($\theta = 0$), the eigenstate can be written as
\begin{equation}
    |{\psi _{n,k}}\rangle  = \sum\limits_{\sigma=1,2,3}  {{u^\sigma_{n,k}} |k,\sigma \rangle }  = \sum\limits_{\sigma=1,2,3}  {{u^\sigma_{n,k}} {e^{ikx}}|x,\sigma \rangle } .
\end{equation}
For the three-band system with space-time inversion symmetry, the matrix element of $u$ and $v$ matrices can be easily found according to Eq.~\eqref{eqn:eigenvector_3bands}:
\begin{eqnarray}
\langle x, 1|{\psi _{n,k}}\rangle  &=& {\alpha _k}{e^{ - i{\xi _k}}}{e^{ikx}},\nonumber \\
\langle x, 3|{\psi _{n,k}}\rangle  &=& {\alpha _k}^*{e^{ikx}}
\end{eqnarray}
In this condition, it can be readily seen that all columns of matrix $u$ or $v$ are mutually orthogonal (but not normalized) provided that the band is gapped and $\alpha _k \ne 0$.
Since the number of these columns is equal to the dimension of the matrix, they form a complete set in this subspace.
Thus, we can conclude that matrices $u$ and $v$ are both non-singular under PBC.
When the TBC is imposed ($\theta \ne 0$), this property should remains as long as the band gap is still finite since the twist angle only acts as a $1/L$-order perturbation \citep{PhysRevB.98.155137, PhysRevB.103.224208, PhysRevB.108.174204}.
Next, we multiply the boundary parameter $r\in[0,1]$ on the tunnelings across the boundary, as introduced above.
The inversion winding number in Eq.~\eqref{eqn:SM_TBC_winding_number} is equivalent to the following closed integral on a complex plane:
\begin{equation}
\nu  = \frac{1}{{2\pi }}\oint_{r=1} {\frac{{dz}}{z}} ,
\end{equation}
in which we define $z\equiv \det q(r,\theta) \in \mathbb{C}$.
Non-trivial winding number under PBC implies that there are residues of the complex function $f(z)=z^{-1}$ inside the $r=1$ circle.
This means that the matrix $q(r,\theta)$ becomes singular at some points within $r\in [0,1)$ and $\theta \in [0,2\pi]$.
On the one hand, since the twist angle $\theta$ is only a perturbative parameter in order of $1/L$ for a gapped band, which is an extremely small quantity in the thermodynamic limit, we can conclude that finite singular points should only reside at $r=0$ in the thermodynamic limit.
Otherwise, there are infinite singular points for $\theta\in[0, 2\pi]$ for $r>0$.
Finally, let us briefly discuss the relation between the singularity of the matrix $q(r,\theta)$ and the boundary mode.
Recall that the $q(r,\theta)$ can be written as the product of two matrices $q(r,\theta) = u(r,\theta)^\dagger v(r,\theta) $, in which $u(r,\theta)$ and $v(r,\theta)$ are respectively the $\sigma = 1,3$ sector of target eigenstates.
When $r=0$, the appearance of singularity in $q(r,\theta)$ also means there are singularities in $u(r,\theta)$ or $v(r,\theta)$, since $ \mathrm{det}[q(r,\theta)] = \mathrm{det}[u(r,\theta)^\dagger] \mathrm{det} [v(r,\theta)] =0$.
Assuming that $u(r,\theta)$ is singular, it is always possible to perform a $\mathrm{U}(L)$ transformation for the target states ${{\bf{\Psi }}_{r,\theta }}$ such that some columns in $u(r,\theta)$ are completely zero.
The number of zero columns is equal to the number of zero singular values in $u(r,\theta)$.
Although the target states are usually not eigenstates of the Hamiltonian after the $\mathrm{U}(L)$ transformation, the total density distribution of these target states will not change.
Then, the presence of zero columns in $u(r,\theta)$ means some density distributions in $\sigma=1$ sublattices vanish.
This also applies to the case where $v(r,\theta)$ is singular.
Therefore, it suggests the existence of boundary modes localized in one side.
As the total system is inversion-symmetric, the boundary modes should be even-fold degenerate.

\section{Real-space inversion winding number and robustness}
\label{sec:Robustness_winding_number}
Having established the momentum-space formula of the inversion winding number, we then delve into exploring its real-space counterpart to assess the robustness of band topology against disorder that preserves the space-time inversion symmetry.
In analogy to the chiral-symmetric system \citep{PhysRevB.103.224208,PhysRevLett.128.127601}, the real-space formula of the inversion winding number can be expressed as a Bott index form,
\begin{equation}
\label{eqn:realspace_nu_X_proj}
\nu  = -\frac{1}{{2\pi }}{\mathop{\rm Im}\nolimits} \left\{ {{\rm{Tr}}\left[ {{\rm{log}}\left( {{{\cal X}_1}^\dag {{\cal X}_3}} \right)} \right]} \right\},
\end{equation}
in which ${{\cal X}_\sigma } = {{\bf{\Psi }}^\dag }\exp \left( {i\frac{{2\pi }}{L}\hat x} \right){{\hat P}_\sigma}{\bf{\Psi }}$ is the projected-position matrix of the $\sigma = 1, 3 $ sublattices and ${\bf{\Psi }} = \left( { \cdots ,|{\psi _j}\rangle , \cdots } \right)$ denotes a set of eigenstates on the target band as mentioned above.
Eq.~\eqref{eqn:realspace_nu_X_proj} is equivalent to the momentum-space formula in the presence of translation symmetry; see Appendix.~\ref{appendix:realspace_formula} for detailed discussions. 

\subsection{Effect of inversion-invariant disorder}
The real-space formula enables us to efficiently calculate the inversion winding number in the presence of inversion-invariant disorders.
Here, we introduce disorders to the nearest-neighbor tunneling between the first and third sublattice: ${{\hat H}_{\rm{dis}}} =  - W \sum\nolimits_j {{\delta _x}|x,1\rangle \langle x,3|}  + {\rm{H.c.}}$, where $\delta_x \in [-1/2,1/2]$ represents a uniformly distributed random variable and $W$ is the strength of the disorder.
Here, $|x,\sigma\rangle, x=1,2,\cdots, L$ denotes the position basis.
To preserve the space-time inversion symmetry, we enforce $\hat{\mathcal{I}}{{\hat H}_{\rm{dis}}}\hat{\mathcal{I}}^{-1} = {( {{{\hat H}_{\rm{dis}}}} )^*}$.
Meanwhile, this term only entails the coupling between the sublattices $\sigma = 1$ and $\sigma = 3$. 
As mentioned above, the inversion winding number retains its well-defined nature as long as $t_1\ne 0$ in this system.

\begin{figure}
\includegraphics[width = \columnwidth ]{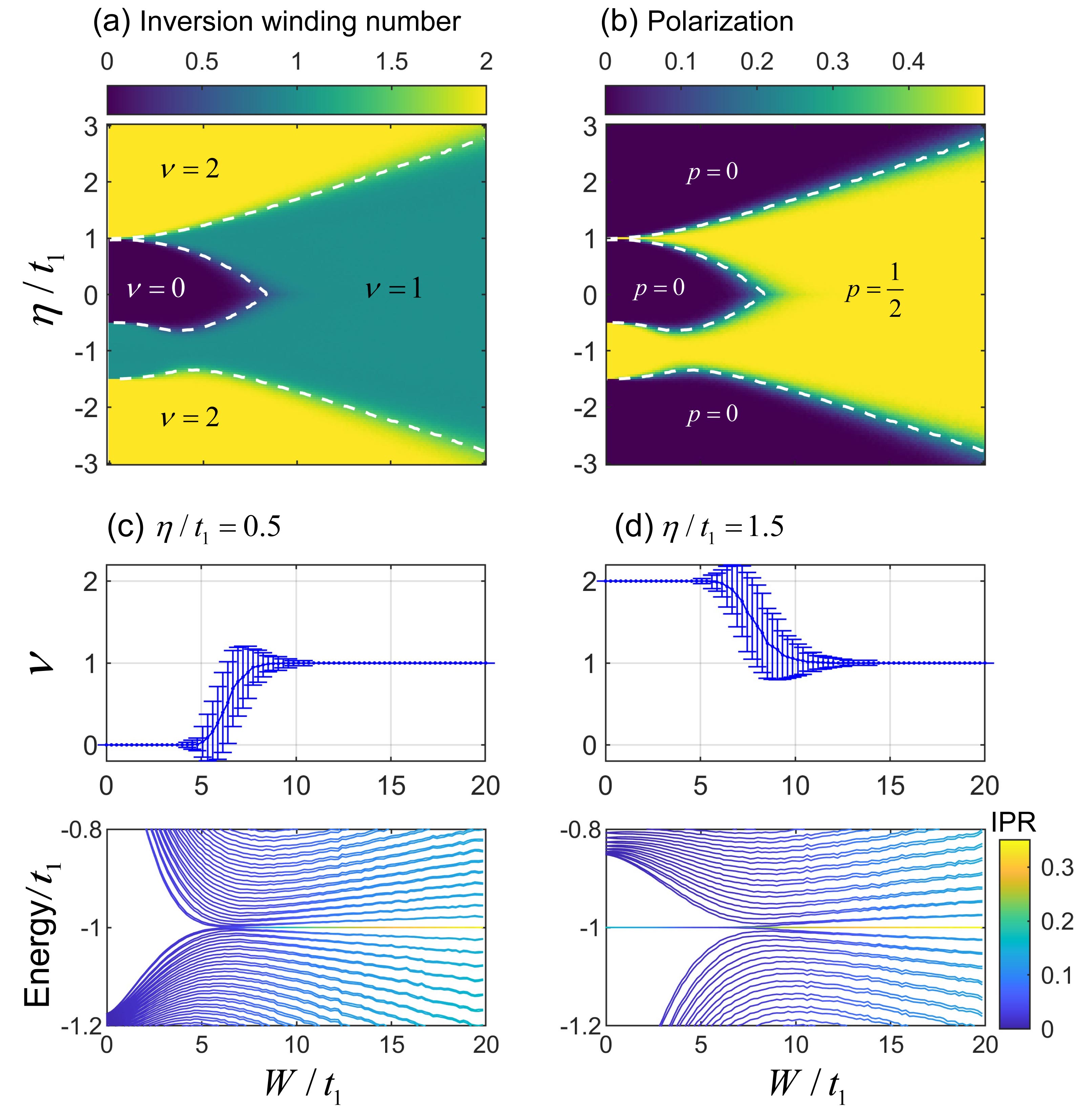}
\caption{\label{fig:FIG_disorder_scan}
Phase diagram of the lowest band in the presence of disorder: (a) the inversion winding number and (b) the polarization.
Dashed lines indicate the phase boundary determined by the half-height between two phases.
(c-d) respectively show the inversion winding number (under PBC) and spectrum (under OBC) near the first gap with $\eta/t_1 = 0.5$ and $\eta/t_1 = 1.5$.
Number of cells is $L = 200$ and other parameters are the same as Fig.~\ref{fig:FIG_three_band_spectrum}.
All results are averaged over $10^3$ random realizations.
}
\end{figure}

Subsequently, by utilizing Eq.~\eqref{eqn:realspace_nu_X_proj}, we calculate the $\eta$-$W$ phase diagram of inversion winding number for the lowest band under PBC, see Fig.~\ref{fig:FIG_disorder_scan} (a).
These phase diagrams show rich phenomena of disorder-induced topological transitions.
More specifically, in Figs.~\ref{fig:FIG_disorder_scan} (c) and (d), we present two examples of the topological transitions $\nu=0 \to \nu=1$ and $\nu=1\to \nu=2$, respectively.
The variations in the inversion winding number align well with the gap-closing process, and the number of these edge states satisfy $N_{\rm{edge}} = 2\nu$.
Particularly, there appears the disorder-induced topological phase ($\nu=0 \to \nu=1$), which is usually identified as the topological Anderson insulating phase \citep{PhysRevLett.102.136806, PhysRevLett.112.206602, Wu_2016,Science362,lin2022observation,PhysRevA.105.063327,PhysRevLett.132.066602}.
Significant statistical fluctuations in the inversion winding number are observed around all phase transitions.
Away from the phase transition points, the inversion winding number remains remarkably stable in broad areas for different random realizations.
These results effectively demonstrate the topological stability of the inversion winding number.
More information, including the phase diagram of statistical fluctuation of inversion winding number and the phase diagram of the $n=2,3$ band, are presented in Appendix.~\ref{appendix:realspace_formula}.
For comparison, we calculate the polarization through the projected position operator methods \citep{PhysRevLett.80.1800,PhysRevB.96.245115,PhysRevB.107.125161}, as shown in Fig.~\ref{fig:FIG_disorder_scan} (b).
Due to the inversion symmetry, the polarization is also quantized.
The polarization $p$ is connected to the inversion winding number $\nu$ through the relation $p=\nu/2 \mod 1$.
As expected, the polarization fails to distinguish the phase of $\nu = 0$ and 2.

\begin{figure}
\includegraphics[width = \columnwidth ]{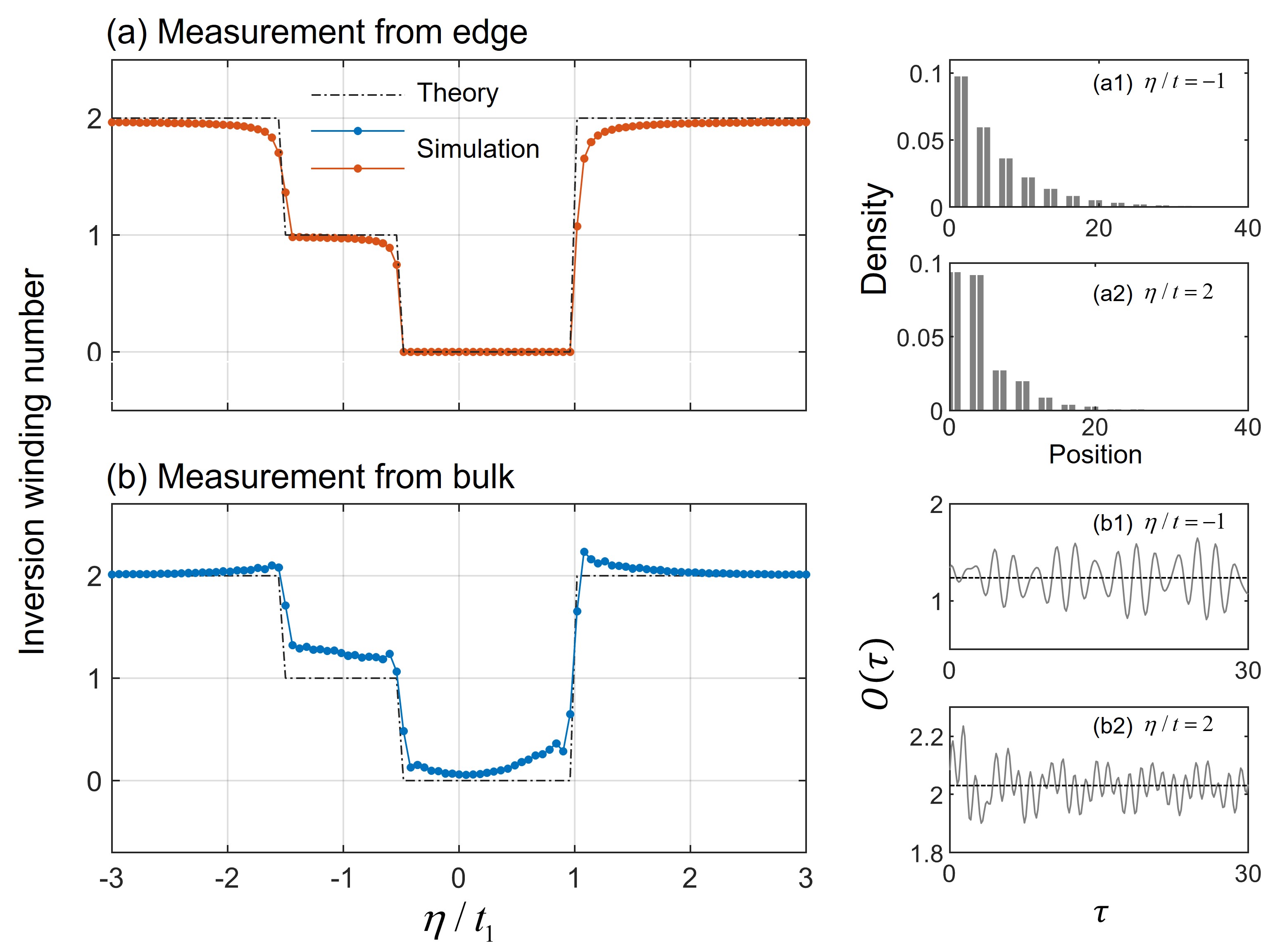}
\caption{\label{fig:FIG_winding_number_Edge}
Simulation of measuring the inversion winding number through dynamical evolution.
(a) Measuring through the density distribution of edge states in the first gap.
(a1-a2) shows two examples of the edge states (only the left end is shown for simplicity).
(b) Measuring through the time-evolution method with the total evolution time being $T=30/t_1$.
The system's size is chosen as $L=200$.
Other parameters are the same as Fig.~\ref{fig:FIG_three_band_spectrum}.
}
\end{figure}

\subsection{Real-space characterization of band topology}
In above, we have found that the inversion winding number is associated with the polarization difference between two sublattices.
This offers the feasibility for experimentally characterizing different topological phases.
Under the OBC, the inversion winding number can be approximately extracted from bulk states or edge states via
\begin{equation}
\label{eqn:nu_real_space_Edge_Bulk}
\nu  = (-1)^{\chi}\frac{2}{L} \sum\limits_{n \in {\rm{target}}} {\frac{{\langle {\psi _n}|{{\hat X}_3} - {{\hat X}_1}|{\psi _n}\rangle }}{{\langle {\psi _n}|{{\hat P}_3} + {{\hat P}_1}|{\psi _n}\rangle }}} 
\end{equation}
in which $\hat{X}_\sigma = \hat{x}\hat{P}_{\sigma}$ is the position operator projected to the $\sigma = \{1,2,3\}$ sublattice.
Eq.~\eqref{eqn:nu_real_space_Edge_Bulk} is a summation over bulk states in a certain band ($\chi = 1$) or edge states in a certain gap ($\chi = 0$).
It implies that one can measure the inversion winding number through the bulk or the edge states.
In experiments, the edge states can be probed via versatile approaches and then the information of inversion winding number can be directly extracted.
As for the bulk state, it is also feasible to measure the winding number through time-evolution \citep{cardano2017detection, Science362, PhysRevA.98.013835, xie2019topological, PhysRevLett.122.193903,PhysRevLett.123.080501,PhysRevLett.124.050502}: $\nu  \approx  \mathrm{lim}_{T\to\infty}\frac{1}{T}\int_0^T {d\tau \;O\left( \tau \right)} $ with $O\left( \tau \right) = \frac{{\langle \Psi \left( \tau  \right)|{{\hat X}_3} - {{\hat X}_1}|\Psi \left( \tau  \right)\rangle }}{{\langle \Psi \left( \tau  \right)|{{\hat P}_3} + {{\hat P}_1}|\Psi \left( \tau  \right)\rangle }}$ and $|\Psi \left( \tau  \right)\rangle$ being a superposition of bulk states in a certain band.
In Fig.~\ref{fig:FIG_winding_number_Edge}, we present our numerical simulation for the experimental measurement through edge states or bulk states.
Detailed implementation can be found in Appendix.~\ref{appendix:model_and_experimental_realization}.
Remarkably, both methods produces a very prominent results compared with the theoretical prediction.
The tiny difference between them can be attributed to the finite-size effect and the relatively small spectral gap, especially for bulk measurement.
This again highlights the bulk-edge correspondence principle in this system: the appearance of in-gap edge state is a result of the nontrivial sublattice polarization of a band.
The nontrivial inversion winding number has prominent physical consequences.
The above results can be verified in various systems, such as, electrical circuits \citep{PhysRevB.105.L081107, PhysRevApplied.20.064042}, acoustic systems \citep{PhysRevLett.131.157201,PhysRevB.108.205135} and  other photonic or atomic systems \citep{poddubny2014topological,Song_sciadv_2018,topological_review_SL_Zhu, kruk2019nonlinear, RevModPhys.91.015006, RevModPhys.91.015006, Wang_2019,PhysRevB.108.085116, PhysRevLett.123.080501,guo2021experimental,PhysRevLett.127.147401}.

\section{Summary and discussion}
\label{sec:Summary}
In this work, we show that, under the space-time inversion symmetry, the band topology can be further classified beyond the conventional $\mathbb{Z}_2$ classification in a large family of 1D three-band models upon forbidding some certain kinds of tunneling process.
Based upon the sublattice polarization difference regarding to the space-time inversion symmetry, we introduce a gauge-invariant winding number: the inversion winding number.
It is worth noting that general approach using the polarization (Berry phase) to distinguish the band topology lies on investigating the center-of-mass position of Wannier states within the unit cell, that is $p \propto \langle w\left( R \right)|\hat x|w\left( R \right)\rangle $, with $|w\left( R \right)\rangle$ being the Wannier state localized at $R$th cell.
Thus, it only produces $\mathbb{Z}_2$-quantized value in the presence of inversion symmetry.
As for the inversion winding number, which computes the relative polarization between sublattices, it measures the relative position of different sublattices in the Wannier state: $\nu  \propto \langle w\left( R \right)|\hat x( {{{\hat P}_3} - {{\hat P}_1}} )|w\left( R \right)\rangle $.
Hence, non-zero inversion winding number leads to extra charge accumulation at edges and it is beyond the $\mathbb{Z}_2$ classification.
The bulk topology and the bulk-edge correspondence can be well characterized by the inversion winding number $\nu \in \mathbb{Z}$.
Non-trivial inversion winding under PBC number can be attributed to the appearance of degenerate boundary modes under OBC.
In addition, we derive a real-space formula for the inversion winding number.
The inversion winding number is found to remain effective even in the presence of strong symmetry-invariant disorders.
With the real-space formula for the inversion winding number, it is possible to identify distinct topological phases experimentally.

Our theory introduces an idea to further explore and classify TCIs and many intriguing questions remain to be addressed.
In interacting multi-particle systems, our theory is also expected to be valid provided the space-time inversion symmetry is present.
On the other hand, we note that previously the chiral winding number is extended to the chiral-symmetric higher-order topological insulator \citep{PhysRevLett.128.127601, PhysRevLett.131.157201, PhysRevApplied.20.064042,lin2024probing,PhysRevB.110.L201117, PhysRevB.110.075103,PhysRevB.111.035115}.
It is also of interest to generalize our approach to higher-order topological insulators protected by inversion symmetry or rotation symmetry and study the multipole polarization nature between sublattices.
For example, it is desirable to define the inversion winding number for the Wannier band with regard to the mirror symmetry or the $C_4$ symmetry.
Moreover, one can further explore the sublattice polarization in views of topological pumping by breaking the inversion symmetry, since the change of polarization leads to the adiabatic current.

\acknowledgments
This work is supported by the National Key Research and Development Program of China (Grant No. 2022YFA1404104), the National Natural Science Foundation of China (Grant No. 12025509, Grant No. 12247134 and Grant No. 12275365), and the Natural Science Foundation of Guangdong Province (Grant No.2023A1515012099). 

\appendix

\section{Relation between the inversion winding number and the polarization}
\label{appendix:relation_between_inv_winding_and_polarization}
In the main text, we have defined inversion winding numbers for three-band topological crystalline insulators.
Here, we show that the inversion winding numbers are related to the band polarization such that $p = \nu/2 \mod 1$.
It can be found that
\begin{eqnarray}
&& \int_0^{2\pi } {dk\;\left( {{u _k}^\dag i{\partial _k}{u _k}} \right)}  \nonumber  \\
&& =  \int_0^{2\pi } {dk\;\left( {{\alpha _k}^*i{\partial _k}{\alpha _k}} \right)}  + \int_0^{2\pi } {dk\;\left( {{\beta _k}^*i{\partial _k}{\beta _k}} \right)}  \nonumber  \\
&& + \int_0^{2\pi } {dk\;\left( {{\alpha _k}i{\partial _k}{\alpha _k}^*} \right)}  \nonumber  \\
&& + \int_0^{2\pi } {dk\left[ {\;\left( {{{\left| {{\alpha _k}} \right|}^2} + \frac{1}{2}{{\left| {{\beta _k}} \right|}^2}} \right){\partial _k}{\xi _k}} \right]} .
\end{eqnarray}
Using the normalization relation, we have $2{\left| {{\alpha _k}} \right|^2} + {\left| {{\beta _k}} \right|^2} = 1$ due to the inversion symmetry.
Thus, we find the following relation:
\begin{eqnarray}
\label{eqn_appendix:Berry_phase_psi_k_half_quantized}
p = \frac{1}{2\pi}\int_0^{2\pi } {dk\;\left( {{u _k}^\dag i{\partial _k}{u _k}} \right)}  = \frac{1}{2}\int_0^{2\pi } {dk\;{\partial _k}{\xi _k}} \in \frac{\mathbb{Z}}{2}.
\end{eqnarray}
which again proves the quantization of the Berry phase modulo $2\pi$.
These results reflect the importance of the phase factor $\xi_k$ in the space-time inversion-symmetric topological insulators.
Then, according to Eq.~\eqref{eqn:eigenvector_3bands}, we can expand the formula of the inversion winding number:
\begin{eqnarray}
\nu  &=& \frac{1}{{\pi i}}{\rm{Im}}\int_0^{2\pi } {dk\;\left[ {\frac{{\langle {u_k}|\frac{\partial }{{\partial k}}\left( {{{\hat P}_3} - {{\hat P}_1}} \right){u_k}\rangle }}{{\langle {u_k}|{{\hat P}_1} + {{\hat P}_3}|{u_k}\rangle }}} \right]} \nonumber \\
&=& \frac{1}{{\pi i}}\int\limits_{B.Z.} {dk\frac{{{\alpha _k}\frac{\partial }{{\partial k}}\left( {{\alpha _k}^*} \right) - {{\left( {{\alpha _k}{e^{ - i{\xi _k}}}} \right)}^*}\frac{\partial }{{\partial k}}\left( {{\alpha _k}{e^{ - i{\xi _k}}}} \right)}}{{2|{\alpha _k}{|^2}}}} \nonumber \\
 &=&  \frac{1}{{\pi i}}\int\limits_{B.Z.} {dk\frac{{{\alpha _k}\frac{\partial }{{\partial k}}\left( {{\alpha _k}^*} \right)}}{{|{\alpha _k}{|^2}}}}  + \frac{1}{{2\pi i}}\int\limits_{B.Z.} {dk{{\left( {{e^{i{\xi _k}}}} \right)}^*}\frac{\partial }{{\partial k}}\left( {{e^{i{\xi _k}}}} \right)}   \nonumber \\
 &=& 2{\nu _\alpha } + {\nu _\xi },
\end{eqnarray}
where we have denoted two integer numbers
\begin{eqnarray}
{\nu _\alpha }  &=& \frac{1}{{2\pi i}}\int\limits_{B.Z.} {dk\;\frac{{{\alpha _k}\frac{\partial }{{\partial k}}\left( {{\alpha _k}^*} \right)}}{{|{\alpha _k}{|^2}}}} \quad  \in \mathbb{Z} \nonumber \\
{\nu _\xi }  &=& \frac{1}{{2\pi i}}\int\limits_{B.Z.} {dk\;{{\left( {{e^{i{\xi _k}}}} \right)}^*}\frac{\partial }{{\partial k}}\left( {{e^{i{\xi _k}}}} \right)} \nonumber \\
&=& \int_0^{2\pi } {dk\;{\partial _k}{\xi _k}} \quad  \in  \mathbb{Z}.
\end{eqnarray}
Compared to Eq.~\eqref{eqn_appendix:Berry_phase_psi_k_half_quantized}, we can immediately obtain the relation between the inversion winding number $\nu$ and the polarization $p$: $
\frac{\nu }{2} \mod \,1 = p $.
It should be noted that, even though the inversion winding number may become ill-defined if $\alpha_k = 0$ while the spectral gap is not closed, we can always add some perturbative terms to let $\alpha_k \ne 0$.
In this sense, the inversion winding number is at least a $\mathbb{Z}_2$ topological invariant.
As mentioned in the main text, by further imposing constraints on the form of Hamiltonian, the inversion winding number becomes the topologically stable $\mathbb{Z}$ topological invariant.

\section{Real-space formula of the inversion winding number}
\label{appendix:realspace_formula}
\subsection{Equivalence to the momentum-space formula}
Here, we prove that the Bott index form of the real-space inversion winding number formula is equivalent to the momentum-space formula in the presence of translation symmetry.
First of all, we write the momentum-space formula as an equivalent discretized form:
\begin{equation}
\label{eqn:SM_winding_momentum_discretized}
\nu  = \frac{1}{{2\pi }}{\mathop{\rm Im}\nolimits} \sum\limits_k {{\rm{Tr}}\left[ {\log \left( {{F_k}{{^{\left[ 1 \right]}}^\dag }{F_k}^{\left[ 3 \right]}} \right)} \right]} ,
\end{equation}
in which ${F_k}^{\left[ \sigma  \right]} = \langle {u_k}|{{\hat P}_\sigma }|{u_{k + \delta k}}\rangle $ $\sigma=1,2,3$ and $\delta k = 2\pi/L$ is the minimal increment of the (quasi) momentum in finite systems.
Eq.~\eqref{eqn:SM_winding_momentum_discretized} is similar to the Wilson loop form and is convenient for numerical computations in finite systems.
Notably, the normalization factor is dropped here, as it is a real positive number $ \langle {u_k}|{{\hat P}_\sigma }|{u_{k}}\rangle \in \mathbb{R}^+$ and does not contribute to the result after taking the imaginary part.
For convenience, we only focus on single gapped band and omit the band index.
Then, noting that the projected position matrix ${\cal{X}_\sigma } = {{\bf{\Psi }}^\dag }\exp \left( {i\frac{{2\pi }}{L}\hat x} \right){{\hat P}_\sigma }{\bf{\Psi }}$ and the Bloch state satisfy:
\begin{eqnarray}
 {\left[ {{{\cal X}_\sigma }} \right]_{k',k}} &= &\langle {\psi _{k'}}|\exp \left( {i\frac{{2\pi }}{L}\hat x} \right){{\hat P}_\sigma }|{\psi _{k}}\rangle \nonumber \\
  &=& {\delta _{k',k + \delta k}}\langle {u_{k'}}|{{\hat P}_\sigma }|{u_k}\rangle ,
\end{eqnarray}
where we have assumed that ${\bf{\Psi }} = \left( { \cdots ,|{\psi _k}\rangle ,|{\psi _{k + \delta k}}\rangle , \cdots } \right)$ targets the whole band with $k= 2m\pi/L, m=1,2,\cdots, L$.
Thus, it can be found that
\begin{eqnarray}
{\left[ {{{\cal X}_1}^\dag {{\cal X}_3}} \right]_{k',k''}} &=& \sum\limits_k {{{\left[ {{{\cal X}_1}^\dag } \right]}_{k',k}}{{\left[ {{{\cal X}_3}} \right]}_{k,k''}}} \nonumber \\
 &=& {\delta _{k',k''}}\sum\limits_k {\langle {u_{k - \delta k}}|{{\hat P}_1}|{u_k}\rangle \langle {u_k}|{{\hat P}_3 }|{u_{k - \delta k}}\rangle }  \nonumber \\
 &=& {\delta _{k',k''}}\sum\limits_k {{F_{k - \delta k}}^{\left[ 1 \right]}{{\left( {{F_{k - \delta k}}^{\left[ 3 \right]}} \right)}^\dag }} 
\end{eqnarray}
which means the matrix ${{{\cal X}_1}^\dag {{\cal X}_3}}$ is a (block) diagonal matrix and the diagonal term is associated to Eq.~\eqref{eqn:SM_winding_momentum_discretized}.
Thus, by taking the matrix logarithm, we then finish the proof:
\begin{eqnarray}
\nu  &=& \frac{1}{{2\pi }}{\mathop{\rm Im}\nolimits} \sum\limits_k {{\rm{Tr}}\left[ {\log \left( {{F_k}{{^{\left[ 1 \right]}}^\dag }{F_k}^{\left[ 3 \right]}} \right)} \right]} \nonumber \\
&=& -\frac{1}{{2\pi }}{\mathop{\rm Im}\nolimits} \left\{ {{\rm{Tr}}\left[ {{\rm{log}}\left( {{{\cal X}_1}^\dag {{\cal X}_3}} \right)} \right]} \right\}.
\end{eqnarray}

\subsection{Derivation via the twisted boundary condition}
Next, let us introduce the twisted boundary condition (TBC), which serves as a magnetic flux inserting to the system under PBC.
Similar to the momentum-space formula in Eq.~\eqref{eqn:winding_number_3bands_momentum}, we can express the inversion winding number under the TBC as:
\begin{equation}
\label{eqn:windingNumber_TBC_projected}
\nu  = -\frac{1}{{\pi }}{\rm{Im}}\int_0^{2\pi } {d\theta \;{\rm{Tr}}\left\{ {\frac{{{{\bf{\Psi }}_\theta }^\dag \frac{\partial }{{\partial \theta }}\left[ {\left( {{{\hat P}_3} - {{\hat P}_1}} \right){{\bf{\Psi }}_\theta }} \right]}}{{{{\bf{\Psi }}_\theta }^\dag {{\left( {{{\hat P}_1} + {{\hat P}_3}} \right)}}{{\bf{\Psi }}_\theta }}}} \right\}} .
\end{equation}
Alternatively, one can use the following discrete form of parallel transport for the state following from Eq.~\eqref{eqn:SM_TBC_winding_number}:
\begin{equation}
\label{eqn:windingNumber_TBC_projected_product}
   \nu  \approx \frac{{ i}}{{2\pi}}{\rm{Tr}}\left\{ {\log \left[ {\prod\limits_n {\left( {{{\bf{\Psi }}_{{\theta _n}}}^\dag {{\hat P}_3}{{\bf{\Psi }}_{{\theta _{n + 1}}}}} \right){{\left( {{{\bf{\Psi }}_{{\theta _n}}}^\dag {{\hat P}_1}{{\bf{\Psi }}_{{\theta _{n + 1}}}}} \right)}^\dag }} } \right]} \right\} ,
\end{equation}
in which we have discretized the twist angle such that $\theta_1 = 0$ and $\theta_N = 2\pi$, and the normalization factor is omitted for simplicity as it does not contribute to the result here.
In the presence of translation symmetry, the above formulae are equivalent to the momentum-space form.
Moreover, the formula defined through TBC can still work if the translation symmetry is broken while the space-time inversion symmetry is preserved.
Then, to obtain the real-space formula in Eq.~\eqref{eqn:realspace_nu_X_proj}, we need to use the perturbative property of the twist angle in the twisted boundary condition (TBC).
As shown in Refs.~\citep{PhysRevB.103.224208, PhysRevB.107.125161, PhysRevB.108.174204}, there is gauge freedom in the TBC, as the TBC is equivalent to the insertion of magnetic flux to the system in ring geometry.
The twist angle attached to the tunneling process across the boundary can be regarded as the effect of vector potential induced by the magnetic field.
By appropriately choosing the gauge (called the uniform gauge), the twist angle will distributed uniformly throughout the system.
In this case, the twist angle will be threaded to all tunneling process with a strength of $\theta/L$, which is an extremely small quantity if $L\to \infty$.
To obtain the uniform gauge, we can introduce the following twist operator
\begin{equation}
{{\hat U}_\theta } = \exp \left( {i\frac{\theta }{L}\hat x} \right),
\end{equation}
where $\hat{x}$ is the position operator.
The Hamiltonian under conventional TBC can be transformed to the uniform-gauge form (labelled by tilde notation):
\begin{equation}
\tilde H\left( \theta  \right) = {{\hat U}_\theta }\hat H\left( \theta  \right){{\hat U}_\theta }^{ - 1}.
\end{equation}
The system satisfies the gauge-invariant condition: 
\begin{eqnarray}
\label{eqn_appendix:Ham_TBC_gauge_invariant_relation}
\hat H\left( {\theta  + 2\pi } \right)& =& \hat H\left( \theta  \right) \nonumber \\
\tilde H\left( {\theta  + 2\pi } \right) &=& {{\hat U}_{2\pi }}\tilde H\left( \theta  \right){{\hat U}_{2\pi }}^{ - 1}.
\end{eqnarray}
Under the uniform gauge, we can treat $\theta/L$ as a perturbation and expand the Hamiltonians and the eigenstates.
There is 
\begin{equation}
{\tilde{\bf{\Psi }}_\theta } = {\tilde{\bf{\Psi }}_0} + \frac{\theta }{L}{\left( {\frac{\partial }{{\partial \left( {\theta /L} \right)}}{\tilde{\bf{\Psi }}_\theta }} \right)_{\theta  = 0}} + O\left( {\frac{1}{{{L^2}}}} \right).
\end{equation}
By substituting it into the expression of inversion winding number and keeping up to the leading order of $1/L$, we have
\begin{eqnarray}
\nu  &=&  - \frac{1}{{\pi }}{\rm{Im}}\int_0^{2\pi } {d\theta \;{\rm{Tr}}\left\{ {\frac{{{{{\bf{\tilde \Psi }}}_\theta }^\dag \frac{\partial }{{\partial \theta }}\left[ {\left( {{{\hat P}_3} - {{\hat P}_1}} \right){{{\bf{\tilde \Psi }}}_\theta }} \right]}}{{{{{\bf{\tilde \Psi }}}_\theta }^\dag \left( {{{\hat P}_1} + {{\hat P}_3}} \right){{{\bf{\tilde \Psi }}}_\theta }}}} \right\}} \nonumber \\
& \approx&  - \frac{1}{{\pi }}{\rm{Im}}\int_0^{2\pi } {d\theta \;{\rm{Tr}}\left[ {\frac{{{{{\bf{\tilde \Psi }}}_0}^\dag \left( {{{\hat P}_3} - {{\hat P}_1}} \right){{\left( {\frac{\partial }{{\partial \theta }}{{{\bf{\tilde \Psi }}}_\theta }} \right)}_{\theta  = 0}}}}{{{{{\bf{\tilde \Psi }}}_0}^\dag \left( {{{\hat P}_1} + {{\hat P}_3}} \right){{{\bf{\tilde \Psi }}}_0}}}} \right]}  \nonumber  \\
&=&  -2{\mathop{\rm Im}\nolimits} \;\left\{ {{\rm{Tr}}\left[ {\frac{{{{{\bf{\tilde \Psi }}}_0}^\dag \left( {{{\hat P}_3} - {{\hat P}_1}} \right){{\left( {\frac{\partial }{{\partial \theta }}{{{\bf{\tilde \Psi }}}_\theta }} \right)}_{\theta  = 0}}}}{{{{{\bf{\tilde \Psi }}}_0}^\dag \left( {{{\hat P}_1} + {{\hat P}_3}} \right){{{\bf{\tilde \Psi }}}_0}}}} \right]} \right\}
\end{eqnarray}
According to Eq.~\eqref{eqn_appendix:Ham_TBC_gauge_invariant_relation}, the eigenstates will follow the same relations:
\begin{equation}
{\tilde{\bf{\Psi }}_{2\pi }} = {{\hat U}_{2\pi }}{\tilde{\bf{\Psi }}_0}
\end{equation}
given that the target eigenstates are gapped.
Thus, we can approximate the partial derivative
\begin{eqnarray}
{\left( {\frac{\partial }{{\partial {\theta } }}{\tilde{\bf{\Psi }}_\theta }} \right)_{\theta  = 0}} & \approx& \frac{1}{{2\pi }}\left( {{\tilde{\bf{\Psi }}_{2\pi }} - {\tilde{\bf{\Psi }}_0}} \right) \nonumber \\
 &=& \frac{1}{{2\pi }}\left( {{{\hat U}_{2\pi }}{\tilde{\bf{\Psi }}_0} - {\tilde{\bf{\Psi }}_0}} \right).
\end{eqnarray}
Therefore the expression of the inversion winding number can be further approximated to:
\begin{eqnarray}
\nu \approx -\frac{1}{\pi }{\rm{Im}}\left\{ {{\rm{Tr}}\left[ {\frac{{{{\bf{\Psi }}_0}^\dag \left( {{{\hat P}_3} - {{\hat P}_1}} \right)\left( {{{\hat U}_{2\pi }}{{\bf{\Psi }}_0} - {{\bf{\Psi }}_0}} \right)}}{{{{{\bf{\tilde \Psi }}}_0}^\dag \left( {{{\hat P}_1} + {{\hat P}_3}} \right){{{\bf{\tilde \Psi }}}_0}}}} \right]} \right\}\nonumber \\
\end{eqnarray}
where we have used the fact that ${\rm{Im}}\left\{ {{\rm{Tr}}\left[ {{{\bf{\Psi }}_0}^\dag \left( {{{\hat P}_3} - {{\hat P}_1}} \right){{\bf{\Psi }}_0}} \right]} \right\} = 0$.
Then, we further expand the twisted operator such that
\begin{equation}
{{\hat U}_{2\pi }} = \exp \left( {i\frac{{2\pi }}{L}\hat x} \right) \approx \hat I + i\frac{{2\pi }}{L}\hat x.
\end{equation}
Finally, after these approximations, we obtain the following real-space formula 
\begin{equation}
\label{eqn:SM_Winding_X1_X3_realspace}
\nu  \approx -\frac{2}{L}{\rm{Tr}}\left[ {\frac{{{{\bf{\Psi }}_0}^\dag \left( {{{\hat X}_3} - {{\hat X}_1}} \right)\hat x{{\bf{\Psi }}_0}}}{{{{\bf{\Psi }}_0}^\dag \left( {{{\hat P}_1} + {{\hat P}_3}} \right){{\bf{\Psi }}_0}}}} \right]
\end{equation}
which clearly shows that the inversion winding number describes the polarization difference between sublattices upon normalization.
It is worth noting that Eq.~\eqref{eqn:SM_Winding_X1_X3_realspace} works for bulk states in the gapped band.
In fact, this approximation is poor under PBC, as the convergence is slow.
Nevertheless, when the system is under the OBC, the edge state and the bulk state are separated.
By applying Eq.~\eqref{eqn:SM_Winding_X1_X3_realspace} respectively for the bulk states and the edge states yield satisfactory results, as already shown in the main text.
Interestingly, the bulk states and the edge states approximately have the opposite polarization profile in real space.
We would like to point out that the edge state is degenerate under OBC and they are equally contributed from the lower and upper bands.
Thus, when targeting the edge states, we need to distinguish them.
This yields the expression for the edge state in the main text.
Notably, the real-space formula is affected by the band gap, since we have treated the twist angle as a perturbation.
This explains why the real-space formula produces a result deviating from the quantized value.

On the other hand, the Bott index form of the real-space inversion winding number can be also derived through the TBC.
The derivation is reminiscent to that in Ref.~\citep{PhysRevB.103.224208}.
With Eq.~\eqref{eqn:windingNumber_TBC_projected_product}, we leverage the perturbative nature of the twist angle and approximate the product as
\begin{eqnarray}
\nu  &\approx& - \frac{{ 1}}{{2\pi i}}{\rm{Tr}}\left\{ {\log \left[ {\left( {{{\widetilde {\bf{\Psi }}}_0}^\dag {{\hat P}_3}{{\widetilde {\bf{\Psi }}}_{2\pi }}} \right){{\left( {{{\widetilde {\bf{\Psi }}}_0}^\dag {{\hat P}_1}{{\widetilde {\bf{\Psi }}}_{2\pi }}} \right)}^\dag }} \right]} \right\} \nonumber \\
 &=& -\frac{{  1}}{{2\pi }}{\mathop{\rm Im}\nolimits} \left\{ {{\rm{Tr}}\left[ {\log \left( {{{\cal X}_1}^\dag {{\cal X}_3}} \right)} \right]} \right\},
\end{eqnarray}
which is the real-space formula of the inversion winding number appearing in the main text.

\subsection{More information on the phase diagram in the presence of disorder}
\begin{figure}
\includegraphics[width = 0.9\columnwidth ]{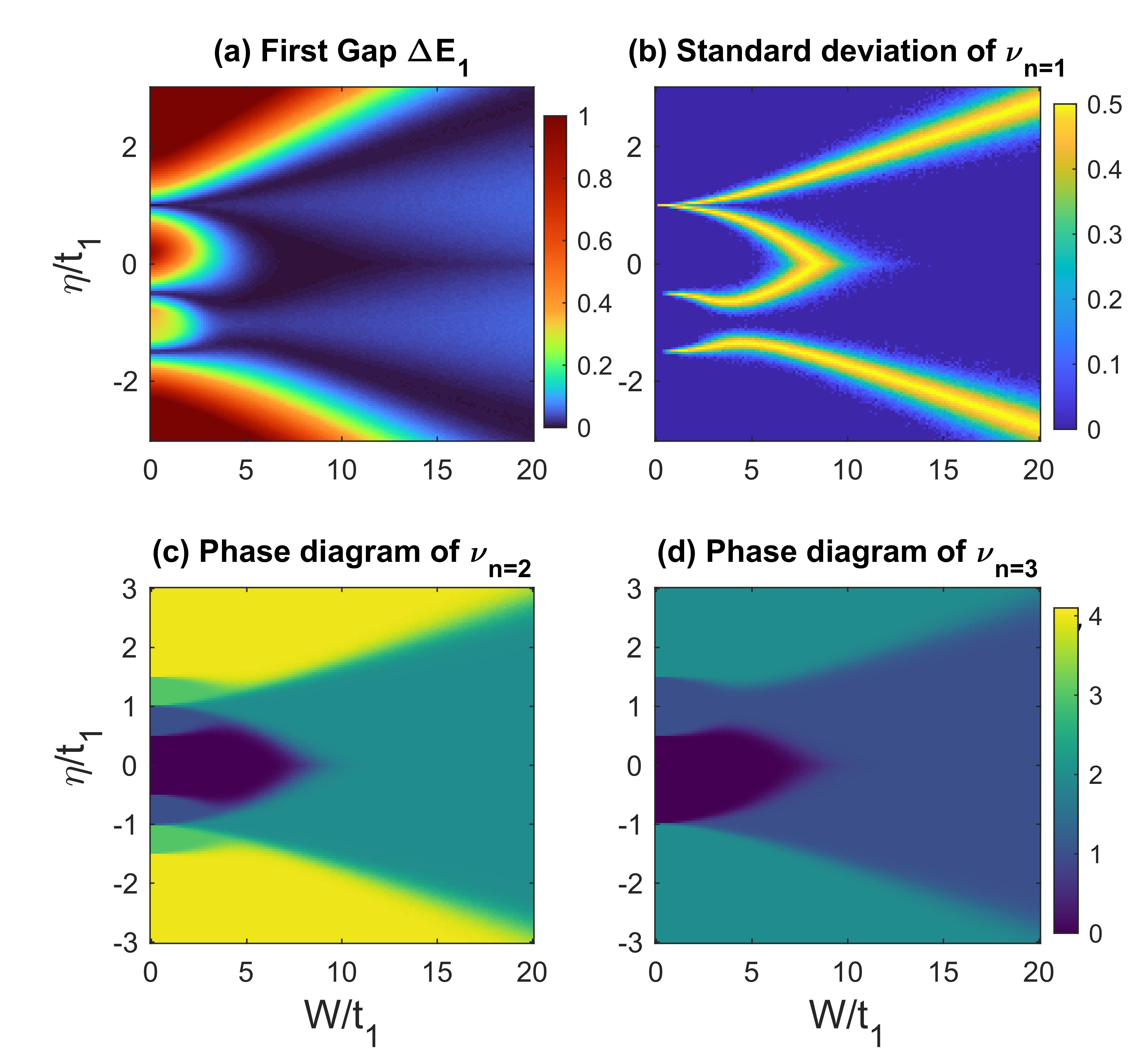}
\caption{\label{fig:SM_FIG_std_windingNumber}
$\eta-W$ phase diagram of: (a) first spectral gap, (b) the standard deviation of the inversion winding number among different realizations: $\Delta \nu  = \sqrt {{{\sum\nolimits_{i = 1}^N {{{\left( {{\nu ^{(i)}} - \bar \nu } \right)}^2}} }}/{N}} $.
Dashed lines indicate the phase boundary.
(c-d) inversion winding number of $n=2$ and $n=3$ bands.
The number of cell is chosen as $L=200$ and each data is averaged over $10^3$ random realizations.
}
\end{figure}

Here, we supplement some more data of the phase diagram in the presence of inversion-invariant disorder.
In Fig.~\ref{fig:SM_FIG_std_windingNumber}, we present the first gap (a) and the standard deviation of the inversion winding number (b) numerically.
It can be seen that the finite spectral gap only survives for weak disorder ($W/t_1 \ll 1$).
For strong disorder ($W/t_1  \gg  1$), the energy gap between the first and the second band is vague near the phase transition area.
However, in those areas where the inversion winding number fluctuate less, we can still distinguish a small finite energy gap.
According to the diagram of the standard deviation, one can clearly find wide and stable areas with extremely small fluctuation even if the spectral gap is small.
The disorder-induced topological phase transition is signified by large fluctuations of the inversion winding number, which highlights the phase boundary.
%

For completeness, we present the phase diagram of the inversion winding number for $n=2,3$ bands in Fig.~\ref{fig:SM_FIG_std_windingNumber} (c-d) with the same parameters.
It can be clearly seen that inversion winding numbers of the three bands still satisfy the relation: $\nu_{n=2} = \nu_{n=1} + \nu_{n=3}$.
The phase transition boundaries for two adjacent bands are well consistent.
This means that the topological phases in this three-band system protected by the space-time inversion symmetry are stable for strong symmetry-invariant disorders.
These results further evidence that the inversion winding number is capable to depict the band topology in such kind of systems.

\section{Evidence for the bulk-edge correspondence}
\label{appendix:Bulk_edge_correspondence}
\subsection{Singularity of $q(\theta, r)$ matrix}
%

\begin{figure}
\includegraphics[width = \columnwidth ]{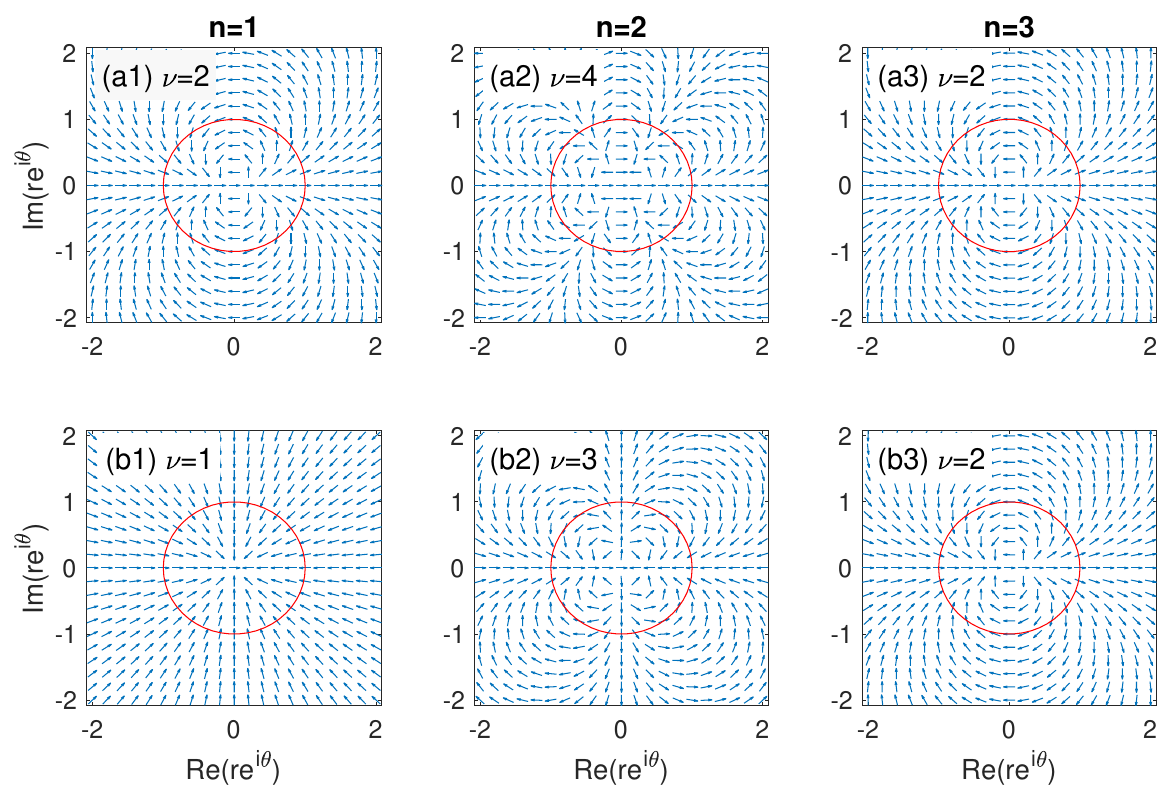}
\caption{\label{fig:FIG_SM_winding_number_r_theta}
Distribution of the complex number $\det q(\theta, r)$ on the complex plane for $n=1$ to $n=3$ bands.
(a) and (b) respectively corresponds to $\eta/t_1 = 2$ and $\eta/t_1 = -1.2$.
Red line indicates the $r=1$ circle, corresponding to normal TBC.
Other parameters are set to: $t_2/t_1 = 0.5$ and $L = 50$.
}
\end{figure}


In the main text, we argue that the appearance of the degenerate boundary modes is associated with the singularity of $q(\theta, r)$ matrix under OBC ($r=0$).
Here, we present the numerical calculation for $\det q(\theta, r)$ in Fig.~\ref{fig:FIG_SM_winding_number_r_theta}.
The inversion winding number under TBC can be easily extracted from the winding of $\det q(\theta, r)$.
Here we have set $t_2/t_1=1/2$ and set $\eta/t_1 = 2$ in (a1-a3) and $\eta/t_1 = -1.2$ in (b1-b3).
In both instances, the inversion winding number among these three bands are $\nu_{n=2} = \nu_{n=1} + \nu_{n=3}$.
Moreover, it can be identified that the singular point resides at the $r=0$.
This is consistent to our previous conclusion that the appearance of degenerate edge states is the result of non-trivial band topology.

\subsection{Adding coupling between the same type of sublattices}
\begin{figure}
\includegraphics[width = \columnwidth ]{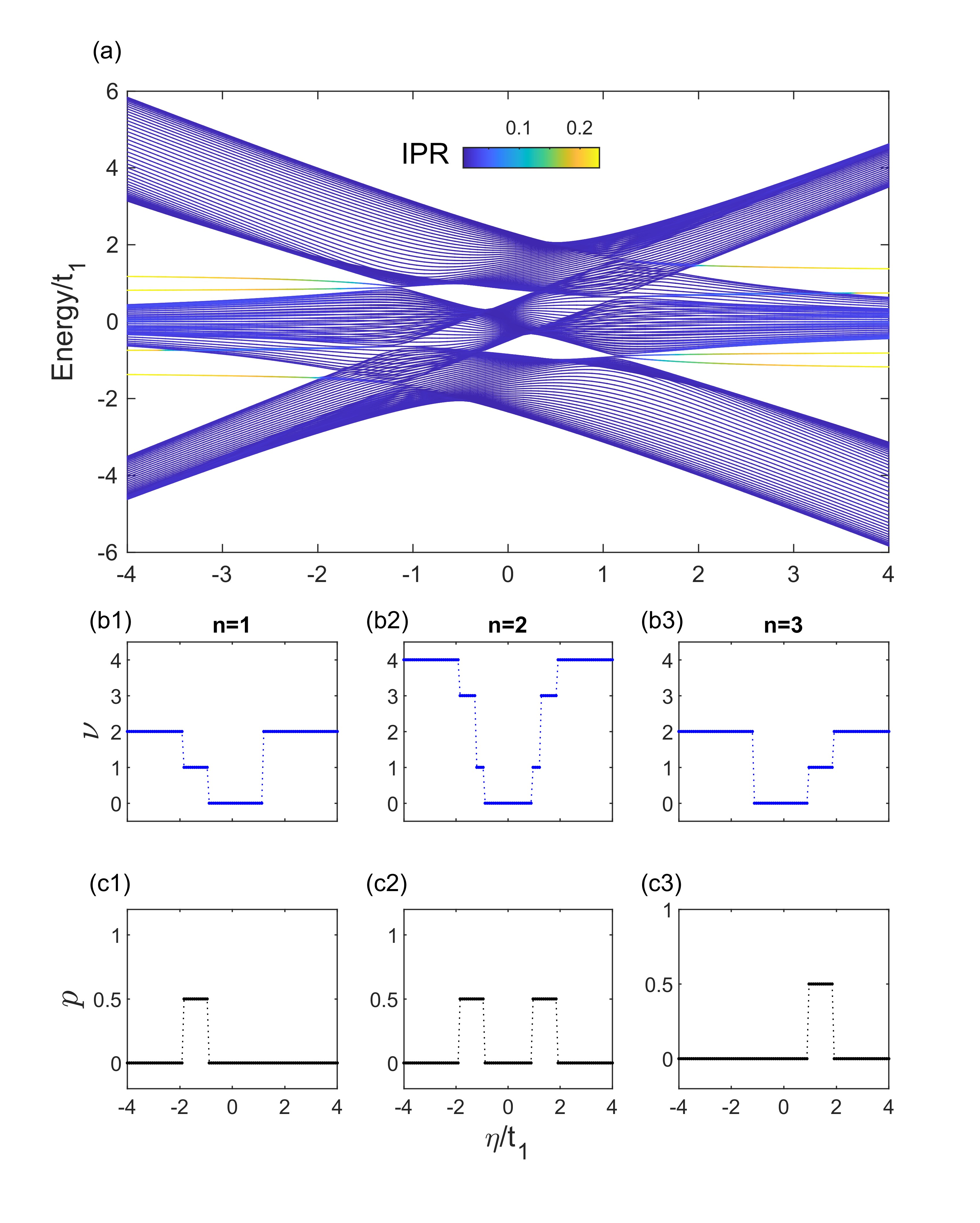}
\caption{\label{fig:FIG_SM_Spectrum_SameSublattices}
(a) Energy spectrum as a function of $\eta/t_1$.
The couplings between the same type of sublattices are added with the strength being $J/t_1=0.5$.
Other parameters are the same as Fig.~\ref{fig:FIG_three_band_spectrum}.
(b1-b3) and (c1-c3) respectively correspond to the inversion winding number and the polarization of the three bands.
}
\end{figure}

Here we discuss the case where couplings between the same sublattices are present.
This is different from the chiral-symmetric system, where this kind of coupling is forbidden.
It appears as the diagonal entry in the Bloch Hamiltonian.
In real space, this kind of term will couple different edge states, leading to the breaking of degeneracy.
As an example, we consider a couplings between sublattices $\sigma=1,3$ in the nearest cells such that $\Delta_k^{(1)} = -2J\cos (k), \Delta_k^{(2)}=0$ in the Bloch Hamiltonian, with $J$ denoting the coupling strength.
We add this term and plot the spectrum in Fig.~\ref{fig:FIG_SM_Spectrum_SameSublattices}.
It can be seen that the results are much more complicated than the $J=0$ case.
The previous four-fold degenerate edge states are now turned into two pairs of two-fold degenerate edge states.
The presence of couplings between the same sublattices lifts this degeneracy.
Moreover, some edge states merge in the bulk spectrum, which can be still characterized by relatively large IPR.
According to the discussion in  Sec.~\ref{sec:three_band_model_and_inversion_windng_number}, this newly added coupling term keeps the inversion winding number well-defined.
In Fig.~\ref{fig:FIG_SM_Spectrum_SameSublattices} (b-c), we compute the inversion winding number and polarization for each band.
The results show that the change of inversion winding number is associated with band closing, which is consistent to the discussion on the topological stability of inversion winding number in the main text.
%
%
%

\begin{figure}
\includegraphics[width = \columnwidth ]{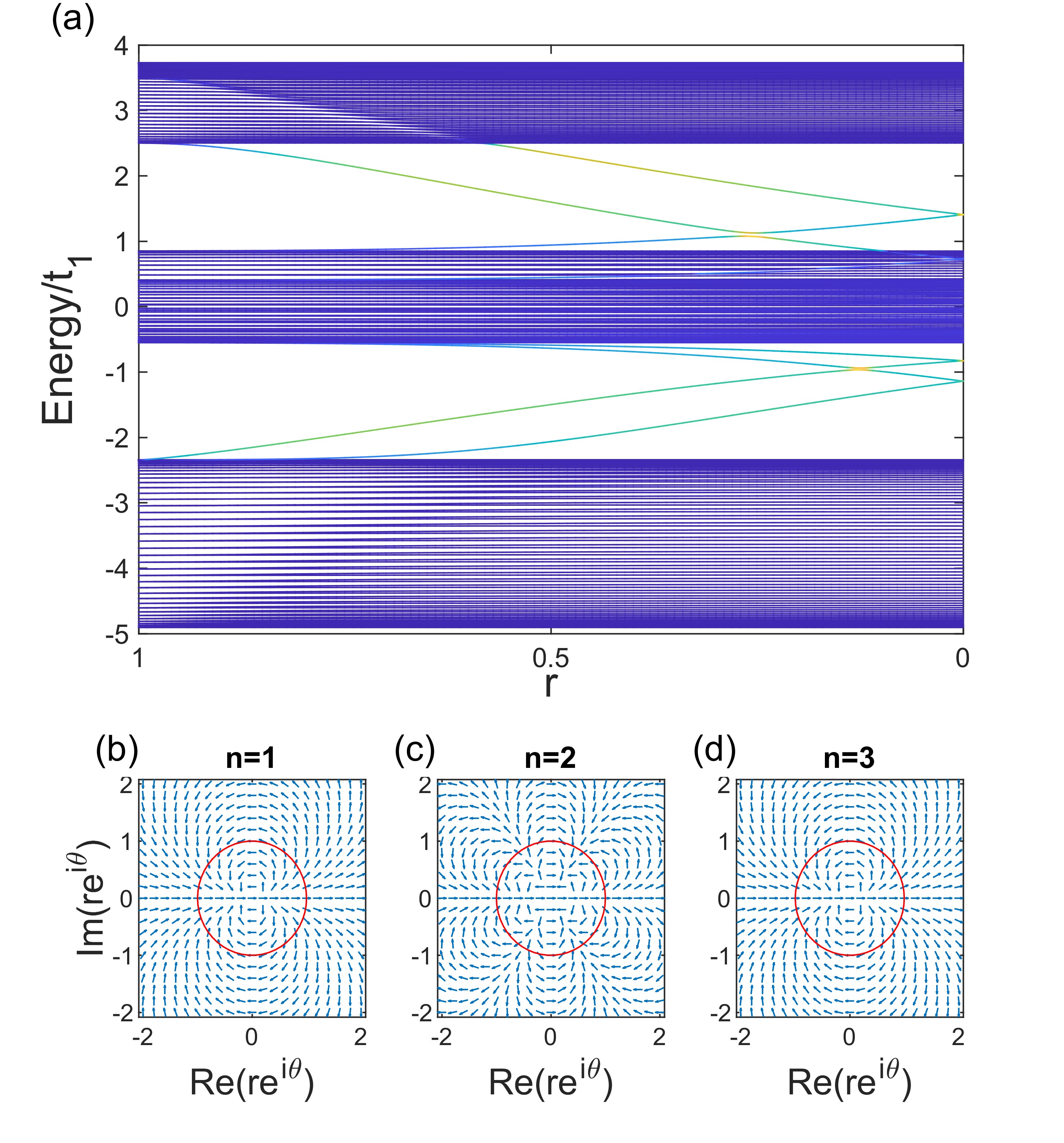}
\caption{\label{fig:FIG_SM_winding_same_sublattices}
(a) Energy spectrum as a function of boundary parameter $r\in [1,0]$.
(b-d) Distribution of the complex number $\mathrm{det} q(\theta, r)$ on the complex plane for $n=1,2,3$ band respectively.
Parameters are chosen as $t_1 = 1$, $t_2=0.5$, $\eta = 3$ and $J/t_1 = 0.5$, $L=50$.
}
\end{figure}

On the other hand, it can be noted that the bulk-edge correspondence becomes vague in this case, as some boundary modes reside in the bulk spectrum.
To see the bulk-edge correspondence more clearly, we plot the spectrum as a function of boundary parameter $r$ in Fig.~\ref{fig:FIG_SM_winding_same_sublattices}.
Particularly, we note that some boundary modes arise from different bands merge in the bulk spectrum, which can be identified through the IPR and the flow of spectrum in Fig.~\ref{fig:FIG_SM_winding_same_sublattices} (a) .
As discussed in Sec.~\ref{subsec:bulk_edge_correspondence}, non-trivial inversion winding number under PBC leads to boundary modes under OBC, although they may not be in the band gap.
The number of these boundary modes rising from the bulk spectrum is equal to the inversion winding number.
%
Under the OBC, the edge states are two-fold degenerate. 
It is remarkable that the emergence of boundary modes is consistent to the non-trivial inversion winding number in Fig.~\ref{fig:FIG_SM_winding_same_sublattices} (b-d) .
%

\section{Real-space model and experimental realization}
\label{appendix:model_and_experimental_realization}
\begin{figure}
\includegraphics[width =  0.9\columnwidth ]{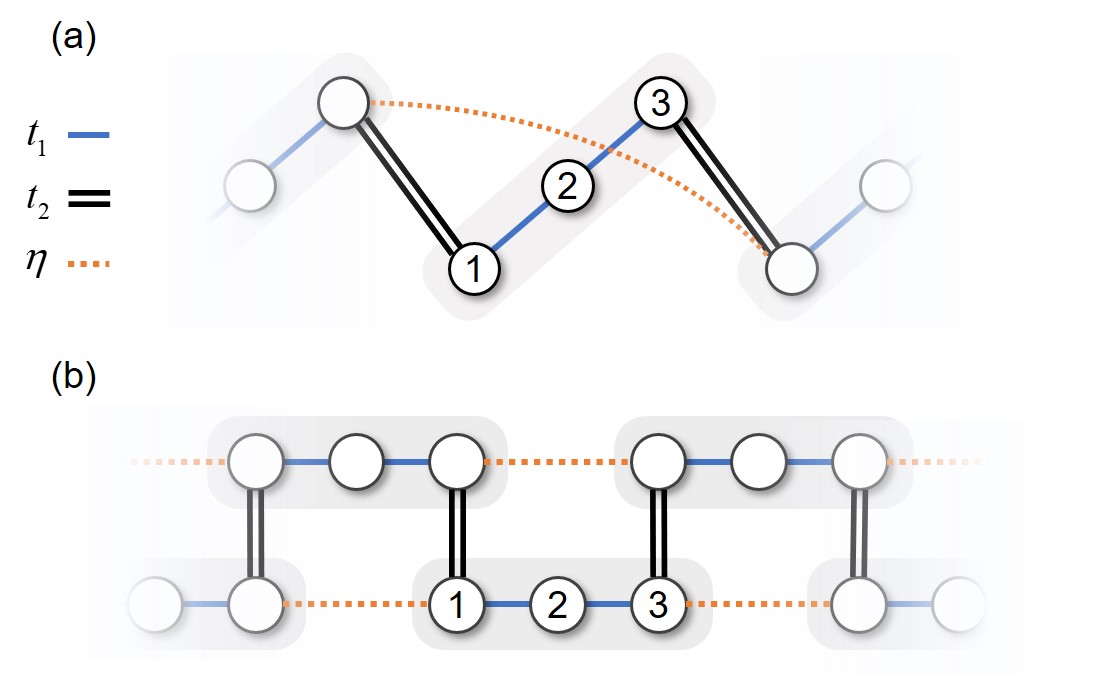}
\caption{\label{fig:FIG_SM_Model}
Schematic illustration of the 1D three-band model in real space.
(a) and (b) are equivalent.
}
\end{figure}

In this section, we discuss the three-band model in 1D presented in the main text.
By transforming the Bloch Hamiltonian to real space, we obtain the following tight-binding second-quantized form
\begin{eqnarray}
\hat H &=&  - {t_1}\sum\limits_j {\left( {\hat a_{2,j}^\dag {{\hat a}_{1,j}} + \hat a_{3,j}^\dag {{\hat a}_{1,j}}} \right)}  - {t_2}\sum\limits_j {\hat a_{3,j}^\dag {{\hat a}_{1,j + 1}}}  \nonumber \\
&&- \eta \sum\limits_j {\hat a_{3,j}^\dag {{\hat a}_{1,j + 2}}}  + {\rm{H}}.{\rm{c}}. ,
\end{eqnarray}
where $\hat a_{\sigma ,j}^\dag $ represents the particle creation operation at $\sigma = 1,2,3$ sublattice in the $j$th cell.
We present two equivalent real-space illustrations in Fig.~\ref{fig:FIG_SM_Model}.
Fig.~\ref{fig:FIG_SM_Model} (a) shows a general configuration of 1D zigzag chain.
$t_1$ and $t_2$ terms manifest as the nearest-neighbor tunneling while the $\eta$ terms correspond to the long-range tunneling.
%
We can modify the configuration and consider a equivalent ladder-like model, see Fig.~\ref{fig:FIG_SM_Model} (b).
In this ladder-like model, the long-range tunneling is arranged in a relatively short-range manner, which provides a more convenient approach for experimental realizations.

\subsection{Measurement from the edge states}
Since the in-gap edge states are degenerate, their arbitrary linear combinations are still the eigenstates of the system.
Nevertheless, the overall density distribution of these in-gap edge states is unique and it is possible to observe experimentally.
For example, the edge state can be excited selectively in acoustic systems \citep{PhysRevLett.131.157201,PhysRevB.108.205135} by varying the energy.
To detect the inversion winding number through the density distribution of the edge state, we can write the real-space formula for the edge state presented in the main text as the following convenient form
\begin{equation}
    \nu  = \frac{\mathcal{D}_{\rm{edge}}}{L}\frac{{\sum\limits_x {x\left( {{{\bar n}_{x,3}} - {{\bar n}_{x,1}}} \right)} }}{{\sum\limits_x {\left( {{{\bar n}_{x,3}} + {{\bar n}_{x,1}}} \right)} }},
\end{equation}
where $\bar{n}_{x,\sigma}$ represents the density distribution at $\sigma$ lattice of the $x$th cell and $\mathcal{D}_{\rm{edge}}$ is the degeneracy of the edge state.
Here we have assumed that the density distribution is normalized to identity: $\sum_{x,\sigma} \bar{n}_{x,\sigma} = 1$.
It is worth noting that, in experiment, the degeneracy of the edge state $\mathcal{D}_{\rm{edge}}$ is not obvious when measuring the density distribution.
In finite systems, the degeneracy may be lifted due to the large localization length of edge states, which offers possibility for resolving them.
On the other hand, even though the degeneracy is not lifted, we can manually add some extra terms to slightly break the degeneracy and then we can measure the winding number for different edge states.

\subsection{Measurement from the bulk states}
As for the measurement through the bulk state, it is generally difficult to directly probe the density distribution of them.
However, it is possible to realize it via the dynamical evolution.
Suppose that we can prepare a state being the superposition of the bulk state in a given band, the time evolution of this state yields $
    |\Psi \left( t \right)\rangle  = \sum\nolimits_{n \in {\rm{target}}} {{e^{ - i{E_n}t}}{c_n}|{\psi _n}\rangle } .$
The expectation of the operator $\hat{O}$ reads as
\begin{equation}
    \bar O(t) = \sum\limits_{n,n' \in {\rm{target}}} {{e^{ - i\left( {{E_n} - {E_{n'}}} \right)t}}{c_{n'}}^*{c_n}\langle {\psi _{n'}}|\hat O|{\psi _n}\rangle } .
\end{equation}
By averaging over time, we have
\begin{eqnarray}
   &&   \frac{1}{T}\int_0^T {dt\;\bar O(t)} =\nonumber \\
  &&   \sum\limits_{n,n' \in {\rm{target}}} { {\frac{1}{T}\int_0^T {dt\;{e^{ - i\left( {{E_n} - {E_{n'}}} \right)t}}} }  }    {c_{n'}}^*{c_n}\langle {\psi _{n'}}|\hat O|{\psi _n}\rangle \nonumber \\
\end{eqnarray}
If the evolution time is sufficiently long $T\to \infty$, we approximately have
\begin{equation}
    \frac{1}{T}\int_0^T {dt\;\bar O(t)}  \approx \sum\limits_{n \in {\rm{target}}} {|{c_n}{|^2}\langle {\psi _{n'}}|\hat O|{\psi _n}\rangle } ,
\end{equation}
which means that the long-time average leads to the ensemble average.
Thus, by selecting the appropriate initial state, we can accomplish the measurement of the topological invariant from bulk states.
It can be seen that the most important step is to prepare the proper initial state on the certain band.
To realize it, we consider a adiabatic preparation.
By adding the following extra potential to break the inversion symmetry:
\begin{equation}
    {{\hat H}_\Delta } = \Delta \sum\limits_x {\left( {{{\hat n}_{x,3}} - {{\hat n}_{x,1}}} \right)} ,
\end{equation}
the three bands will be well separated by large energy difference $\Delta  \gg  0$ and the system is approximately in the atomic limit.
Thus, it is quite easy to prepare the desired state as the superposition of the bulk states on certain band by using the Fock state localized in certain site.
Then, we can slowly reduce the extra potential to zero.
Finally, this state will greatly overlap with the target band.
Then we let this state evolve and measure according to the formula of the inversion winding number, which leads to the results in Fig.~\ref{fig:FIG_winding_number_Edge} (b) in the main text.

\bibliography{Inversion_Winding_BIB}

\begin{thebibliography}{83}%
\makeatletter
\providecommand \@ifxundefined [1]{%
 \@ifx{#1\undefined}
}%
\providecommand \@ifnum [1]{%
 \ifnum #1\expandafter \@firstoftwo
 \else \expandafter \@secondoftwo
 \fi
}%
\providecommand \@ifx [1]{%
 \ifx #1\expandafter \@firstoftwo
 \else \expandafter \@secondoftwo
 \fi
}%
\providecommand \natexlab [1]{#1}%
\providecommand \enquote  [1]{``#1''}%
\providecommand \bibnamefont  [1]{#1}%
\providecommand \bibfnamefont [1]{#1}%
\providecommand \citenamefont [1]{#1}%
\providecommand \href@noop [0]{\@secondoftwo}%
\providecommand \href [0]{\begingroup \@sanitize@url \@href}%
\providecommand \@href[1]{\@@startlink{#1}\@@href}%
\providecommand \@@href[1]{\endgroup#1\@@endlink}%
\providecommand \@sanitize@url [0]{\catcode `\\12\catcode `\$12\catcode
  `\&12\catcode `\#12\catcode `\^12\catcode `\_12\catcode `\%12\relax}%
\providecommand \@@startlink[1]{}%
\providecommand \@@endlink[0]{}%
\providecommand \url  [0]{\begingroup\@sanitize@url \@url }%
\providecommand \@url [1]{\endgroup\@href {#1}{\urlprefix }}%
\providecommand \urlprefix  [0]{URL }%
\providecommand \Eprint [0]{\href }%
\providecommand \doibase [0]{https://doi.org/}%
\providecommand \selectlanguage [0]{\@gobble}%
\providecommand \bibinfo  [0]{\@secondoftwo}%
\providecommand \bibfield  [0]{\@secondoftwo}%
\providecommand \translation [1]{[#1]}%
\providecommand \BibitemOpen [0]{}%
\providecommand \bibitemStop [0]{}%
\providecommand \bibitemNoStop [0]{.\EOS\space}%
\providecommand \EOS [0]{\spacefactor3000\relax}%
\providecommand \BibitemShut  [1]{\csname bibitem#1\endcsname}%
\let\auto@bib@innerbib\@empty
\bibitem [{\citenamefont {Fu}(2011)}]{PhysRevLett.106.106802}%
  \BibitemOpen
  \bibfield  {author} {\bibinfo {author} {\bibfnamefont {L.}~\bibnamefont
  {Fu}},\ }\bibfield  {title} {\bibinfo {title} {Topological crystalline
  insulators},\ }\href {https://doi.org/10.1103/PhysRevLett.106.106802}
  {\bibfield  {journal} {\bibinfo  {journal} {Phys. Rev. Lett.}\ }\textbf
  {\bibinfo {volume} {106}},\ \bibinfo {pages} {106802} (\bibinfo {year}
  {2011})}\BibitemShut {NoStop}%
\bibitem [{\citenamefont {Slager}\ \emph {et~al.}(2013)\citenamefont {Slager},
  \citenamefont {Mesaros}, \citenamefont {Juri{\v{c}}i{\'c}},\ and\
  \citenamefont {Zaanen}}]{slager2013space}%
  \BibitemOpen
  \bibfield  {author} {\bibinfo {author} {\bibfnamefont {R.-J.}\ \bibnamefont
  {Slager}}, \bibinfo {author} {\bibfnamefont {A.}~\bibnamefont {Mesaros}},
  \bibinfo {author} {\bibfnamefont {V.}~\bibnamefont {Juri{\v{c}}i{\'c}}},\
  and\ \bibinfo {author} {\bibfnamefont {J.}~\bibnamefont {Zaanen}},\
  }\bibfield  {title} {\bibinfo {title} {The space group classification of
  topological band-insulators},\ }\href {https://doi.org/10.1038/nphys2513}
  {\bibfield  {journal} {\bibinfo  {journal} {Nature Physics}\ }\textbf
  {\bibinfo {volume} {9}},\ \bibinfo {pages} {98} (\bibinfo {year}
  {2013})}\BibitemShut {NoStop}%
\bibitem [{\citenamefont {Kruthoff}\ \emph {et~al.}(2017)\citenamefont
  {Kruthoff}, \citenamefont {de~Boer}, \citenamefont {van Wezel}, \citenamefont
  {Kane},\ and\ \citenamefont {Slager}}]{PhysRevX.7.041069}%
  \BibitemOpen
  \bibfield  {author} {\bibinfo {author} {\bibfnamefont {J.}~\bibnamefont
  {Kruthoff}}, \bibinfo {author} {\bibfnamefont {J.}~\bibnamefont {de~Boer}},
  \bibinfo {author} {\bibfnamefont {J.}~\bibnamefont {van Wezel}}, \bibinfo
  {author} {\bibfnamefont {C.~L.}\ \bibnamefont {Kane}},\ and\ \bibinfo
  {author} {\bibfnamefont {R.-J.}\ \bibnamefont {Slager}},\ }\bibfield  {title}
  {\bibinfo {title} {Topological classification of crystalline insulators
  through band structure combinatorics},\ }\href
  {https://doi.org/10.1103/PhysRevX.7.041069} {\bibfield  {journal} {\bibinfo
  {journal} {Phys. Rev. X}\ }\textbf {\bibinfo {volume} {7}},\ \bibinfo {pages}
  {041069} (\bibinfo {year} {2017})}\BibitemShut {NoStop}%
\bibitem [{\citenamefont {Bradlyn}\ \emph {et~al.}(2017)\citenamefont
  {Bradlyn}, \citenamefont {Elcoro}, \citenamefont {Cano}, \citenamefont
  {Vergniory}, \citenamefont {Wang}, \citenamefont {Felser}, \citenamefont
  {Aroyo},\ and\ \citenamefont {Bernevig}}]{bradlyn2017topological}%
  \BibitemOpen
  \bibfield  {author} {\bibinfo {author} {\bibfnamefont {B.}~\bibnamefont
  {Bradlyn}}, \bibinfo {author} {\bibfnamefont {L.}~\bibnamefont {Elcoro}},
  \bibinfo {author} {\bibfnamefont {J.}~\bibnamefont {Cano}}, \bibinfo {author}
  {\bibfnamefont {M.~G.}\ \bibnamefont {Vergniory}}, \bibinfo {author}
  {\bibfnamefont {Z.}~\bibnamefont {Wang}}, \bibinfo {author} {\bibfnamefont
  {C.}~\bibnamefont {Felser}}, \bibinfo {author} {\bibfnamefont {M.~I.}\
  \bibnamefont {Aroyo}},\ and\ \bibinfo {author} {\bibfnamefont {B.~A.}\
  \bibnamefont {Bernevig}},\ }\bibfield  {title} {\bibinfo {title} {Topological
  quantum chemistry},\ }\href {https://doi.org/10.1038/nature23268} {\bibfield
  {journal} {\bibinfo  {journal} {Nature}\ }\textbf {\bibinfo {volume} {547}},\
  \bibinfo {pages} {298} (\bibinfo {year} {2017})}\BibitemShut {NoStop}%
\bibitem [{\citenamefont {Po}\ \emph {et~al.}(2017)\citenamefont {Po},
  \citenamefont {Vishwanath},\ and\ \citenamefont {Watanabe}}]{po2017symmetry}%
  \BibitemOpen
  \bibfield  {author} {\bibinfo {author} {\bibfnamefont {H.~C.}\ \bibnamefont
  {Po}}, \bibinfo {author} {\bibfnamefont {A.}~\bibnamefont {Vishwanath}},\
  and\ \bibinfo {author} {\bibfnamefont {H.}~\bibnamefont {Watanabe}},\
  }\bibfield  {title} {\bibinfo {title} {Symmetry-based indicators of band
  topology in the 230 space groups},\ }\href
  {https://doi.org/10.1038/s41467-017-00133-2} {\bibfield  {journal} {\bibinfo
  {journal} {Nature Communications}\ }\textbf {\bibinfo {volume} {8}},\
  \bibinfo {pages} {50} (\bibinfo {year} {2017})}\BibitemShut {NoStop}%
\bibitem [{\citenamefont {Fu}\ and\ \citenamefont
  {Kane}(2007)}]{PhysRevB.76.045302}%
  \BibitemOpen
  \bibfield  {author} {\bibinfo {author} {\bibfnamefont {L.}~\bibnamefont
  {Fu}}\ and\ \bibinfo {author} {\bibfnamefont {C.~L.}\ \bibnamefont {Kane}},\
  }\bibfield  {title} {\bibinfo {title} {Topological insulators with inversion
  symmetry},\ }\href {https://doi.org/10.1103/PhysRevB.76.045302} {\bibfield
  {journal} {\bibinfo  {journal} {Phys. Rev. B}\ }\textbf {\bibinfo {volume}
  {76}},\ \bibinfo {pages} {045302} (\bibinfo {year} {2007})}\BibitemShut
  {NoStop}%
\bibitem [{\citenamefont {Cornfeld}\ and\ \citenamefont
  {Chapman}(2019)}]{PhysRevB.99.075105}%
  \BibitemOpen
  \bibfield  {author} {\bibinfo {author} {\bibfnamefont {E.}~\bibnamefont
  {Cornfeld}}\ and\ \bibinfo {author} {\bibfnamefont {A.}~\bibnamefont
  {Chapman}},\ }\bibfield  {title} {\bibinfo {title} {Classification of
  crystalline topological insulators and superconductors with point group
  symmetries},\ }\href {https://doi.org/10.1103/PhysRevB.99.075105} {\bibfield
  {journal} {\bibinfo  {journal} {Phys. Rev. B}\ }\textbf {\bibinfo {volume}
  {99}},\ \bibinfo {pages} {075105} (\bibinfo {year} {2019})}\BibitemShut
  {NoStop}%
\bibitem [{\citenamefont {Wu}\ \emph {et~al.}(2019)\citenamefont {Wu},
  \citenamefont {Soluyanov},\ and\ \citenamefont {Bzdu{\v{s}}ek}}]{wu2019non}%
  \BibitemOpen
  \bibfield  {author} {\bibinfo {author} {\bibfnamefont {Q.}~\bibnamefont
  {Wu}}, \bibinfo {author} {\bibfnamefont {A.~A.}\ \bibnamefont {Soluyanov}},\
  and\ \bibinfo {author} {\bibfnamefont {T.}~\bibnamefont {Bzdu{\v{s}}ek}},\
  }\bibfield  {title} {\bibinfo {title} {{Non-Abelian band topology in
  noninteracting metals}},\ }\href {https://doi.org/10.1126/science.aau8740}
  {\bibfield  {journal} {\bibinfo  {journal} {Science}\ }\textbf {\bibinfo
  {volume} {365}},\ \bibinfo {pages} {1273} (\bibinfo {year}
  {2019})}\BibitemShut {NoStop}%
\bibitem [{\citenamefont {Slager}(2019)}]{SLAGER201924}%
  \BibitemOpen
  \bibfield  {author} {\bibinfo {author} {\bibfnamefont {R.-J.}\ \bibnamefont
  {Slager}},\ }\bibfield  {title} {\bibinfo {title} {The translational side of
  topological band insulators},\ }\href
  {https://doi.org/https://doi.org/10.1016/j.jpcs.2018.01.023} {\bibfield
  {journal} {\bibinfo  {journal} {Journal of Physics and Chemistry of Solids}\
  }\textbf {\bibinfo {volume} {128}},\ \bibinfo {pages} {24} (\bibinfo {year}
  {2019})},\ \bibinfo {note} {spin-Orbit Coupled Materials}\BibitemShut
  {NoStop}%
\bibitem [{\citenamefont {\"Unal}\ \emph {et~al.}(2020)\citenamefont {\"Unal},
  \citenamefont {Bouhon},\ and\ \citenamefont
  {Slager}}]{PhysRevLett.125.053601}%
  \BibitemOpen
  \bibfield  {author} {\bibinfo {author} {\bibfnamefont {F.~N.}\ \bibnamefont
  {\"Unal}}, \bibinfo {author} {\bibfnamefont {A.}~\bibnamefont {Bouhon}},\
  and\ \bibinfo {author} {\bibfnamefont {R.-J.}\ \bibnamefont {Slager}},\
  }\bibfield  {title} {\bibinfo {title} {{Topological Euler Class as a
  Dynamical Observable in Optical Lattices}},\ }\href
  {https://doi.org/10.1103/PhysRevLett.125.053601} {\bibfield  {journal}
  {\bibinfo  {journal} {Phys. Rev. Lett.}\ }\textbf {\bibinfo {volume} {125}},\
  \bibinfo {pages} {053601} (\bibinfo {year} {2020})}\BibitemShut {NoStop}%
\bibitem [{\citenamefont {Ando}\ and\ \citenamefont
  {Fu}(2015)}]{ando2015topological}%
  \BibitemOpen
  \bibfield  {author} {\bibinfo {author} {\bibfnamefont {Y.}~\bibnamefont
  {Ando}}\ and\ \bibinfo {author} {\bibfnamefont {L.}~\bibnamefont {Fu}},\
  }\bibfield  {title} {\bibinfo {title} {Topological crystalline insulators and
  topological superconductors: From concepts to materials},\ }\href
  {https://doi.org/https://doi.org/10.1146/annurev-conmatphys-031214-014501}
  {\bibfield  {journal} {\bibinfo  {journal} {Annual Review of Condensed Matter
  Physics}\ }\textbf {\bibinfo {volume} {6}},\ \bibinfo {pages} {361} (\bibinfo
  {year} {2015})}\BibitemShut {NoStop}%
\bibitem [{\citenamefont {Breach}\ \emph {et~al.}(2024)\citenamefont {Breach},
  \citenamefont {Slager},\ and\ \citenamefont
  {\"Unal}}]{PhysRevLett.133.093404}%
  \BibitemOpen
  \bibfield  {author} {\bibinfo {author} {\bibfnamefont {O.}~\bibnamefont
  {Breach}}, \bibinfo {author} {\bibfnamefont {R.-J.}\ \bibnamefont {Slager}},\
  and\ \bibinfo {author} {\bibfnamefont {F.~N.}\ \bibnamefont {\"Unal}},\
  }\bibfield  {title} {\bibinfo {title} {{Interferometry of Non-Abelian Band
  Singularities and Euler Class Topology}},\ }\href
  {https://doi.org/10.1103/PhysRevLett.133.093404} {\bibfield  {journal}
  {\bibinfo  {journal} {Phys. Rev. Lett.}\ }\textbf {\bibinfo {volume} {133}},\
  \bibinfo {pages} {093404} (\bibinfo {year} {2024})}\BibitemShut {NoStop}%
\bibitem [{\citenamefont {Bouhon}\ \emph {et~al.}(2019)\citenamefont {Bouhon},
  \citenamefont {Black-Schaffer},\ and\ \citenamefont
  {Slager}}]{PhysRevB.100.195135}%
  \BibitemOpen
  \bibfield  {author} {\bibinfo {author} {\bibfnamefont {A.}~\bibnamefont
  {Bouhon}}, \bibinfo {author} {\bibfnamefont {A.~M.}\ \bibnamefont
  {Black-Schaffer}},\ and\ \bibinfo {author} {\bibfnamefont {R.-J.}\
  \bibnamefont {Slager}},\ }\bibfield  {title} {\bibinfo {title} {{Wilson loop
  approach to fragile topology of split elementary band representations and
  topological crystalline insulators with time-reversal symmetry}},\ }\href
  {https://doi.org/10.1103/PhysRevB.100.195135} {\bibfield  {journal} {\bibinfo
   {journal} {Phys. Rev. B}\ }\textbf {\bibinfo {volume} {100}},\ \bibinfo
  {pages} {195135} (\bibinfo {year} {2019})}\BibitemShut {NoStop}%
\bibitem [{\citenamefont {Liu}\ and\ \citenamefont
  {Wakabayashi}(2017)}]{Feng2017}%
  \BibitemOpen
  \bibfield  {author} {\bibinfo {author} {\bibfnamefont {F.}~\bibnamefont
  {Liu}}\ and\ \bibinfo {author} {\bibfnamefont {K.}~\bibnamefont
  {Wakabayashi}},\ }\bibfield  {title} {\bibinfo {title} {{Novel Topological
  Phase with a Zero Berry Curvature}},\ }\href
  {https://doi.org/10.1103/PhysRevLett.118.076803} {\bibfield  {journal}
  {\bibinfo  {journal} {Phys. Rev. Lett.}\ }\textbf {\bibinfo {volume} {118}},\
  \bibinfo {pages} {076803} (\bibinfo {year} {2017})}\BibitemShut {NoStop}%
\bibitem [{\citenamefont {Benalcazar}\ \emph
  {et~al.}(2017{\natexlab{a}})\citenamefont {Benalcazar}, \citenamefont
  {Bernevig},\ and\ \citenamefont {Hughes}}]{benalcazar2017quantized}%
  \BibitemOpen
  \bibfield  {author} {\bibinfo {author} {\bibfnamefont {W.~A.}\ \bibnamefont
  {Benalcazar}}, \bibinfo {author} {\bibfnamefont {B.~A.}\ \bibnamefont
  {Bernevig}},\ and\ \bibinfo {author} {\bibfnamefont {T.~L.}\ \bibnamefont
  {Hughes}},\ }\bibfield  {title} {\bibinfo {title} {Quantized electric
  multipole insulators},\ }\href {https://doi.org/10.1126/science.aah6442}
  {\bibfield  {journal} {\bibinfo  {journal} {Science}\ }\textbf {\bibinfo
  {volume} {357}},\ \bibinfo {pages} {61} (\bibinfo {year}
  {2017}{\natexlab{a}})}\BibitemShut {NoStop}%
\bibitem [{\citenamefont {Benalcazar}\ \emph
  {et~al.}(2017{\natexlab{b}})\citenamefont {Benalcazar}, \citenamefont
  {Bernevig},\ and\ \citenamefont {Hughes}}]{PhysRevB.96.245115}%
  \BibitemOpen
  \bibfield  {author} {\bibinfo {author} {\bibfnamefont {W.~A.}\ \bibnamefont
  {Benalcazar}}, \bibinfo {author} {\bibfnamefont {B.~A.}\ \bibnamefont
  {Bernevig}},\ and\ \bibinfo {author} {\bibfnamefont {T.~L.}\ \bibnamefont
  {Hughes}},\ }\bibfield  {title} {\bibinfo {title} {Electric multipole
  moments, topological multipole moment pumping, and chiral hinge states in
  crystalline insulators},\ }\href {https://doi.org/10.1103/PhysRevB.96.245115}
  {\bibfield  {journal} {\bibinfo  {journal} {Phys. Rev. B}\ }\textbf {\bibinfo
  {volume} {96}},\ \bibinfo {pages} {245115} (\bibinfo {year}
  {2017}{\natexlab{b}})}\BibitemShut {NoStop}%
\bibitem [{\citenamefont {Bouhon}\ \emph {et~al.}(2020)\citenamefont {Bouhon},
  \citenamefont {Wu}, \citenamefont {Slager}, \citenamefont {Weng},
  \citenamefont {Yazyev},\ and\ \citenamefont {Bzdu{\v{s}}ek}}]{bouhon2020non}%
  \BibitemOpen
  \bibfield  {author} {\bibinfo {author} {\bibfnamefont {A.}~\bibnamefont
  {Bouhon}}, \bibinfo {author} {\bibfnamefont {Q.}~\bibnamefont {Wu}}, \bibinfo
  {author} {\bibfnamefont {R.-J.}\ \bibnamefont {Slager}}, \bibinfo {author}
  {\bibfnamefont {H.}~\bibnamefont {Weng}}, \bibinfo {author} {\bibfnamefont
  {O.~V.}\ \bibnamefont {Yazyev}},\ and\ \bibinfo {author} {\bibfnamefont
  {T.}~\bibnamefont {Bzdu{\v{s}}ek}},\ }\bibfield  {title} {\bibinfo {title}
  {Non-abelian reciprocal braiding of weyl points and its manifestation in
  zrte},\ }\href {https://doi.org/10.1038/s41567-020-0967-9} {\bibfield
  {journal} {\bibinfo  {journal} {Nature Physics}\ }\textbf {\bibinfo {volume}
  {16}},\ \bibinfo {pages} {1137} (\bibinfo {year} {2020})}\BibitemShut
  {NoStop}%
\bibitem [{\citenamefont {Jiang}\ \emph {et~al.}(2021)\citenamefont {Jiang},
  \citenamefont {Bouhon}, \citenamefont {Lin}, \citenamefont {Zhou},
  \citenamefont {Hou}, \citenamefont {Li}, \citenamefont {Slager},\ and\
  \citenamefont {Jiang}}]{jiang2021experimental}%
  \BibitemOpen
  \bibfield  {author} {\bibinfo {author} {\bibfnamefont {B.}~\bibnamefont
  {Jiang}}, \bibinfo {author} {\bibfnamefont {A.}~\bibnamefont {Bouhon}},
  \bibinfo {author} {\bibfnamefont {Z.-K.}\ \bibnamefont {Lin}}, \bibinfo
  {author} {\bibfnamefont {X.}~\bibnamefont {Zhou}}, \bibinfo {author}
  {\bibfnamefont {B.}~\bibnamefont {Hou}}, \bibinfo {author} {\bibfnamefont
  {F.}~\bibnamefont {Li}}, \bibinfo {author} {\bibfnamefont {R.-J.}\
  \bibnamefont {Slager}},\ and\ \bibinfo {author} {\bibfnamefont {J.-H.}\
  \bibnamefont {Jiang}},\ }\bibfield  {title} {\bibinfo {title} {{Experimental
  observation of non-Abelian topological acoustic semimetals and their phase
  transitions}},\ }\href {https://doi.org/10.1038/s41567-021-01340-x}
  {\bibfield  {journal} {\bibinfo  {journal} {Nature Physics}\ }\textbf
  {\bibinfo {volume} {17}},\ \bibinfo {pages} {1239} (\bibinfo {year}
  {2021})}\BibitemShut {NoStop}%
\bibitem [{\citenamefont {Guo}\ \emph {et~al.}(2021)\citenamefont {Guo},
  \citenamefont {Jiang}, \citenamefont {Zhang}, \citenamefont {Zhang},
  \citenamefont {Zhang}, \citenamefont {Yang}, \citenamefont {Zhang},\ and\
  \citenamefont {Chan}}]{guo2021experimental}%
  \BibitemOpen
  \bibfield  {author} {\bibinfo {author} {\bibfnamefont {Q.}~\bibnamefont
  {Guo}}, \bibinfo {author} {\bibfnamefont {T.}~\bibnamefont {Jiang}}, \bibinfo
  {author} {\bibfnamefont {R.-Y.}\ \bibnamefont {Zhang}}, \bibinfo {author}
  {\bibfnamefont {L.}~\bibnamefont {Zhang}}, \bibinfo {author} {\bibfnamefont
  {Z.-Q.}\ \bibnamefont {Zhang}}, \bibinfo {author} {\bibfnamefont
  {B.}~\bibnamefont {Yang}}, \bibinfo {author} {\bibfnamefont {S.}~\bibnamefont
  {Zhang}},\ and\ \bibinfo {author} {\bibfnamefont {C.~T.}\ \bibnamefont
  {Chan}},\ }\bibfield  {title} {\bibinfo {title} {{Experimental observation of
  non-Abelian topological charges and edge states}},\ }\href
  {https://doi.org/10.1038/s41586-021-03521-3} {\bibfield  {journal} {\bibinfo
  {journal} {Nature}\ }\textbf {\bibinfo {volume} {594}},\ \bibinfo {pages}
  {195} (\bibinfo {year} {2021})}\BibitemShut {NoStop}%
\bibitem [{\citenamefont {Peng}\ \emph {et~al.}(2022)\citenamefont {Peng},
  \citenamefont {Bouhon}, \citenamefont {Monserrat},\ and\ \citenamefont
  {Slager}}]{peng2022phonons}%
  \BibitemOpen
  \bibfield  {author} {\bibinfo {author} {\bibfnamefont {B.}~\bibnamefont
  {Peng}}, \bibinfo {author} {\bibfnamefont {A.}~\bibnamefont {Bouhon}},
  \bibinfo {author} {\bibfnamefont {B.}~\bibnamefont {Monserrat}},\ and\
  \bibinfo {author} {\bibfnamefont {R.-J.}\ \bibnamefont {Slager}},\ }\bibfield
   {title} {\bibinfo {title} {{Phonons as a platform for non-Abelian braiding
  and its manifestation in layered silicates}},\ }\href
  {https://doi.org/10.1038/s41467-022-28046-9} {\bibfield  {journal} {\bibinfo
  {journal} {Nature Communications}\ }\textbf {\bibinfo {volume} {13}},\
  \bibinfo {pages} {423} (\bibinfo {year} {2022})}\BibitemShut {NoStop}%
\bibitem [{\citenamefont {Li}\ and\ \citenamefont {Hu}(2023)}]{li2023floquet}%
  \BibitemOpen
  \bibfield  {author} {\bibinfo {author} {\bibfnamefont {T.}~\bibnamefont
  {Li}}\ and\ \bibinfo {author} {\bibfnamefont {H.}~\bibnamefont {Hu}},\
  }\bibfield  {title} {\bibinfo {title} {{Floquet non-Abelian topological
  insulator and multifold bulk-edge correspondence}},\ }\href
  {https://doi.org/10.1038/s41467-023-42139-z} {\bibfield  {journal} {\bibinfo
  {journal} {Nature Communications}\ }\textbf {\bibinfo {volume} {14}},\
  \bibinfo {pages} {6418} (\bibinfo {year} {2023})}\BibitemShut {NoStop}%
\bibitem [{\citenamefont {Wang}\ \emph {et~al.}(2024)\citenamefont {Wang},
  \citenamefont {Zhu}, \citenamefont {Zhu},\ and\ \citenamefont
  {Zheng}}]{PhysRevA.110.023321}%
  \BibitemOpen
  \bibfield  {author} {\bibinfo {author} {\bibfnamefont {Q.-D.}\ \bibnamefont
  {Wang}}, \bibinfo {author} {\bibfnamefont {Y.-Q.}\ \bibnamefont {Zhu}},
  \bibinfo {author} {\bibfnamefont {S.-L.}\ \bibnamefont {Zhu}},\ and\ \bibinfo
  {author} {\bibfnamefont {Z.}~\bibnamefont {Zheng}},\ }\bibfield  {title}
  {\bibinfo {title} {{Synthetic non-Abelian topological charges in ultracold
  atomic gases}},\ }\href {https://doi.org/10.1103/PhysRevA.110.023321}
  {\bibfield  {journal} {\bibinfo  {journal} {Phys. Rev. A}\ }\textbf {\bibinfo
  {volume} {110}},\ \bibinfo {pages} {023321} (\bibinfo {year}
  {2024})}\BibitemShut {NoStop}%
\bibitem [{\citenamefont {Slager}\ \emph {et~al.}(2024)\citenamefont {Slager},
  \citenamefont {Bouhon},\ and\ \citenamefont {{\"U}nal}}]{slager2024non}%
  \BibitemOpen
  \bibfield  {author} {\bibinfo {author} {\bibfnamefont {R.-J.}\ \bibnamefont
  {Slager}}, \bibinfo {author} {\bibfnamefont {A.}~\bibnamefont {Bouhon}},\
  and\ \bibinfo {author} {\bibfnamefont {F.~N.}\ \bibnamefont {{\"U}nal}},\
  }\bibfield  {title} {\bibinfo {title} {{Non-Abelian Floquet braiding and
  anomalous Dirac string phase in periodically driven systems}},\ }\href
  {https://doi.org/10.1038/s41467-024-45302-2} {\bibfield  {journal} {\bibinfo
  {journal} {Nature Communications}\ }\textbf {\bibinfo {volume} {15}},\
  \bibinfo {pages} {1144} (\bibinfo {year} {2024})}\BibitemShut {NoStop}%
\bibitem [{\citenamefont {H\"oller}\ and\ \citenamefont
  {Alexandradinata}(2018)}]{Holler2018}%
  \BibitemOpen
  \bibfield  {author} {\bibinfo {author} {\bibfnamefont {J.}~\bibnamefont
  {H\"oller}}\ and\ \bibinfo {author} {\bibfnamefont {A.}~\bibnamefont
  {Alexandradinata}},\ }\bibfield  {title} {\bibinfo {title} {Topological bloch
  oscillations},\ }\href {https://doi.org/10.1103/PhysRevB.98.024310}
  {\bibfield  {journal} {\bibinfo  {journal} {Phys. Rev. B}\ }\textbf {\bibinfo
  {volume} {98}},\ \bibinfo {pages} {024310} (\bibinfo {year}
  {2018})}\BibitemShut {NoStop}%
\bibitem [{\citenamefont {Di~Liberto}\ \emph {et~al.}(2020)\citenamefont
  {Di~Liberto}, \citenamefont {Goldman},\ and\ \citenamefont
  {Palumbo}}]{di2020non}%
  \BibitemOpen
  \bibfield  {author} {\bibinfo {author} {\bibfnamefont {M.}~\bibnamefont
  {Di~Liberto}}, \bibinfo {author} {\bibfnamefont {N.}~\bibnamefont
  {Goldman}},\ and\ \bibinfo {author} {\bibfnamefont {G.}~\bibnamefont
  {Palumbo}},\ }\bibfield  {title} {\bibinfo {title} {{Non-Abelian Bloch
  oscillations in higher-order topological insulators}},\ }\href
  {https://doi.org/10.1038/s41467-020-19518-x} {\bibfield  {journal} {\bibinfo
  {journal} {Nature communications}\ }\textbf {\bibinfo {volume} {11}},\
  \bibinfo {pages} {5942} (\bibinfo {year} {2020})}\BibitemShut {NoStop}%
\bibitem [{\citenamefont {Hughes}\ \emph {et~al.}(2011)\citenamefont {Hughes},
  \citenamefont {Prodan},\ and\ \citenamefont {Bernevig}}]{PhysRevB.83.245132}%
  \BibitemOpen
  \bibfield  {author} {\bibinfo {author} {\bibfnamefont {T.~L.}\ \bibnamefont
  {Hughes}}, \bibinfo {author} {\bibfnamefont {E.}~\bibnamefont {Prodan}},\
  and\ \bibinfo {author} {\bibfnamefont {B.~A.}\ \bibnamefont {Bernevig}},\
  }\bibfield  {title} {\bibinfo {title} {Inversion-symmetric topological
  insulators},\ }\href {https://doi.org/10.1103/PhysRevB.83.245132} {\bibfield
  {journal} {\bibinfo  {journal} {Phys. Rev. B}\ }\textbf {\bibinfo {volume}
  {83}},\ \bibinfo {pages} {245132} (\bibinfo {year} {2011})}\BibitemShut
  {NoStop}%
\bibitem [{\citenamefont {Alexandradinata}\ \emph {et~al.}(2014)\citenamefont
  {Alexandradinata}, \citenamefont {Dai},\ and\ \citenamefont
  {Bernevig}}]{PhysRevB.89.155114}%
  \BibitemOpen
  \bibfield  {author} {\bibinfo {author} {\bibfnamefont {A.}~\bibnamefont
  {Alexandradinata}}, \bibinfo {author} {\bibfnamefont {X.}~\bibnamefont
  {Dai}},\ and\ \bibinfo {author} {\bibfnamefont {B.~A.}\ \bibnamefont
  {Bernevig}},\ }\bibfield  {title} {\bibinfo {title} {Wilson-loop
  characterization of inversion-symmetric topological insulators},\ }\href
  {https://doi.org/10.1103/PhysRevB.89.155114} {\bibfield  {journal} {\bibinfo
  {journal} {Phys. Rev. B}\ }\textbf {\bibinfo {volume} {89}},\ \bibinfo
  {pages} {155114} (\bibinfo {year} {2014})}\BibitemShut {NoStop}%
\bibitem [{\citenamefont {Shiozaki}\ and\ \citenamefont
  {Sato}(2014)}]{PhysRevB.90.165114}%
  \BibitemOpen
  \bibfield  {author} {\bibinfo {author} {\bibfnamefont {K.}~\bibnamefont
  {Shiozaki}}\ and\ \bibinfo {author} {\bibfnamefont {M.}~\bibnamefont
  {Sato}},\ }\bibfield  {title} {\bibinfo {title} {Topology of crystalline
  insulators and superconductors},\ }\href
  {https://doi.org/10.1103/PhysRevB.90.165114} {\bibfield  {journal} {\bibinfo
  {journal} {Phys. Rev. B}\ }\textbf {\bibinfo {volume} {90}},\ \bibinfo
  {pages} {165114} (\bibinfo {year} {2014})}\BibitemShut {NoStop}%
\bibitem [{\citenamefont {Xiao}\ \emph {et~al.}(2014)\citenamefont {Xiao},
  \citenamefont {Zhang},\ and\ \citenamefont {Chan}}]{PhysRevX.4.021017}%
  \BibitemOpen
  \bibfield  {author} {\bibinfo {author} {\bibfnamefont {M.}~\bibnamefont
  {Xiao}}, \bibinfo {author} {\bibfnamefont {Z.~Q.}\ \bibnamefont {Zhang}},\
  and\ \bibinfo {author} {\bibfnamefont {C.~T.}\ \bibnamefont {Chan}},\
  }\bibfield  {title} {\bibinfo {title} {Surface impedance and bulk band
  geometric phases in one-dimensional systems},\ }\href
  {https://doi.org/10.1103/PhysRevX.4.021017} {\bibfield  {journal} {\bibinfo
  {journal} {Phys. Rev. X}\ }\textbf {\bibinfo {volume} {4}},\ \bibinfo {pages}
  {021017} (\bibinfo {year} {2014})}\BibitemShut {NoStop}%
\bibitem [{\citenamefont {Hwang}\ \emph {et~al.}(2019)\citenamefont {Hwang},
  \citenamefont {Ahn},\ and\ \citenamefont {Yang}}]{PhysRevB.100.205126}%
  \BibitemOpen
  \bibfield  {author} {\bibinfo {author} {\bibfnamefont {Y.}~\bibnamefont
  {Hwang}}, \bibinfo {author} {\bibfnamefont {J.}~\bibnamefont {Ahn}},\ and\
  \bibinfo {author} {\bibfnamefont {B.-J.}\ \bibnamefont {Yang}},\ }\bibfield
  {title} {\bibinfo {title} {{Fragile topology protected by inversion symmetry:
  Diagnosis, bulk-boundary correspondence, and Wilson loop}},\ }\href
  {https://doi.org/10.1103/PhysRevB.100.205126} {\bibfield  {journal} {\bibinfo
   {journal} {Phys. Rev. B}\ }\textbf {\bibinfo {volume} {100}},\ \bibinfo
  {pages} {205126} (\bibinfo {year} {2019})}\BibitemShut {NoStop}%
\bibitem [{\citenamefont {Vaidya}\ \emph {et~al.}(2023)\citenamefont {Vaidya},
  \citenamefont {Ghorashi}, \citenamefont {Christensen}, \citenamefont
  {Rechtsman},\ and\ \citenamefont {Benalcazar}}]{PhysRevB.108.085116}%
  \BibitemOpen
  \bibfield  {author} {\bibinfo {author} {\bibfnamefont {S.}~\bibnamefont
  {Vaidya}}, \bibinfo {author} {\bibfnamefont {A.}~\bibnamefont {Ghorashi}},
  \bibinfo {author} {\bibfnamefont {T.}~\bibnamefont {Christensen}}, \bibinfo
  {author} {\bibfnamefont {M.~C.}\ \bibnamefont {Rechtsman}},\ and\ \bibinfo
  {author} {\bibfnamefont {W.~A.}\ \bibnamefont {Benalcazar}},\ }\bibfield
  {title} {\bibinfo {title} {Topological phases of photonic crystals under
  crystalline symmetries},\ }\href
  {https://doi.org/10.1103/PhysRevB.108.085116} {\bibfield  {journal} {\bibinfo
   {journal} {Phys. Rev. B}\ }\textbf {\bibinfo {volume} {108}},\ \bibinfo
  {pages} {085116} (\bibinfo {year} {2023})}\BibitemShut {NoStop}%
\bibitem [{\citenamefont {Zak}(1980)}]{PhysRevLett.45.1025}%
  \BibitemOpen
  \bibfield  {author} {\bibinfo {author} {\bibfnamefont {J.}~\bibnamefont
  {Zak}},\ }\bibfield  {title} {\bibinfo {title} {Symmetry specification of
  bands in solids},\ }\href {https://doi.org/10.1103/PhysRevLett.45.1025}
  {\bibfield  {journal} {\bibinfo  {journal} {Phys. Rev. Lett.}\ }\textbf
  {\bibinfo {volume} {45}},\ \bibinfo {pages} {1025} (\bibinfo {year}
  {1980})}\BibitemShut {NoStop}%
\bibitem [{\citenamefont {Zak}(1981)}]{PhysRevB.23.2824}%
  \BibitemOpen
  \bibfield  {author} {\bibinfo {author} {\bibfnamefont {J.}~\bibnamefont
  {Zak}},\ }\bibfield  {title} {\bibinfo {title} {Band representations and
  symmetry types of bands in solids},\ }\href
  {https://doi.org/10.1103/PhysRevB.23.2824} {\bibfield  {journal} {\bibinfo
  {journal} {Phys. Rev. B}\ }\textbf {\bibinfo {volume} {23}},\ \bibinfo
  {pages} {2824} (\bibinfo {year} {1981})}\BibitemShut {NoStop}%
\bibitem [{\citenamefont {Zak}(1982)}]{PhysRevLett.48.359}%
  \BibitemOpen
  \bibfield  {author} {\bibinfo {author} {\bibfnamefont {J.}~\bibnamefont
  {Zak}},\ }\bibfield  {title} {\bibinfo {title} {Band center---a conserved
  quantity in solids},\ }\href {https://doi.org/10.1103/PhysRevLett.48.359}
  {\bibfield  {journal} {\bibinfo  {journal} {Phys. Rev. Lett.}\ }\textbf
  {\bibinfo {volume} {48}},\ \bibinfo {pages} {359} (\bibinfo {year}
  {1982})}\BibitemShut {NoStop}%
\bibitem [{\citenamefont {Zak}(1989)}]{PhysRevLett.62.2747}%
  \BibitemOpen
  \bibfield  {author} {\bibinfo {author} {\bibfnamefont {J.}~\bibnamefont
  {Zak}},\ }\bibfield  {title} {\bibinfo {title} {Berry's phase for energy
  bands in solids},\ }\href {https://doi.org/10.1103/PhysRevLett.62.2747}
  {\bibfield  {journal} {\bibinfo  {journal} {Phys. Rev. Lett.}\ }\textbf
  {\bibinfo {volume} {62}},\ \bibinfo {pages} {2747} (\bibinfo {year}
  {1989})}\BibitemShut {NoStop}%
\bibitem [{\citenamefont {Song}\ \emph {et~al.}(2018)\citenamefont {Song},
  \citenamefont {Zhang}, \citenamefont {He}, \citenamefont {Poon},
  \citenamefont {Hajiyev}, \citenamefont {Zhang}, \citenamefont {Liu},\ and\
  \citenamefont {Jo}}]{Song_sciadv_2018}%
  \BibitemOpen
  \bibfield  {author} {\bibinfo {author} {\bibfnamefont {B.}~\bibnamefont
  {Song}}, \bibinfo {author} {\bibfnamefont {L.}~\bibnamefont {Zhang}},
  \bibinfo {author} {\bibfnamefont {C.}~\bibnamefont {He}}, \bibinfo {author}
  {\bibfnamefont {T.~F.~J.}\ \bibnamefont {Poon}}, \bibinfo {author}
  {\bibfnamefont {E.}~\bibnamefont {Hajiyev}}, \bibinfo {author} {\bibfnamefont
  {S.}~\bibnamefont {Zhang}}, \bibinfo {author} {\bibfnamefont {X.-J.}\
  \bibnamefont {Liu}},\ and\ \bibinfo {author} {\bibfnamefont {G.-B.}\
  \bibnamefont {Jo}},\ }\bibfield  {title} {\bibinfo {title} {Observation of
  symmetry-protected topological band with ultracold fermions},\ }\href
  {https://doi.org/10.1126/sciadv.aao4748} {\bibfield  {journal} {\bibinfo
  {journal} {Science Advances}\ }\textbf {\bibinfo {volume} {4}},\ \bibinfo
  {pages} {eaao4748} (\bibinfo {year} {2018})}\BibitemShut {NoStop}%
\bibitem [{\citenamefont {Het\'enyi}(2020)}]{PhysRevResearch.2.023277}%
  \BibitemOpen
  \bibfield  {author} {\bibinfo {author} {\bibfnamefont {B.}~\bibnamefont
  {Het\'enyi}},\ }\bibfield  {title} {\bibinfo {title} {Interaction-driven
  polarization shift in the
  $t\text{\ensuremath{-}}v\text{\ensuremath{-}}{V}^{\ensuremath{'}}$ lattice
  fermion model at half filling: Emergent haldane phase},\ }\href
  {https://doi.org/10.1103/PhysRevResearch.2.023277} {\bibfield  {journal}
  {\bibinfo  {journal} {Phys. Rev. Res.}\ }\textbf {\bibinfo {volume} {2}},\
  \bibinfo {pages} {023277} (\bibinfo {year} {2020})}\BibitemShut {NoStop}%
\bibitem [{\citenamefont {Rhim}\ \emph {et~al.}(2018)\citenamefont {Rhim},
  \citenamefont {Bardarson},\ and\ \citenamefont
  {Slager}}]{PhysRevB.97.115143}%
  \BibitemOpen
  \bibfield  {author} {\bibinfo {author} {\bibfnamefont {J.-W.}\ \bibnamefont
  {Rhim}}, \bibinfo {author} {\bibfnamefont {J.~H.}\ \bibnamefont
  {Bardarson}},\ and\ \bibinfo {author} {\bibfnamefont {R.-J.}\ \bibnamefont
  {Slager}},\ }\bibfield  {title} {\bibinfo {title} {Unified bulk-boundary
  correspondence for band insulators},\ }\href
  {https://doi.org/10.1103/PhysRevB.97.115143} {\bibfield  {journal} {\bibinfo
  {journal} {Phys. Rev. B}\ }\textbf {\bibinfo {volume} {97}},\ \bibinfo
  {pages} {115143} (\bibinfo {year} {2018})}\BibitemShut {NoStop}%
\bibitem [{\citenamefont {Takahashi}\ \emph {et~al.}(2020)\citenamefont
  {Takahashi}, \citenamefont {Tanaka},\ and\ \citenamefont
  {Murakami}}]{PhysRevResearch.2.013300}%
  \BibitemOpen
  \bibfield  {author} {\bibinfo {author} {\bibfnamefont {R.}~\bibnamefont
  {Takahashi}}, \bibinfo {author} {\bibfnamefont {Y.}~\bibnamefont {Tanaka}},\
  and\ \bibinfo {author} {\bibfnamefont {S.}~\bibnamefont {Murakami}},\
  }\bibfield  {title} {\bibinfo {title} {Bulk-edge and bulk-hinge
  correspondence in inversion-symmetric insulators},\ }\href
  {https://doi.org/10.1103/PhysRevResearch.2.013300} {\bibfield  {journal}
  {\bibinfo  {journal} {Phys. Rev. Res.}\ }\textbf {\bibinfo {volume} {2}},\
  \bibinfo {pages} {013300} (\bibinfo {year} {2020})}\BibitemShut {NoStop}%
\bibitem [{\citenamefont {Chiu}\ \emph {et~al.}(2016)\citenamefont {Chiu},
  \citenamefont {Teo}, \citenamefont {Schnyder},\ and\ \citenamefont
  {Ryu}}]{RevModPhys.88.035005}%
  \BibitemOpen
  \bibfield  {author} {\bibinfo {author} {\bibfnamefont {C.-K.}\ \bibnamefont
  {Chiu}}, \bibinfo {author} {\bibfnamefont {J.~C.~Y.}\ \bibnamefont {Teo}},
  \bibinfo {author} {\bibfnamefont {A.~P.}\ \bibnamefont {Schnyder}},\ and\
  \bibinfo {author} {\bibfnamefont {S.}~\bibnamefont {Ryu}},\ }\bibfield
  {title} {\bibinfo {title} {Classification of topological quantum matter with
  symmetries},\ }\href {https://doi.org/10.1103/RevModPhys.88.035005}
  {\bibfield  {journal} {\bibinfo  {journal} {Rev. Mod. Phys.}\ }\textbf
  {\bibinfo {volume} {88}},\ \bibinfo {pages} {035005} (\bibinfo {year}
  {2016})}\BibitemShut {NoStop}%
\bibitem [{\citenamefont {Ryu}\ \emph {et~al.}(2010)\citenamefont {Ryu},
  \citenamefont {Schnyder}, \citenamefont {Furusaki},\ and\ \citenamefont
  {Ludwig}}]{Ryu2010}%
  \BibitemOpen
  \bibfield  {author} {\bibinfo {author} {\bibfnamefont {S.}~\bibnamefont
  {Ryu}}, \bibinfo {author} {\bibfnamefont {A.~P.}\ \bibnamefont {Schnyder}},
  \bibinfo {author} {\bibfnamefont {A.}~\bibnamefont {Furusaki}},\ and\
  \bibinfo {author} {\bibfnamefont {A.~W.~W.}\ \bibnamefont {Ludwig}},\
  }\bibfield  {title} {\bibinfo {title} {{Topological insulators and
  superconductors: tenfold way and dimensional hierarchy}},\ }\href
  {https://doi.org/10.1088/1367-2630/12/6/065010} {\bibfield  {journal}
  {\bibinfo  {journal} {New J. Phys.}\ }\textbf {\bibinfo {volume} {12}},\
  \bibinfo {pages} {065010} (\bibinfo {year} {2010})}\BibitemShut {NoStop}%
\bibitem [{\citenamefont {Mondragon-Shem}\ \emph {et~al.}(2014)\citenamefont
  {Mondragon-Shem}, \citenamefont {Hughes}, \citenamefont {Song},\ and\
  \citenamefont {Prodan}}]{PhysRevLett.113.046802}%
  \BibitemOpen
  \bibfield  {author} {\bibinfo {author} {\bibfnamefont {I.}~\bibnamefont
  {Mondragon-Shem}}, \bibinfo {author} {\bibfnamefont {T.~L.}\ \bibnamefont
  {Hughes}}, \bibinfo {author} {\bibfnamefont {J.}~\bibnamefont {Song}},\ and\
  \bibinfo {author} {\bibfnamefont {E.}~\bibnamefont {Prodan}},\ }\bibfield
  {title} {\bibinfo {title} {{Topological Criticality in the Chiral-Symmetric
  AIII Class at Strong Disorder}},\ }\href
  {https://doi.org/10.1103/PhysRevLett.113.046802} {\bibfield  {journal}
  {\bibinfo  {journal} {Phys. Rev. Lett.}\ }\textbf {\bibinfo {volume} {113}},\
  \bibinfo {pages} {046802} (\bibinfo {year} {2014})}\BibitemShut {NoStop}%
\bibitem [{\citenamefont {Meier}\ \emph {et~al.}(2018)\citenamefont {Meier},
  \citenamefont {An}, \citenamefont {Dauphin}, \citenamefont {Maffei},
  \citenamefont {Massignan}, \citenamefont {Hughes},\ and\ \citenamefont
  {Gadway}}]{Science362}%
  \BibitemOpen
  \bibfield  {author} {\bibinfo {author} {\bibfnamefont {E.~J.}\ \bibnamefont
  {Meier}}, \bibinfo {author} {\bibfnamefont {F.~A.}\ \bibnamefont {An}},
  \bibinfo {author} {\bibfnamefont {A.}~\bibnamefont {Dauphin}}, \bibinfo
  {author} {\bibfnamefont {M.}~\bibnamefont {Maffei}}, \bibinfo {author}
  {\bibfnamefont {P.}~\bibnamefont {Massignan}}, \bibinfo {author}
  {\bibfnamefont {T.~L.}\ \bibnamefont {Hughes}},\ and\ \bibinfo {author}
  {\bibfnamefont {B.}~\bibnamefont {Gadway}},\ }\bibfield  {title} {\bibinfo
  {title} {{Observation of the topological Anderson insulator in disordered
  atomic wires}},\ }\href {https://doi.org/10.1126/science.aat3406} {\bibfield
  {journal} {\bibinfo  {journal} {Science}\ }\textbf {\bibinfo {volume}
  {362}},\ \bibinfo {pages} {929} (\bibinfo {year} {2018})}\BibitemShut
  {NoStop}%
\bibitem [{\citenamefont {Lin}\ \emph {et~al.}(2021)\citenamefont {Lin},
  \citenamefont {Ke},\ and\ \citenamefont {Lee}}]{PhysRevB.103.224208}%
  \BibitemOpen
  \bibfield  {author} {\bibinfo {author} {\bibfnamefont {L.}~\bibnamefont
  {Lin}}, \bibinfo {author} {\bibfnamefont {Y.}~\bibnamefont {Ke}},\ and\
  \bibinfo {author} {\bibfnamefont {C.}~\bibnamefont {Lee}},\ }\bibfield
  {title} {\bibinfo {title} {Real-space representation of the winding number
  for a one-dimensional chiral-symmetric topological insulator},\ }\href
  {https://doi.org/10.1103/PhysRevB.103.224208} {\bibfield  {journal} {\bibinfo
   {journal} {Phys. Rev. B}\ }\textbf {\bibinfo {volume} {103}},\ \bibinfo
  {pages} {224208} (\bibinfo {year} {2021})}\BibitemShut {NoStop}%
\bibitem [{\citenamefont {Lin}\ and\ \citenamefont
  {Lee}(2024)}]{lin2024probing}%
  \BibitemOpen
  \bibfield  {author} {\bibinfo {author} {\bibfnamefont {L.}~\bibnamefont
  {Lin}}\ and\ \bibinfo {author} {\bibfnamefont {C.}~\bibnamefont {Lee}},\
  }\bibfield  {title} {\bibinfo {title} {Probing chiral-symmetric higher-order
  topological insulators with multipole winding number},\ }\href
  {https://doi.org/10.1038/s42005-024-01884-3} {\bibfield  {journal} {\bibinfo
  {journal} {Communications Physics}\ }\textbf {\bibinfo {volume} {7}},\
  \bibinfo {pages} {1} (\bibinfo {year} {2024})}\BibitemShut {NoStop}%
\bibitem [{\citenamefont {Teo}\ \emph {et~al.}(2008)\citenamefont {Teo},
  \citenamefont {Fu},\ and\ \citenamefont {Kane}}]{PhysRevB.78.045426}%
  \BibitemOpen
  \bibfield  {author} {\bibinfo {author} {\bibfnamefont {J.~C.~Y.}\
  \bibnamefont {Teo}}, \bibinfo {author} {\bibfnamefont {L.}~\bibnamefont
  {Fu}},\ and\ \bibinfo {author} {\bibfnamefont {C.~L.}\ \bibnamefont {Kane}},\
  }\bibfield  {title} {\bibinfo {title} {{Surface states and topological
  invariants in three-dimensional topological insulators: Application to
  ${\text{Bi}}_{1\ensuremath{-}x}{\text{Sb}}_{x}$}},\ }\href
  {https://doi.org/10.1103/PhysRevB.78.045426} {\bibfield  {journal} {\bibinfo
  {journal} {Phys. Rev. B}\ }\textbf {\bibinfo {volume} {78}},\ \bibinfo
  {pages} {045426} (\bibinfo {year} {2008})}\BibitemShut {NoStop}%
\bibitem [{\citenamefont {Hsieh}\ \emph {et~al.}(2012)\citenamefont {Hsieh},
  \citenamefont {Lin}, \citenamefont {Liu}, \citenamefont {Duan}, \citenamefont
  {Bansil},\ and\ \citenamefont {Fu}}]{hsieh2012topological}%
  \BibitemOpen
  \bibfield  {author} {\bibinfo {author} {\bibfnamefont {T.~H.}\ \bibnamefont
  {Hsieh}}, \bibinfo {author} {\bibfnamefont {H.}~\bibnamefont {Lin}}, \bibinfo
  {author} {\bibfnamefont {J.}~\bibnamefont {Liu}}, \bibinfo {author}
  {\bibfnamefont {W.}~\bibnamefont {Duan}}, \bibinfo {author} {\bibfnamefont
  {A.}~\bibnamefont {Bansil}},\ and\ \bibinfo {author} {\bibfnamefont
  {L.}~\bibnamefont {Fu}},\ }\bibfield  {title} {\bibinfo {title} {{Topological
  crystalline insulators in the SnTe material class}},\ }\href
  {https://doi.org/10.1038/ncomms1969} {\bibfield  {journal} {\bibinfo
  {journal} {Nature Communications}\ }\textbf {\bibinfo {volume} {3}},\
  \bibinfo {pages} {982} (\bibinfo {year} {2012})}\BibitemShut {NoStop}%
\bibitem [{\citenamefont {Verma}\ and\ \citenamefont
  {Ghosh}(2024)}]{PhysRevB.110.125424}%
  \BibitemOpen
  \bibfield  {author} {\bibinfo {author} {\bibfnamefont {S.}~\bibnamefont
  {Verma}}\ and\ \bibinfo {author} {\bibfnamefont {T.~K.}\ \bibnamefont
  {Ghosh}},\ }\bibfield  {title} {\bibinfo {title} {{Bulk-boundary
  correspondence in extended trimer Su-Schrieffer-Heeger model}},\ }\href
  {https://doi.org/10.1103/PhysRevB.110.125424} {\bibfield  {journal} {\bibinfo
   {journal} {Phys. Rev. B}\ }\textbf {\bibinfo {volume} {110}},\ \bibinfo
  {pages} {125424} (\bibinfo {year} {2024})}\BibitemShut {NoStop}%
\bibitem [{\citenamefont {King-Smith}\ and\ \citenamefont
  {Vanderbilt}(1993)}]{PhysRevB.47.1651}%
  \BibitemOpen
  \bibfield  {author} {\bibinfo {author} {\bibfnamefont {R.~D.}\ \bibnamefont
  {King-Smith}}\ and\ \bibinfo {author} {\bibfnamefont {D.}~\bibnamefont
  {Vanderbilt}},\ }\bibfield  {title} {\bibinfo {title} {Theory of polarization
  of crystalline solids},\ }\href {https://doi.org/10.1103/PhysRevB.47.1651}
  {\bibfield  {journal} {\bibinfo  {journal} {Phys. Rev. B}\ }\textbf {\bibinfo
  {volume} {47}},\ \bibinfo {pages} {1651} (\bibinfo {year}
  {1993})}\BibitemShut {NoStop}%
\bibitem [{\citenamefont {Resta}(1994)}]{RevModPhys.66.899}%
  \BibitemOpen
  \bibfield  {author} {\bibinfo {author} {\bibfnamefont {R.}~\bibnamefont
  {Resta}},\ }\bibfield  {title} {\bibinfo {title} {Macroscopic polarization in
  crystalline dielectrics: the geometric phase approach},\ }\href
  {https://doi.org/10.1103/RevModPhys.66.899} {\bibfield  {journal} {\bibinfo
  {journal} {Rev. Mod. Phys.}\ }\textbf {\bibinfo {volume} {66}},\ \bibinfo
  {pages} {899} (\bibinfo {year} {1994})}\BibitemShut {NoStop}%
\bibitem [{\citenamefont {Marzari}\ \emph {et~al.}(2012)\citenamefont
  {Marzari}, \citenamefont {Mostofi}, \citenamefont {Yates}, \citenamefont
  {Souza},\ and\ \citenamefont {Vanderbilt}}]{RevModPhys.84.1419}%
  \BibitemOpen
  \bibfield  {author} {\bibinfo {author} {\bibfnamefont {N.}~\bibnamefont
  {Marzari}}, \bibinfo {author} {\bibfnamefont {A.~A.}\ \bibnamefont
  {Mostofi}}, \bibinfo {author} {\bibfnamefont {J.~R.}\ \bibnamefont {Yates}},
  \bibinfo {author} {\bibfnamefont {I.}~\bibnamefont {Souza}},\ and\ \bibinfo
  {author} {\bibfnamefont {D.}~\bibnamefont {Vanderbilt}},\ }\bibfield  {title}
  {\bibinfo {title} {{Maximally localized Wannier functions: Theory and
  applications}},\ }\href {https://doi.org/10.1103/RevModPhys.84.1419}
  {\bibfield  {journal} {\bibinfo  {journal} {Rev. Mod. Phys.}\ }\textbf
  {\bibinfo {volume} {84}},\ \bibinfo {pages} {1419} (\bibinfo {year}
  {2012})}\BibitemShut {NoStop}%
\bibitem [{\citenamefont {Ahn}\ \emph {et~al.}(2019)\citenamefont {Ahn},
  \citenamefont {Park}, \citenamefont {Kim}, \citenamefont {Kim},\ and\
  \citenamefont {Yang}}]{Ahn_2019}%
  \BibitemOpen
  \bibfield  {author} {\bibinfo {author} {\bibfnamefont {J.}~\bibnamefont
  {Ahn}}, \bibinfo {author} {\bibfnamefont {S.}~\bibnamefont {Park}}, \bibinfo
  {author} {\bibfnamefont {D.}~\bibnamefont {Kim}}, \bibinfo {author}
  {\bibfnamefont {Y.}~\bibnamefont {Kim}},\ and\ \bibinfo {author}
  {\bibfnamefont {B.-J.}\ \bibnamefont {Yang}},\ }\bibfield  {title} {\bibinfo
  {title} {{Stiefel–Whitney classes and topological phases in band theory}},\
  }\href {https://doi.org/10.1088/1674-1056/ab4d3b} {\bibfield  {journal}
  {\bibinfo  {journal} {Chinese Physics B}\ }\textbf {\bibinfo {volume} {28}},\
  \bibinfo {pages} {117101} (\bibinfo {year} {2019})}\BibitemShut {NoStop}%
\bibitem [{\citenamefont {Qi}\ \emph {et~al.}(2006)\citenamefont {Qi},
  \citenamefont {Wu},\ and\ \citenamefont {Zhang}}]{PhysRevB.74.045125}%
  \BibitemOpen
  \bibfield  {author} {\bibinfo {author} {\bibfnamefont {X.-L.}\ \bibnamefont
  {Qi}}, \bibinfo {author} {\bibfnamefont {Y.-S.}\ \bibnamefont {Wu}},\ and\
  \bibinfo {author} {\bibfnamefont {S.-C.}\ \bibnamefont {Zhang}},\ }\bibfield
  {title} {\bibinfo {title} {General theorem relating the bulk topological
  number to edge states in two-dimensional insulators},\ }\href
  {https://doi.org/10.1103/PhysRevB.74.045125} {\bibfield  {journal} {\bibinfo
  {journal} {Phys. Rev. B}\ }\textbf {\bibinfo {volume} {74}},\ \bibinfo
  {pages} {045125} (\bibinfo {year} {2006})}\BibitemShut {NoStop}%
\bibitem [{\citenamefont {Watanabe}(2018)}]{PhysRevB.98.155137}%
  \BibitemOpen
  \bibfield  {author} {\bibinfo {author} {\bibfnamefont {H.}~\bibnamefont
  {Watanabe}},\ }\bibfield  {title} {\bibinfo {title} {Insensitivity of bulk
  properties to the twisted boundary condition},\ }\href
  {https://doi.org/10.1103/PhysRevB.98.155137} {\bibfield  {journal} {\bibinfo
  {journal} {Phys. Rev. B}\ }\textbf {\bibinfo {volume} {98}},\ \bibinfo
  {pages} {155137} (\bibinfo {year} {2018})}\BibitemShut {NoStop}%
\bibitem [{\citenamefont {Lin}\ \emph {et~al.}(2023{\natexlab{a}})\citenamefont
  {Lin}, \citenamefont {Ke}, \citenamefont {Zhang},\ and\ \citenamefont
  {Lee}}]{PhysRevB.108.174204}%
  \BibitemOpen
  \bibfield  {author} {\bibinfo {author} {\bibfnamefont {L.}~\bibnamefont
  {Lin}}, \bibinfo {author} {\bibfnamefont {Y.}~\bibnamefont {Ke}}, \bibinfo
  {author} {\bibfnamefont {L.}~\bibnamefont {Zhang}},\ and\ \bibinfo {author}
  {\bibfnamefont {C.}~\bibnamefont {Lee}},\ }\bibfield  {title} {\bibinfo
  {title} {Calculations of the chern number: Equivalence of real-space and
  twisted-boundary-condition formulas},\ }\href
  {https://doi.org/10.1103/PhysRevB.108.174204} {\bibfield  {journal} {\bibinfo
   {journal} {Phys. Rev. B}\ }\textbf {\bibinfo {volume} {108}},\ \bibinfo
  {pages} {174204} (\bibinfo {year} {2023}{\natexlab{a}})}\BibitemShut
  {NoStop}%
\bibitem [{\citenamefont {Benalcazar}\ and\ \citenamefont
  {Cerjan}(2022)}]{PhysRevLett.128.127601}%
  \BibitemOpen
  \bibfield  {author} {\bibinfo {author} {\bibfnamefont {W.~A.}\ \bibnamefont
  {Benalcazar}}\ and\ \bibinfo {author} {\bibfnamefont {A.}~\bibnamefont
  {Cerjan}},\ }\bibfield  {title} {\bibinfo {title} {Chiral-symmetric
  higher-order topological phases of matter},\ }\href
  {https://doi.org/10.1103/PhysRevLett.128.127601} {\bibfield  {journal}
  {\bibinfo  {journal} {Phys. Rev. Lett.}\ }\textbf {\bibinfo {volume} {128}},\
  \bibinfo {pages} {127601} (\bibinfo {year} {2022})}\BibitemShut {NoStop}%
\bibitem [{\citenamefont {Li}\ \emph {et~al.}(2009)\citenamefont {Li},
  \citenamefont {Chu}, \citenamefont {Jain},\ and\ \citenamefont
  {Shen}}]{PhysRevLett.102.136806}%
  \BibitemOpen
  \bibfield  {author} {\bibinfo {author} {\bibfnamefont {J.}~\bibnamefont
  {Li}}, \bibinfo {author} {\bibfnamefont {R.-L.}\ \bibnamefont {Chu}},
  \bibinfo {author} {\bibfnamefont {J.~K.}\ \bibnamefont {Jain}},\ and\
  \bibinfo {author} {\bibfnamefont {S.-Q.}\ \bibnamefont {Shen}},\ }\bibfield
  {title} {\bibinfo {title} {{Topological Anderson Insulator}},\ }\href
  {https://doi.org/10.1103/PhysRevLett.102.136806} {\bibfield  {journal}
  {\bibinfo  {journal} {Phys. Rev. Lett.}\ }\textbf {\bibinfo {volume} {102}},\
  \bibinfo {pages} {136806} (\bibinfo {year} {2009})}\BibitemShut {NoStop}%
\bibitem [{\citenamefont {Altland}\ \emph {et~al.}(2014)\citenamefont
  {Altland}, \citenamefont {Bagrets}, \citenamefont {Fritz}, \citenamefont
  {Kamenev},\ and\ \citenamefont {Schmiedt}}]{PhysRevLett.112.206602}%
  \BibitemOpen
  \bibfield  {author} {\bibinfo {author} {\bibfnamefont {A.}~\bibnamefont
  {Altland}}, \bibinfo {author} {\bibfnamefont {D.}~\bibnamefont {Bagrets}},
  \bibinfo {author} {\bibfnamefont {L.}~\bibnamefont {Fritz}}, \bibinfo
  {author} {\bibfnamefont {A.}~\bibnamefont {Kamenev}},\ and\ \bibinfo {author}
  {\bibfnamefont {H.}~\bibnamefont {Schmiedt}},\ }\bibfield  {title} {\bibinfo
  {title} {{Quantum Criticality of Quasi-One-Dimensional Topological Anderson
  Insulators}},\ }\href {https://doi.org/10.1103/PhysRevLett.112.206602}
  {\bibfield  {journal} {\bibinfo  {journal} {Phys. Rev. Lett.}\ }\textbf
  {\bibinfo {volume} {112}},\ \bibinfo {pages} {206602} (\bibinfo {year}
  {2014})}\BibitemShut {NoStop}%
\bibitem [{\citenamefont {Wu}\ \emph {et~al.}(2016)\citenamefont {Wu},
  \citenamefont {Song}, \citenamefont {Zhou},\ and\ \citenamefont
  {Jiang}}]{Wu_2016}%
  \BibitemOpen
  \bibfield  {author} {\bibinfo {author} {\bibfnamefont {B.}~\bibnamefont
  {Wu}}, \bibinfo {author} {\bibfnamefont {J.}~\bibnamefont {Song}}, \bibinfo
  {author} {\bibfnamefont {J.}~\bibnamefont {Zhou}},\ and\ \bibinfo {author}
  {\bibfnamefont {H.}~\bibnamefont {Jiang}},\ }\bibfield  {title} {\bibinfo
  {title} {Disorder effects in topological states: Brief review of the recent
  developments*},\ }\href {https://doi.org/10.1088/1674-1056/25/11/117311}
  {\bibfield  {journal} {\bibinfo  {journal} {Chinese Physics B}\ }\textbf
  {\bibinfo {volume} {25}},\ \bibinfo {pages} {117311} (\bibinfo {year}
  {2016})}\BibitemShut {NoStop}%
\bibitem [{\citenamefont {Lin}\ \emph {et~al.}(2022)\citenamefont {Lin},
  \citenamefont {Li}, \citenamefont {Xiao}, \citenamefont {Wang}, \citenamefont
  {Yi},\ and\ \citenamefont {Xue}}]{lin2022observation}%
  \BibitemOpen
  \bibfield  {author} {\bibinfo {author} {\bibfnamefont {Q.}~\bibnamefont
  {Lin}}, \bibinfo {author} {\bibfnamefont {T.}~\bibnamefont {Li}}, \bibinfo
  {author} {\bibfnamefont {L.}~\bibnamefont {Xiao}}, \bibinfo {author}
  {\bibfnamefont {K.}~\bibnamefont {Wang}}, \bibinfo {author} {\bibfnamefont
  {W.}~\bibnamefont {Yi}},\ and\ \bibinfo {author} {\bibfnamefont
  {P.}~\bibnamefont {Xue}},\ }\bibfield  {title} {\bibinfo {title}
  {{Observation of non-Hermitian topological Anderson insulator in quantum
  dynamics}},\ }\href {https://doi.org/10.1038/s41467-022-30938-9} {\bibfield
  {journal} {\bibinfo  {journal} {Nature Communications}\ }\textbf {\bibinfo
  {volume} {13}},\ \bibinfo {pages} {3229} (\bibinfo {year}
  {2022})}\BibitemShut {NoStop}%
\bibitem [{\citenamefont {Tang}\ \emph {et~al.}(2022)\citenamefont {Tang},
  \citenamefont {Liu}, \citenamefont {Zhang},\ and\ \citenamefont
  {Zhang}}]{PhysRevA.105.063327}%
  \BibitemOpen
  \bibfield  {author} {\bibinfo {author} {\bibfnamefont {L.-Z.}\ \bibnamefont
  {Tang}}, \bibinfo {author} {\bibfnamefont {S.-N.}\ \bibnamefont {Liu}},
  \bibinfo {author} {\bibfnamefont {G.-Q.}\ \bibnamefont {Zhang}},\ and\
  \bibinfo {author} {\bibfnamefont {D.-W.}\ \bibnamefont {Zhang}},\ }\bibfield
  {title} {\bibinfo {title} {{Topological Anderson insulators with different
  bulk states in quasiperiodic chains}},\ }\href
  {https://doi.org/10.1103/PhysRevA.105.063327} {\bibfield  {journal} {\bibinfo
   {journal} {Phys. Rev. A}\ }\textbf {\bibinfo {volume} {105}},\ \bibinfo
  {pages} {063327} (\bibinfo {year} {2022})}\BibitemShut {NoStop}%
\bibitem [{\citenamefont {Ren}\ \emph {et~al.}(2024)\citenamefont {Ren},
  \citenamefont {Yu}, \citenamefont {Wu}, \citenamefont {Qi}, \citenamefont
  {Wang}, \citenamefont {Yao}, \citenamefont {Ren}, \citenamefont {Guo},
  \citenamefont {Jiang}, \citenamefont {Chen}, \citenamefont {Liu},
  \citenamefont {Chen},\ and\ \citenamefont {Sun}}]{PhysRevLett.132.066602}%
  \BibitemOpen
  \bibfield  {author} {\bibinfo {author} {\bibfnamefont {M.}~\bibnamefont
  {Ren}}, \bibinfo {author} {\bibfnamefont {Y.}~\bibnamefont {Yu}}, \bibinfo
  {author} {\bibfnamefont {B.}~\bibnamefont {Wu}}, \bibinfo {author}
  {\bibfnamefont {X.}~\bibnamefont {Qi}}, \bibinfo {author} {\bibfnamefont
  {Y.}~\bibnamefont {Wang}}, \bibinfo {author} {\bibfnamefont {X.}~\bibnamefont
  {Yao}}, \bibinfo {author} {\bibfnamefont {J.}~\bibnamefont {Ren}}, \bibinfo
  {author} {\bibfnamefont {Z.}~\bibnamefont {Guo}}, \bibinfo {author}
  {\bibfnamefont {H.}~\bibnamefont {Jiang}}, \bibinfo {author} {\bibfnamefont
  {H.}~\bibnamefont {Chen}}, \bibinfo {author} {\bibfnamefont {X.-J.}\
  \bibnamefont {Liu}}, \bibinfo {author} {\bibfnamefont {Z.}~\bibnamefont
  {Chen}},\ and\ \bibinfo {author} {\bibfnamefont {Y.}~\bibnamefont {Sun}},\
  }\bibfield  {title} {\bibinfo {title} {{Realization of Gapped and Ungapped
  Photonic Topological Anderson Insulators}},\ }\href
  {https://doi.org/10.1103/PhysRevLett.132.066602} {\bibfield  {journal}
  {\bibinfo  {journal} {Phys. Rev. Lett.}\ }\textbf {\bibinfo {volume} {132}},\
  \bibinfo {pages} {066602} (\bibinfo {year} {2024})}\BibitemShut {NoStop}%
\bibitem [{\citenamefont {Resta}(1998)}]{PhysRevLett.80.1800}%
  \BibitemOpen
  \bibfield  {author} {\bibinfo {author} {\bibfnamefont {R.}~\bibnamefont
  {Resta}},\ }\bibfield  {title} {\bibinfo {title} {Quantum-mechanical position
  operator in extended systems},\ }\href
  {https://doi.org/10.1103/PhysRevLett.80.1800} {\bibfield  {journal} {\bibinfo
   {journal} {Phys. Rev. Lett.}\ }\textbf {\bibinfo {volume} {80}},\ \bibinfo
  {pages} {1800} (\bibinfo {year} {1998})}\BibitemShut {NoStop}%
\bibitem [{\citenamefont {Lin}\ \emph {et~al.}(2023{\natexlab{b}})\citenamefont
  {Lin}, \citenamefont {Ke},\ and\ \citenamefont {Lee}}]{PhysRevB.107.125161}%
  \BibitemOpen
  \bibfield  {author} {\bibinfo {author} {\bibfnamefont {L.}~\bibnamefont
  {Lin}}, \bibinfo {author} {\bibfnamefont {Y.}~\bibnamefont {Ke}},\ and\
  \bibinfo {author} {\bibfnamefont {C.}~\bibnamefont {Lee}},\ }\bibfield
  {title} {\bibinfo {title} {Topological invariants for interacting systems:
  From twisted boundary conditions to center-of-mass momentum},\ }\href
  {https://doi.org/10.1103/PhysRevB.107.125161} {\bibfield  {journal} {\bibinfo
   {journal} {Phys. Rev. B}\ }\textbf {\bibinfo {volume} {107}},\ \bibinfo
  {pages} {125161} (\bibinfo {year} {2023}{\natexlab{b}})}\BibitemShut
  {NoStop}%
\bibitem [{\citenamefont {Cardano}\ \emph {et~al.}(2017)\citenamefont
  {Cardano}, \citenamefont {D’Errico}, \citenamefont {Dauphin}, \citenamefont
  {Maffei}, \citenamefont {Piccirillo}, \citenamefont {de~Lisio}, \citenamefont
  {De~Filippis}, \citenamefont {Cataudella}, \citenamefont {Santamato},
  \citenamefont {Marrucci} \emph {et~al.}}]{cardano2017detection}%
  \BibitemOpen
  \bibfield  {author} {\bibinfo {author} {\bibfnamefont {F.}~\bibnamefont
  {Cardano}}, \bibinfo {author} {\bibfnamefont {A.}~\bibnamefont {D’Errico}},
  \bibinfo {author} {\bibfnamefont {A.}~\bibnamefont {Dauphin}}, \bibinfo
  {author} {\bibfnamefont {M.}~\bibnamefont {Maffei}}, \bibinfo {author}
  {\bibfnamefont {B.}~\bibnamefont {Piccirillo}}, \bibinfo {author}
  {\bibfnamefont {C.}~\bibnamefont {de~Lisio}}, \bibinfo {author}
  {\bibfnamefont {G.}~\bibnamefont {De~Filippis}}, \bibinfo {author}
  {\bibfnamefont {V.}~\bibnamefont {Cataudella}}, \bibinfo {author}
  {\bibfnamefont {E.}~\bibnamefont {Santamato}}, \bibinfo {author}
  {\bibfnamefont {L.}~\bibnamefont {Marrucci}}, \emph {et~al.},\ }\bibfield
  {title} {\bibinfo {title} {{Detection of Zak phases and topological
  invariants in a chiral quantum walk of twisted photons}},\ }\href
  {https://doi.org/10.1038/ncomms15516} {\bibfield  {journal} {\bibinfo
  {journal} {Nature communications}\ }\textbf {\bibinfo {volume} {8}},\
  \bibinfo {pages} {15516} (\bibinfo {year} {2017})}\BibitemShut {NoStop}%
\bibitem [{\citenamefont {Wang}\ \emph {et~al.}(2018)\citenamefont {Wang},
  \citenamefont {Xiao}, \citenamefont {Qiu}, \citenamefont {Wang},
  \citenamefont {Yi},\ and\ \citenamefont {Xue}}]{PhysRevA.98.013835}%
  \BibitemOpen
  \bibfield  {author} {\bibinfo {author} {\bibfnamefont {X.}~\bibnamefont
  {Wang}}, \bibinfo {author} {\bibfnamefont {L.}~\bibnamefont {Xiao}}, \bibinfo
  {author} {\bibfnamefont {X.}~\bibnamefont {Qiu}}, \bibinfo {author}
  {\bibfnamefont {K.}~\bibnamefont {Wang}}, \bibinfo {author} {\bibfnamefont
  {W.}~\bibnamefont {Yi}},\ and\ \bibinfo {author} {\bibfnamefont
  {P.}~\bibnamefont {Xue}},\ }\bibfield  {title} {\bibinfo {title} {Detecting
  topological invariants and revealing topological phase transitions in
  discrete-time photonic quantum walks},\ }\href
  {https://doi.org/10.1103/PhysRevA.98.013835} {\bibfield  {journal} {\bibinfo
  {journal} {Phys. Rev. A}\ }\textbf {\bibinfo {volume} {98}},\ \bibinfo
  {pages} {013835} (\bibinfo {year} {2018})}\BibitemShut {NoStop}%
\bibitem [{\citenamefont {Xie}\ \emph {et~al.}(2019)\citenamefont {Xie},
  \citenamefont {Gou}, \citenamefont {Xiao}, \citenamefont {Gadway},\ and\
  \citenamefont {Yan}}]{xie2019topological}%
  \BibitemOpen
  \bibfield  {author} {\bibinfo {author} {\bibfnamefont {D.}~\bibnamefont
  {Xie}}, \bibinfo {author} {\bibfnamefont {W.}~\bibnamefont {Gou}}, \bibinfo
  {author} {\bibfnamefont {T.}~\bibnamefont {Xiao}}, \bibinfo {author}
  {\bibfnamefont {B.}~\bibnamefont {Gadway}},\ and\ \bibinfo {author}
  {\bibfnamefont {B.}~\bibnamefont {Yan}},\ }\bibfield  {title} {\bibinfo
  {title} {Topological characterizations of an extended su--schrieffer--heeger
  model},\ }\href {https://doi.org/10.1038/s41534-019-0159-6} {\bibfield
  {journal} {\bibinfo  {journal} {npj Quantum Information}\ }\textbf {\bibinfo
  {volume} {5}},\ \bibinfo {pages} {55} (\bibinfo {year} {2019})}\BibitemShut
  {NoStop}%
\bibitem [{\citenamefont {Wang}\ \emph
  {et~al.}(2019{\natexlab{a}})\citenamefont {Wang}, \citenamefont {Lu},
  \citenamefont {Mei}, \citenamefont {Gao}, \citenamefont {Li}, \citenamefont
  {Tang}, \citenamefont {Zhu}, \citenamefont {Jia},\ and\ \citenamefont
  {Jin}}]{PhysRevLett.122.193903}%
  \BibitemOpen
  \bibfield  {author} {\bibinfo {author} {\bibfnamefont {Y.}~\bibnamefont
  {Wang}}, \bibinfo {author} {\bibfnamefont {Y.-H.}\ \bibnamefont {Lu}},
  \bibinfo {author} {\bibfnamefont {F.}~\bibnamefont {Mei}}, \bibinfo {author}
  {\bibfnamefont {J.}~\bibnamefont {Gao}}, \bibinfo {author} {\bibfnamefont
  {Z.-M.}\ \bibnamefont {Li}}, \bibinfo {author} {\bibfnamefont
  {H.}~\bibnamefont {Tang}}, \bibinfo {author} {\bibfnamefont {S.-L.}\
  \bibnamefont {Zhu}}, \bibinfo {author} {\bibfnamefont {S.}~\bibnamefont
  {Jia}},\ and\ \bibinfo {author} {\bibfnamefont {X.-M.}\ \bibnamefont {Jin}},\
  }\bibfield  {title} {\bibinfo {title} {Direct observation of topology from
  single-photon dynamics},\ }\href
  {https://doi.org/10.1103/PhysRevLett.122.193903} {\bibfield  {journal}
  {\bibinfo  {journal} {Phys. Rev. Lett.}\ }\textbf {\bibinfo {volume} {122}},\
  \bibinfo {pages} {193903} (\bibinfo {year} {2019}{\natexlab{a}})}\BibitemShut
  {NoStop}%
\bibitem [{\citenamefont {Cai}\ \emph {et~al.}(2019)\citenamefont {Cai},
  \citenamefont {Han}, \citenamefont {Mei}, \citenamefont {Xu}, \citenamefont
  {Ma}, \citenamefont {Li}, \citenamefont {Wang}, \citenamefont {Song},
  \citenamefont {Xue}, \citenamefont {Yin}, \citenamefont {Jia},\ and\
  \citenamefont {Sun}}]{PhysRevLett.123.080501}%
  \BibitemOpen
  \bibfield  {author} {\bibinfo {author} {\bibfnamefont {W.}~\bibnamefont
  {Cai}}, \bibinfo {author} {\bibfnamefont {J.}~\bibnamefont {Han}}, \bibinfo
  {author} {\bibfnamefont {F.}~\bibnamefont {Mei}}, \bibinfo {author}
  {\bibfnamefont {Y.}~\bibnamefont {Xu}}, \bibinfo {author} {\bibfnamefont
  {Y.}~\bibnamefont {Ma}}, \bibinfo {author} {\bibfnamefont {X.}~\bibnamefont
  {Li}}, \bibinfo {author} {\bibfnamefont {H.}~\bibnamefont {Wang}}, \bibinfo
  {author} {\bibfnamefont {Y.~P.}\ \bibnamefont {Song}}, \bibinfo {author}
  {\bibfnamefont {Z.-Y.}\ \bibnamefont {Xue}}, \bibinfo {author} {\bibfnamefont
  {Z.-q.}\ \bibnamefont {Yin}}, \bibinfo {author} {\bibfnamefont
  {S.}~\bibnamefont {Jia}},\ and\ \bibinfo {author} {\bibfnamefont
  {L.}~\bibnamefont {Sun}},\ }\bibfield  {title} {\bibinfo {title} {Observation
  of topological magnon insulator states in a superconducting circuit},\ }\href
  {https://doi.org/10.1103/PhysRevLett.123.080501} {\bibfield  {journal}
  {\bibinfo  {journal} {Phys. Rev. Lett.}\ }\textbf {\bibinfo {volume} {123}},\
  \bibinfo {pages} {080501} (\bibinfo {year} {2019})}\BibitemShut {NoStop}%
\bibitem [{\citenamefont {Xie}\ \emph {et~al.}(2020)\citenamefont {Xie},
  \citenamefont {Deng}, \citenamefont {Xiao}, \citenamefont {Gou},
  \citenamefont {Chen}, \citenamefont {Yi},\ and\ \citenamefont
  {Yan}}]{PhysRevLett.124.050502}%
  \BibitemOpen
  \bibfield  {author} {\bibinfo {author} {\bibfnamefont {D.}~\bibnamefont
  {Xie}}, \bibinfo {author} {\bibfnamefont {T.-S.}\ \bibnamefont {Deng}},
  \bibinfo {author} {\bibfnamefont {T.}~\bibnamefont {Xiao}}, \bibinfo {author}
  {\bibfnamefont {W.}~\bibnamefont {Gou}}, \bibinfo {author} {\bibfnamefont
  {T.}~\bibnamefont {Chen}}, \bibinfo {author} {\bibfnamefont {W.}~\bibnamefont
  {Yi}},\ and\ \bibinfo {author} {\bibfnamefont {B.}~\bibnamefont {Yan}},\
  }\bibfield  {title} {\bibinfo {title} {{Topological Quantum Walks in Momentum
  Space with a Bose-Einstein Condensate}},\ }\href
  {https://doi.org/10.1103/PhysRevLett.124.050502} {\bibfield  {journal}
  {\bibinfo  {journal} {Phys. Rev. Lett.}\ }\textbf {\bibinfo {volume} {124}},\
  \bibinfo {pages} {050502} (\bibinfo {year} {2020})}\BibitemShut {NoStop}%
\bibitem [{\citenamefont {Olekhno}\ \emph {et~al.}(2022)\citenamefont
  {Olekhno}, \citenamefont {Rozenblit}, \citenamefont {Kachin}, \citenamefont
  {Dmitriev}, \citenamefont {Burmistrov}, \citenamefont {Seregin},
  \citenamefont {Zhirihin},\ and\ \citenamefont
  {Gorlach}}]{PhysRevB.105.L081107}%
  \BibitemOpen
  \bibfield  {author} {\bibinfo {author} {\bibfnamefont {N.~A.}\ \bibnamefont
  {Olekhno}}, \bibinfo {author} {\bibfnamefont {A.~D.}\ \bibnamefont
  {Rozenblit}}, \bibinfo {author} {\bibfnamefont {V.~I.}\ \bibnamefont
  {Kachin}}, \bibinfo {author} {\bibfnamefont {A.~A.}\ \bibnamefont
  {Dmitriev}}, \bibinfo {author} {\bibfnamefont {O.~I.}\ \bibnamefont
  {Burmistrov}}, \bibinfo {author} {\bibfnamefont {P.~S.}\ \bibnamefont
  {Seregin}}, \bibinfo {author} {\bibfnamefont {D.~V.}\ \bibnamefont
  {Zhirihin}},\ and\ \bibinfo {author} {\bibfnamefont {M.~A.}\ \bibnamefont
  {Gorlach}},\ }\bibfield  {title} {\bibinfo {title} {Experimental realization
  of topological corner states in long-range-coupled electrical circuits},\
  }\href {https://doi.org/10.1103/PhysRevB.105.L081107} {\bibfield  {journal}
  {\bibinfo  {journal} {Phys. Rev. B}\ }\textbf {\bibinfo {volume} {105}},\
  \bibinfo {pages} {L081107} (\bibinfo {year} {2022})}\BibitemShut {NoStop}%
\bibitem [{\citenamefont {Li}\ \emph {et~al.}(2023{\natexlab{a}})\citenamefont
  {Li}, \citenamefont {Zhang}, \citenamefont {Mei}, \citenamefont {Xie},
  \citenamefont {Lu}, \citenamefont {Ma}, \citenamefont {Xiao},\ and\
  \citenamefont {Jia}}]{PhysRevApplied.20.064042}%
  \BibitemOpen
  \bibfield  {author} {\bibinfo {author} {\bibfnamefont {Y.}~\bibnamefont
  {Li}}, \bibinfo {author} {\bibfnamefont {J.-H.}\ \bibnamefont {Zhang}},
  \bibinfo {author} {\bibfnamefont {F.}~\bibnamefont {Mei}}, \bibinfo {author}
  {\bibfnamefont {B.}~\bibnamefont {Xie}}, \bibinfo {author} {\bibfnamefont
  {M.-H.}\ \bibnamefont {Lu}}, \bibinfo {author} {\bibfnamefont
  {J.}~\bibnamefont {Ma}}, \bibinfo {author} {\bibfnamefont {L.}~\bibnamefont
  {Xiao}},\ and\ \bibinfo {author} {\bibfnamefont {S.}~\bibnamefont {Jia}},\
  }\bibfield  {title} {\bibinfo {title} {Large-chiral-number corner modes in
  $\mathbb{Z}$-class higher-order topolectrical circuits},\ }\href
  {https://doi.org/10.1103/PhysRevApplied.20.064042} {\bibfield  {journal}
  {\bibinfo  {journal} {Phys. Rev. Appl.}\ }\textbf {\bibinfo {volume} {20}},\
  \bibinfo {pages} {064042} (\bibinfo {year} {2023}{\natexlab{a}})}\BibitemShut
  {NoStop}%
\bibitem [{\citenamefont {Wang}\ \emph {et~al.}(2023)\citenamefont {Wang},
  \citenamefont {Deng}, \citenamefont {Ji}, \citenamefont {Oudich},
  \citenamefont {Benalcazar}, \citenamefont {Ma},\ and\ \citenamefont
  {Jing}}]{PhysRevLett.131.157201}%
  \BibitemOpen
  \bibfield  {author} {\bibinfo {author} {\bibfnamefont {D.}~\bibnamefont
  {Wang}}, \bibinfo {author} {\bibfnamefont {Y.}~\bibnamefont {Deng}}, \bibinfo
  {author} {\bibfnamefont {J.}~\bibnamefont {Ji}}, \bibinfo {author}
  {\bibfnamefont {M.}~\bibnamefont {Oudich}}, \bibinfo {author} {\bibfnamefont
  {W.~A.}\ \bibnamefont {Benalcazar}}, \bibinfo {author} {\bibfnamefont
  {G.}~\bibnamefont {Ma}},\ and\ \bibinfo {author} {\bibfnamefont
  {Y.}~\bibnamefont {Jing}},\ }\bibfield  {title} {\bibinfo {title}
  {Realization of a $\mathbb{Z}$-classified chiral-symmetric higher-order
  topological insulator in a coupling-inverted acoustic crystal},\ }\href
  {https://doi.org/10.1103/PhysRevLett.131.157201} {\bibfield  {journal}
  {\bibinfo  {journal} {Phys. Rev. Lett.}\ }\textbf {\bibinfo {volume} {131}},\
  \bibinfo {pages} {157201} (\bibinfo {year} {2023})}\BibitemShut {NoStop}%
\bibitem [{\citenamefont {Li}\ \emph {et~al.}(2023{\natexlab{b}})\citenamefont
  {Li}, \citenamefont {Qiu}, \citenamefont {Zhang},\ and\ \citenamefont
  {Qiu}}]{PhysRevB.108.205135}%
  \BibitemOpen
  \bibfield  {author} {\bibinfo {author} {\bibfnamefont {Y.}~\bibnamefont
  {Li}}, \bibinfo {author} {\bibfnamefont {H.}~\bibnamefont {Qiu}}, \bibinfo
  {author} {\bibfnamefont {Q.}~\bibnamefont {Zhang}},\ and\ \bibinfo {author}
  {\bibfnamefont {C.}~\bibnamefont {Qiu}},\ }\bibfield  {title} {\bibinfo
  {title} {Acoustic higher-order topological insulators protected by multipole
  chiral numbers},\ }\href {https://doi.org/10.1103/PhysRevB.108.205135}
  {\bibfield  {journal} {\bibinfo  {journal} {Phys. Rev. B}\ }\textbf {\bibinfo
  {volume} {108}},\ \bibinfo {pages} {205135} (\bibinfo {year}
  {2023}{\natexlab{b}})}\BibitemShut {NoStop}%
\bibitem [{\citenamefont {Poddubny}\ \emph {et~al.}(2014)\citenamefont
  {Poddubny}, \citenamefont {Miroshnichenko}, \citenamefont {Slobozhanyuk},\
  and\ \citenamefont {Kivshar}}]{poddubny2014topological}%
  \BibitemOpen
  \bibfield  {author} {\bibinfo {author} {\bibfnamefont {A.}~\bibnamefont
  {Poddubny}}, \bibinfo {author} {\bibfnamefont {A.}~\bibnamefont
  {Miroshnichenko}}, \bibinfo {author} {\bibfnamefont {A.}~\bibnamefont
  {Slobozhanyuk}},\ and\ \bibinfo {author} {\bibfnamefont {Y.}~\bibnamefont
  {Kivshar}},\ }\bibfield  {title} {\bibinfo {title} {Topological majorana
  states in zigzag chains of plasmonic nanoparticles},\ }\href
  {https://doi.org/doi/pdf/10.1021/ph4000949} {\bibfield  {journal} {\bibinfo
  {journal} {ACS Photonics}\ }\textbf {\bibinfo {volume} {1}},\ \bibinfo
  {pages} {101} (\bibinfo {year} {2014})}\BibitemShut {NoStop}%
\bibitem [{\citenamefont {Zhang}\ \emph {et~al.}(2018)\citenamefont {Zhang},
  \citenamefont {Zhu}, \citenamefont {Zhao}, \citenamefont {Yan},\ and\
  \citenamefont {Zhu}}]{topological_review_SL_Zhu}%
  \BibitemOpen
  \bibfield  {author} {\bibinfo {author} {\bibfnamefont {D.-W.}\ \bibnamefont
  {Zhang}}, \bibinfo {author} {\bibfnamefont {Y.-Q.}\ \bibnamefont {Zhu}},
  \bibinfo {author} {\bibfnamefont {Y.~X.}\ \bibnamefont {Zhao}}, \bibinfo
  {author} {\bibfnamefont {H.}~\bibnamefont {Yan}},\ and\ \bibinfo {author}
  {\bibfnamefont {S.-L.}\ \bibnamefont {Zhu}},\ }\bibfield  {title} {\bibinfo
  {title} {Topological quantum matter with cold atoms},\ }\href
  {https://doi.org/10.1080/00018732.2019.1594094} {\bibfield  {journal}
  {\bibinfo  {journal} {Advances in Physics}\ }\textbf {\bibinfo {volume}
  {67}},\ \bibinfo {pages} {253} (\bibinfo {year} {2018})}\BibitemShut
  {NoStop}%
\bibitem [{\citenamefont {Kruk}\ \emph {et~al.}(2019)\citenamefont {Kruk},
  \citenamefont {Poddubny}, \citenamefont {Smirnova}, \citenamefont {Wang},
  \citenamefont {Slobozhanyuk}, \citenamefont {Shorokhov}, \citenamefont
  {Kravchenko}, \citenamefont {Luther-Davies},\ and\ \citenamefont
  {Kivshar}}]{kruk2019nonlinear}%
  \BibitemOpen
  \bibfield  {author} {\bibinfo {author} {\bibfnamefont {S.}~\bibnamefont
  {Kruk}}, \bibinfo {author} {\bibfnamefont {A.}~\bibnamefont {Poddubny}},
  \bibinfo {author} {\bibfnamefont {D.}~\bibnamefont {Smirnova}}, \bibinfo
  {author} {\bibfnamefont {L.}~\bibnamefont {Wang}}, \bibinfo {author}
  {\bibfnamefont {A.}~\bibnamefont {Slobozhanyuk}}, \bibinfo {author}
  {\bibfnamefont {A.}~\bibnamefont {Shorokhov}}, \bibinfo {author}
  {\bibfnamefont {I.}~\bibnamefont {Kravchenko}}, \bibinfo {author}
  {\bibfnamefont {B.}~\bibnamefont {Luther-Davies}},\ and\ \bibinfo {author}
  {\bibfnamefont {Y.}~\bibnamefont {Kivshar}},\ }\bibfield  {title} {\bibinfo
  {title} {Nonlinear light generation in topological nanostructures},\ }\href
  {https://doi.org/10.1038/s41565-018-0324-7} {\bibfield  {journal} {\bibinfo
  {journal} {Nature Nanotechnology}\ }\textbf {\bibinfo {volume} {14}},\
  \bibinfo {pages} {126} (\bibinfo {year} {2019})}\BibitemShut {NoStop}%
\bibitem [{\citenamefont {Ozawa}\ \emph {et~al.}(2019)\citenamefont {Ozawa},
  \citenamefont {Price}, \citenamefont {Amo}, \citenamefont {Goldman},
  \citenamefont {Hafezi}, \citenamefont {Lu}, \citenamefont {Rechtsman},
  \citenamefont {Schuster}, \citenamefont {Simon}, \citenamefont {Zilberberg},\
  and\ \citenamefont {Carusotto}}]{RevModPhys.91.015006}%
  \BibitemOpen
  \bibfield  {author} {\bibinfo {author} {\bibfnamefont {T.}~\bibnamefont
  {Ozawa}}, \bibinfo {author} {\bibfnamefont {H.~M.}\ \bibnamefont {Price}},
  \bibinfo {author} {\bibfnamefont {A.}~\bibnamefont {Amo}}, \bibinfo {author}
  {\bibfnamefont {N.}~\bibnamefont {Goldman}}, \bibinfo {author} {\bibfnamefont
  {M.}~\bibnamefont {Hafezi}}, \bibinfo {author} {\bibfnamefont
  {L.}~\bibnamefont {Lu}}, \bibinfo {author} {\bibfnamefont {M.~C.}\
  \bibnamefont {Rechtsman}}, \bibinfo {author} {\bibfnamefont {D.}~\bibnamefont
  {Schuster}}, \bibinfo {author} {\bibfnamefont {J.}~\bibnamefont {Simon}},
  \bibinfo {author} {\bibfnamefont {O.}~\bibnamefont {Zilberberg}},\ and\
  \bibinfo {author} {\bibfnamefont {I.}~\bibnamefont {Carusotto}},\ }\bibfield
  {title} {\bibinfo {title} {Topological photonics},\ }\href
  {https://doi.org/10.1103/RevModPhys.91.015006} {\bibfield  {journal}
  {\bibinfo  {journal} {Rev. Mod. Phys.}\ }\textbf {\bibinfo {volume} {91}},\
  \bibinfo {pages} {015006} (\bibinfo {year} {2019})}\BibitemShut {NoStop}%
\bibitem [{\citenamefont {Wang}\ \emph
  {et~al.}(2019{\natexlab{b}})\citenamefont {Wang}, \citenamefont {Guo},\ and\
  \citenamefont {Jiang}}]{Wang_2019}%
  \BibitemOpen
  \bibfield  {author} {\bibinfo {author} {\bibfnamefont {H.-X.}\ \bibnamefont
  {Wang}}, \bibinfo {author} {\bibfnamefont {G.-Y.}\ \bibnamefont {Guo}},\ and\
  \bibinfo {author} {\bibfnamefont {J.-H.}\ \bibnamefont {Jiang}},\ }\bibfield
  {title} {\bibinfo {title} {Band topology in classical waves: Wilson-loop
  approach to topological numbers and fragile topology},\ }\href
  {https://doi.org/10.1088/1367-2630/ab3f71} {\bibfield  {journal} {\bibinfo
  {journal} {New Journal of Physics}\ }\textbf {\bibinfo {volume} {21}},\
  \bibinfo {pages} {093029} (\bibinfo {year} {2019}{\natexlab{b}})}\BibitemShut
  {NoStop}%
\bibitem [{\citenamefont {Jiao}\ \emph {et~al.}(2021)\citenamefont {Jiao},
  \citenamefont {Longhi}, \citenamefont {Wang}, \citenamefont {Gao},
  \citenamefont {Zhou}, \citenamefont {Wang}, \citenamefont {Fu}, \citenamefont
  {Wang}, \citenamefont {Ren}, \citenamefont {Qiao},\ and\ \citenamefont
  {Jin}}]{PhysRevLett.127.147401}%
  \BibitemOpen
  \bibfield  {author} {\bibinfo {author} {\bibfnamefont {Z.-Q.}\ \bibnamefont
  {Jiao}}, \bibinfo {author} {\bibfnamefont {S.}~\bibnamefont {Longhi}},
  \bibinfo {author} {\bibfnamefont {X.-W.}\ \bibnamefont {Wang}}, \bibinfo
  {author} {\bibfnamefont {J.}~\bibnamefont {Gao}}, \bibinfo {author}
  {\bibfnamefont {W.-H.}\ \bibnamefont {Zhou}}, \bibinfo {author}
  {\bibfnamefont {Y.}~\bibnamefont {Wang}}, \bibinfo {author} {\bibfnamefont
  {Y.-X.}\ \bibnamefont {Fu}}, \bibinfo {author} {\bibfnamefont
  {L.}~\bibnamefont {Wang}}, \bibinfo {author} {\bibfnamefont {R.-J.}\
  \bibnamefont {Ren}}, \bibinfo {author} {\bibfnamefont {L.-F.}\ \bibnamefont
  {Qiao}},\ and\ \bibinfo {author} {\bibfnamefont {X.-M.}\ \bibnamefont
  {Jin}},\ }\bibfield  {title} {\bibinfo {title} {{Experimentally Detecting
  Quantized Zak Phases without Chiral Symmetry in Photonic Lattices}},\ }\href
  {https://doi.org/10.1103/PhysRevLett.127.147401} {\bibfield  {journal}
  {\bibinfo  {journal} {Phys. Rev. Lett.}\ }\textbf {\bibinfo {volume} {127}},\
  \bibinfo {pages} {147401} (\bibinfo {year} {2021})}\BibitemShut {NoStop}%
\bibitem [{\citenamefont {Jia}\ \emph {et~al.}(2024)\citenamefont {Jia},
  \citenamefont {Wang}, \citenamefont {Gao},\ and\ \citenamefont
  {An}}]{PhysRevB.110.L201117}%
  \BibitemOpen
  \bibfield  {author} {\bibinfo {author} {\bibfnamefont {W.}~\bibnamefont
  {Jia}}, \bibinfo {author} {\bibfnamefont {B.-Z.}\ \bibnamefont {Wang}},
  \bibinfo {author} {\bibfnamefont {M.-J.}\ \bibnamefont {Gao}},\ and\ \bibinfo
  {author} {\bibfnamefont {J.-H.}\ \bibnamefont {An}},\ }\bibfield  {title}
  {\bibinfo {title} {Unveiling higher-order topology via polarized topological
  charges},\ }\href {https://doi.org/10.1103/PhysRevB.110.L201117} {\bibfield
  {journal} {\bibinfo  {journal} {Phys. Rev. B}\ }\textbf {\bibinfo {volume}
  {110}},\ \bibinfo {pages} {L201117} (\bibinfo {year} {2024})}\BibitemShut
  {NoStop}%
\bibitem [{\citenamefont {Zhu}(2024)}]{PhysRevB.110.075103}%
  \BibitemOpen
  \bibfield  {author} {\bibinfo {author} {\bibfnamefont {X.}~\bibnamefont
  {Zhu}},\ }\bibfield  {title} {\bibinfo {title} {Direct demonstration of
  bulk-boundary correspondence in higher-order topological superconductors with
  chiral symmetry},\ }\href {https://doi.org/10.1103/PhysRevB.110.075103}
  {\bibfield  {journal} {\bibinfo  {journal} {Phys. Rev. B}\ }\textbf {\bibinfo
  {volume} {110}},\ \bibinfo {pages} {075103} (\bibinfo {year}
  {2024})}\BibitemShut {NoStop}%
\bibitem [{\citenamefont {Luo}\ \emph {et~al.}(2025)\citenamefont {Luo},
  \citenamefont {Li}, \citenamefont {Xiao},\ and\ \citenamefont
  {Wu}}]{PhysRevB.111.035115}%
  \BibitemOpen
  \bibfield  {author} {\bibinfo {author} {\bibfnamefont {X.-J.}\ \bibnamefont
  {Luo}}, \bibinfo {author} {\bibfnamefont {J.-Z.}\ \bibnamefont {Li}},
  \bibinfo {author} {\bibfnamefont {M.}~\bibnamefont {Xiao}},\ and\ \bibinfo
  {author} {\bibfnamefont {F.}~\bibnamefont {Wu}},\ }\bibfield  {title}
  {\bibinfo {title} {{Family of third-order topological insulators from
  Su-Schrieffer-Heeger stacking}},\ }\href
  {https://doi.org/10.1103/PhysRevB.111.035115} {\bibfield  {journal} {\bibinfo
   {journal} {Phys. Rev. B}\ }\textbf {\bibinfo {volume} {111}},\ \bibinfo
  {pages} {035115} (\bibinfo {year} {2025})}\BibitemShut {NoStop}%
\end{thebibliography}%

\end{document}